\documentclass[prb,nobalancelastpage,twocolumn,nofootinbib,superscriptaddress,showpacs]{revtex4-1}

\usepackage{color,amsthm,amsmath,amsfonts,graphicx,bm}
\usepackage{bm}
\usepackage{bm,eufrak}
\usepackage[export]{adjustbox}
\usepackage{amsfonts}
\usepackage{amssymb}
\usepackage{amsmath,mathrsfs}
\usepackage{epsfig}
\usepackage{graphicx}
\usepackage{enumerate}
\usepackage{multirow}
\usepackage{verbatim}
\usepackage{subfigure}
\usepackage{bbold}
\usepackage{soul}
\usepackage{scrextend}
\usepackage{tikz}
\usepackage{comment}
\allowdisplaybreaks

\newcommand{\ket}[1]{\vert#1\rangle}
\newcommand{\bra}[1]{\langle#1\vert}

\newcommand{\mr}[1]{\mathrm{#1}}
\newcommand{\dg}{\dagger}

\newcommand{\mb}{\mathbf}

\newcounter{notes}
\stepcounter{notes}

\DeclareFontFamily{OT1}{pzc}{}
\DeclareFontShape{OT1}{pzc}{m}{it}{<-> s * [1.10] pzcmi7t}{}
\DeclareMathAlphabet{\mathpzc}{OT1}{pzc}{m}{it}

\begin{document}

\title{Classification of symmetry-protected topological phases in two-dimensional many-body localized systems}

\date{\today}

\author{Joey Li}
\affiliation{Rudolf Peierls Centre for Theoretical Physics, Oxford, 1 Keble Road, OX1 3NP, United Kingdom.}
\affiliation{Department of Physics, University of Illinois at Urbana-Champaign, Urbana, Illinois 61801, USA.}

\author{Amos Chan}
\affiliation{Rudolf Peierls Centre for Theoretical Physics, Oxford, 1 Keble Road, OX1 3NP, United Kingdom.}
\affiliation{Physics Department, Princeton University, Princeton, New Jersey 08544, USA.}

\author{Thorsten B. Wahl}
\affiliation{Rudolf Peierls Centre for Theoretical Physics, Oxford, 1 Keble Road, OX1 3NP, United Kingdom.}
\affiliation{DAMTP, University of Cambridge, Wilberforce Road, Cambridge, CB3 0WA, United Kingdom.}

\begin{abstract}
We use low-depth quantum circuits, a specific type of tensor networks, to classify two-dimensional symmetry-protected topological many-body localized phases. For (anti-)unitary on-site symmetries we show that the (generalized) third cohomology class of the symmetry group is a topological invariant; however our approach leaves room for the existence of additional topological indices. 
 We argue that our classification applies to quasi-periodic systems in two dimensions and systems with true random disorder within times which scale superexponentially with the inverse interaction strength. Our technique might be adapted to supply arguments suggesting the same classification for two-dimensional symmetry-protected topological ground states with a rigorous proof.

\end{abstract}

\maketitle


\section{Introduction} 

Many-body localization (MBL)~\cite{AltmanReview,Luitz2017,Abanin2017,ImbrieLIOMreview2017,Alet2017} occurs in isolated strongly disordered systems and is characterized by a lack of thermalization. This phenomenon was first conjectured by Anderson in 1958 as an interacting analogue of Anderson localization~\cite{anderson1958absence}. Theoretical support was lacking until less than fifteen years ago, when perturbation theory analyses~\cite{gornyi2005interacting,basko2006metal}, various numerical studies~\cite{oganesyan2007localization,znidaric2008many,pal2010mb,Bardason2012} and a rigorous proof~\cite{imbrie2016many} put the phenomenon in one-dimensional lattice systems on a rigorous footing. In recent years, MBL was also observed in experiments of one-dimensional ultracold atomic gases~\cite{Schreiber842,Lukin2018} and chains of trapped ions~\cite{Smith_MBL}, superconducting qubits~\cite{Roushan2017} and NV-centers~\cite{Choi2017}. Approaches to realizing MBL in solid state systems are currently being pursued~\cite{Ovadia2015,Silevitch2017,Wei2018}. 

In higher dimensions, truly randomly disordered systems have been suggested to thermalize for arbitrarily large disorder via an avalanche effect due to rare regions~\cite{deRoeck2017Stability}, though assumptions underlying this argument have been contested~\cite{Altman2018stability}. Furthermore, the avalanche effect is expected to take place on very long time scales~\cite{chandran2016higherD}, at least in the limit of small interaction strengths~\cite{Gopalakrishnan2019}. This would reconcile the avalanche scenario with very recent ultracold gas experiments, where two-dimensional MBL is observed~\cite{Choi1547,2D_quantum_bath}. The notion of MBL-like behavior on experimental time scales has since been supported by theoretical studies~\cite{Thomson2018,2DMBL,Kennes2018,Bertoli2019,DeTomasi2019,Theveniaut2019,Kshetrimayum2019,Geissler2019}, with recent progress in tensor network methods~\cite{Paeckel2019,Kloss2020,Nieter2020} raising hopes for further insights in the near future. Quasi-periodic potentials in two dimensions lack rare regions and might thus give rise to a stable MBL phase~\cite{bordia2017quasiperiodic2D}.

MBL systems are potentially technologically relevant for the storage and manipulation of quantum information~\cite{2013Bauer_Nayak,Huse2013LPQO,Chandran2014SPT,2015Potter}: In one dimension, MBL systems with on-site symmetries are able to topologically protect qubits from decoherence caused by local noise at finite energy density~\cite{bahri2015localization,Goihl2020}. Two dimensional MBL-like systems may display a similar robustness and furthermore be used to manipulate the stored quantum information~\cite{Parameswaran2018}. 

One-dimensional MBL systems with an (anti-)unitary on-site symmetry can be classified into different symmetry-protected topological (SPT) MBL phases~\cite{Thorsten,1DSPTMBL}. The different topological classes can be labeled by the elements of the (generalized) second cohomology group of the symmetry group. Note that the symmetry group must be abelian to be compatible with a stable MBL phase~\cite{2016Potter_Vasseur}. In two dimensions, the expectation is thus that SPT MBL phases are classified by the elements of the third cohomology group, similarly to SPT ground states in two dimensions~\cite{Chen2013}.

In this work, we use quantum circuits to carry out such a classification in two dimensions. Quantum circuits are a specific type of tensor networks~\cite{PerezGarcia2007,PEPS,verstraete2008matrix,Orus2014} and approximate the unitary diagonalizing the MBL Hamiltonian efficiently in one dimension, as indicated by numerical evidence and analytical considerations~\cite{Pollmann2016TNS,Wahl2017PRX}. Specifically, the error of the approximation decreases like an inverse polynomial of the computational time (and number of parameters of the approximation). The underlying reason is that all eigenstates of MBL systems fulfill the area law of entanglement~\cite{Friesdorf2015} and can thus be efficiently approximated by tensor network states (TNS)~\cite{MPS_faithful,Pekker2017MPO,Devakul2017,Yu2017,Pollmann2016TNS,Wahl2017PRX}. Under the above assumption on the error bound, it is possible to show rigorously that SPT MBL phases are robust to arbitrary symmetry-preserving perturbations and that topologically distinct phases cannot be connected without delocalizing the system~\cite{Thorsten,1DSPTMBL}. Furthermore, it follows that all eigenstates of SPT MBL systems have the same topological label as defined for ground states. Here, we use two-dimensional quantum circuits with four layers of unitaries to describe two-dimensional strongly disordered systems. If there is true MBL in two dimensions, our results will apply for all observation times. If instead the avalanche scenario is correct, as we argue below, our classification applies for observation times which are superexponential in the interaction strength for true random disorder. For quasi-periodic disorder, our classification is likely to hold for arbitrarily long observation times in either case.   

Concretely, we show that two-dimensional MBL phases invariant under a symmetry can be labeled by the elements of the third cohomology group of the symmetry group. However, we cannot rule out the existence of additional topological indices with our approach. Furthermore, we show that the topological labels we find are robust to symmetry-preserving perturbations and cannot be connected without destroying MBL-like behavior. Again, it follows that all eigenstates must have the same topological label. We anticipate that our two-dimensional quantum circuit approach might be adapted to carry out a rigorous classification of two-dimensional SPT ground states, which is currently an outstanding problem~\cite{ciracOpenProblems}. Note that our classification does not apply to topologically ordered MBL systems~\cite{Parameswaran2018}, as their Hamiltonians cannot be diagonalized by short-depth quantum circuits~\cite{topMBL}. 

This article is structured as follows: In Section~\ref{sec:SPTMBL} we give a more formal introduction to the theoretical description of MBL systems in one and two dimensions, their SPT phases and tensor networks. Section~\ref{sec:nontechnical} contains a non-technical summary of our results with the technical part provided in Sections~\ref{sec:main} (unitary on-site symmetries) and~\ref{sec:anti} (anti-unitary on-site symmetries). Section~\ref{sec:robustness} discusses the robustness of the obtained topological phases to symmetry-preserving perturbations and demonstrates that the only way of connecting topologically distinct MBL phases is by either breaking the symmetry or making the perturbation strong enough to destroy MBL-like behavior. In Section~\ref{sec:conclusion}, we summarize our results and present directions for future work. In the Appendix, we provide technical details on the interpretation of the elements of the second and third cohomology group in terms of projective and gerbal representations, respectively.

\section{Symmetry-protected topological many-body localized phases and tensor networks}~\label{sec:SPTMBL}


Here we briefly review the central ideas about many-body localization and symmetry-protected topological phases and introduce tensor network language.  Readers already familiar with these topics may easily skip this Section.  For a similar but slightly more complete review of SPT and MBL, see Section II of Ref.~\onlinecite{1DSPTMBL}.

\subsection{Many-body localization in one dimension}\label{sec:1DMBL}


Here, we briefly review MBL in one dimension before commenting on the two-dimensional case.  
The canonical model of strongly disordered Hamiltonians that exhibits MBL in one dimension is the random field Heisenberg model~\cite{oganesyan2007localization,pal2010mb},
\begin{equation}\label{eq:canonmodel}
    H = J \sum_{i=1}^{N-1} \mb{S}_i\cdot\mb{S}_{i+1} + \sum_{i=1}^N h_i S^z_i,
\end{equation}
where $J > 0$, and $h_i$ is sampled from a uniform distribution $[-W,W]$. \eqref{eq:canonmodel} displays a transition from the ergodic phase to the MBL phase as a function of the disorder strength controlled by $W$. Numerical studies indicate a phase transition at around $W_c \approx 3.5 J$~\cite{pal2010mb,Luitz2015}.

Below but close to the phase transition 
\eqref{eq:canonmodel} exhibits a mobility edge\cite{Luitz2015}: Eigenstates in an energy window in the middle of the spectrum are volume law entangled, while eigenstates outside of this window are area law entangled. For SPT phases we are interested in the fully many-body localized (FMBL) phase ($W\gtrsim 3.5 J$ for \eqref{eq:canonmodel}), where all eigenstates are area law entangled.  The FMBL phase is described by a complete set of local integrals of motion (LIOMs)\cite{Serbyn2013local, huse2014phenomenology} $\tau_i^z$. This remains true after adding small but \textit{non-zero} arbitrary local perturbations, even in the thermodynamic limit. Any resonances of distant spins with similar energies are captured by those LIOMs (which would in that case be particularly wide). We do not consider the case of resonances spreading across the whole system in the thermodynamic limit. In that case, there are volume law entangled eigenstates, which would correspond to a disorder strength below the phase transition point as defined above (where a mobility edge is present and the LIOM picture does not apply). Here we refer to the \textit{actual} MBL-to-thermal phase transition point in the thermodynamic limit, which might be significantly higher than the value quoted above due to finite size effects~\cite{Doggen2018,Abanin2019}. (However, the effect of rare regions on the transition point in the thermodynamic limit has also been questioned in one dimension~\cite{Goihl2019}. Moreover, MBL systems coupled to thermal baths have been argued to delocalize only if the latter take a finite fraction of the overall system size~\cite{Sparaciari2019}.)

LIOMs are local operators which commute with the Hamiltonian and with each other, and therefore form an emergent notion of integrability, 
\begin{align}
    [H,\tau_i^z] = [\tau_i^z, \tau_j^z] = 0 \label{eq:lbits}
\end{align}
for all $i, j = 1, 2, \ldots, N$. 
Hence, all eigenstates $|\psi_{l_1 l_2 \ldots l_N}\rangle$ of the Hamiltonian can be uniquely labeled by the expectation values (say $l_i = \pm 1$, also known as l-bits) of the corresponding $\tau_i^z$ operators. (Here we consider the case of spin-$1/2$ Hamiltonians, though the notion of LIOMs can be straightforwardly generalized to higher spin systems.) According to Eq.~\eqref{eq:lbits}, the LIOMs and the Hamiltonian can all be simultaneously diagonalized by a unitary $U$, that is,
\begin{align}
H &= U E U^\dg, \\
\tau_i^z &= U \sigma_i^z U^\dg.
\end{align}
Any Hamiltonian could be used to construct a commuting set of integrals of motion this way. The special feature of FMBL systems is that the unitary $U$ can be chosen such that the $\tau_i^z$ are \textit{local}, i.e., they have exponentially decaying support from site $i$. The corresponding decay length is known as the \textit{localization length} $\xi_i$. The corresponding unitary $U$ has been argued to be efficiently approximable by a short-depth quantum circuit with long gates~\cite{Wahl2017PRX,Thorsten}. The exact distribution of localization lengths $\xi_i$ for a given system size $N$ depends on the disorder realization. The probability of finding localization length within a range $[\xi,\xi+\Delta \xi]$ decays sharply with $\xi$~\cite{Abi2017}. For a system to be considered as FMBL, we have to assume that the probability that the largest localization length $\xi_\mr{max}$ is of order $\mathcal{O}(N)$ goes to zero in the limit $N \rightarrow \infty$ (otherwise, the system would be delocalized). Hence, we assume $\xi_\mr{max} \leq c N^\mu$ for a given disorder realization and model Hamiltonian (such as Eq.~\eqref{eq:canonmodel}) with constants $c > 0$ and $\mu \in [0,1)$.


\subsection{Many-body localization in higher dimensions}\label{sec:higherDMBL}

It is believed that in higher dimensions for true random disorder, regions with anomalously small disorder will eventually thermalize the entire system~\cite{deRoeck2017Stability,Gopalakrishnan2019}, although this picture has to be taken with care~\cite{Altman2018stability}. Regions with anomalously small disorder contain small thermal inclusions, i.e., local expectation values of all eigenstates look thermal in those regions. This phenomenon also arises in one dimension, and, in the above framework, implies a set of particularly wide LIOMs (with large localization lengths $\xi_i$). While in one dimension such a set of wide LIOMs can be stable, it is believed that in higher dimensions sufficiently large thermal regions cannot remain isolated, as they would gradually thermalize surrounding spins and thus grow via an avalanche effect until the whole system becomes thermal~\cite{Doggen2020}. For concreteness, let us consider a $d$-dimensional cubic lattice with $N^d$ spins described by the general Hamiltonian of Ref.~\onlinecite{Gopalakrishnan2019}
\begin{align}
H = \sum_i h_i O_i + J \sum_{i,j} \phi_{ij} P_{ij}.
\end{align}
$h_i$ are random fields chosen from a uniform distribution centered around zero. $\phi_{ij}$ are taken from the same distribution but have to be multiplied by a prefactor which decays at least exponentially as a function of the distance between sites $i$ and $j$. $O_i$ and $P_{ij}$ are single-site and two-site operators, respectively. Those acting on the same site do not commute with each other. $J \geq 0$ acts as a tuning parameter inducing delocalization if it becomes sufficiently large. 
The probability of having a thermal inclusion of sufficient size to initiate an avalanche has been estimated as~\cite{Gopalakrishnan2019}
\begin{align}
p(N,J) \sim N^d \exp(\log^3(J)) \label{eq:probab_N}
\end{align}
where $J < 1$. For a finite system, there is thus a crossover at $J_c$ given by $p(N,J) \sim 1$, i.e., $\log(J_c) \sim - (d \log N)^{1/3}$. In the infinite system size limit, we would thus have $J_c = 0$. However, the avalanche effect is very slow, and it takes time $t \gtrsim \exp(-R \log(J))$ (with $J < 1$) for an initial thermal inclusion to expand to size $R$ from a comparatively small size. Hence, according to Eq.~\eqref{eq:probab_N}, the probability that a typical spin will have been absorbed by such an avalanche after time $t$ is (setting $N \sim R$)
\begin{align}
p(t,J) \sim \left[ -\log(t) / \log(J) \right]^d \exp(\log^3(J))
\end{align}
for $t > 1/J$. $p(t,J) \sim 1$ gives the time scale for thermalization as 
\begin{align}
t &\sim \exp\left[-\log(J) \exp\left(-\frac{1}{d}\log^3(J)\right)\right] \notag \\
&= (1/J)^{(1/J)^{\log^2(1/J)/d}}, \label{eq:time-scale}
\end{align}
which grows rapidly as $J \rightarrow 0$. Note that $J$ has to be sufficiently small to prevent delocalization via resonances. The avalanche effect is thus likely too slow to be seen experimentally, and the MBL-to-thermal transition observed in two-dimensional systems with true random disorder~\cite{Choi1547,2D_quantum_bath} might be due to similar effects as in one dimension. In the following, we refer to Hamiltonians in higher dimensions as FMBL if their only mechanism of thermalization is the above avalanche effect, and if this remains true after arbitrary infinitesimal perturbations. Note that quasi-periodic systems likely do not display the avalanche effect due to the lack of rare regions. Systems with strong quasi-periodic disorder might thus never thermalize (and we also denote them as FMBL).


\subsection{Symmetry-protected topological phases}\label{sec:SPT}

Quantum phases typically have to do with the ground states of gapped systems.  A topological phase consists of the set of gapped local Hamiltonians that can be continuously deformed into each other without closing the energy gap, or equivalently, whose ground states can be evolved into each other with short-ranged quantum circuits with depth constant in the system size.  A symmetry-protected topological phase is defined in the same way with the added constraint that all Hamiltonians along the connecting path must be invariant under the symmetry.

For MBL systems, we are interested in all eigenstates rather than only ground states, since the properties of the eigenstates constrain the dynamics of the system.  
We say that two FMBL Hamiltonians $H_0$ and $H_1$ are in the same MBL SPT phase if there exists a path $H(\lambda)$ such that $H_0 = H(0)$ and $H_1 = H(1)$ and for all $\lambda \in [0,1]$, $H(\lambda)$ preserves the symmetry and is FMBL~\cite{1DSPTMBL}.

Examples of models displaying SPT MBL can be found in Refs.~\onlinecite{bahri2015localization,Kuno2019,Decker2020,Orito2020}.
In the case of on-site symmetries, it was originally conjectured that the ground state SPT phases of $d$-dimensional spin systems are labeled by the $(d+1)$th cohomology group of the symmetry group~\cite{Chen2013}; however, it has been found that for $d \geq 3$ this classification has to be extended~\cite{Xiong2018,Gaiotto2019}. These classifications have also been proposed for the MBL case\cite{chandran2014many}.  In $d=1$ it was shown that the SPT phases are indeed labeled by the elements of the second cohomology group in the ground state\cite{2011Schuch,2011Chen} and MBL\cite{1DSPTMBL} cases. In this paper, we demonstrate that two-dimensional MBL phases with a symmetry can be classified by the elements of the third cohomology group. However, we do not show that MBL Hamiltonians corresponding to the same third cohomology class can be continuously connected without destroying FMBL, i.e., our classification might be incomplete. 


\subsection{Tensor networks}\label{sec:tn}

Tensor networks and the associated diagrammatic formulation are powerful tools for both analytical\cite{2011Chen, 2011Schuch} and numerical\cite{white1992a, white1992b} studies of quantum many body physics. A tensor is an $n$-dimensional array of (complex) numbers, and is diagrammatically represented by a geometric shape with indices represented by outgoing legs.  For example, 
\begin{equation}
 A_{ijk}\; = \;\;\includegraphics[width = 0.25\linewidth,valign=c]{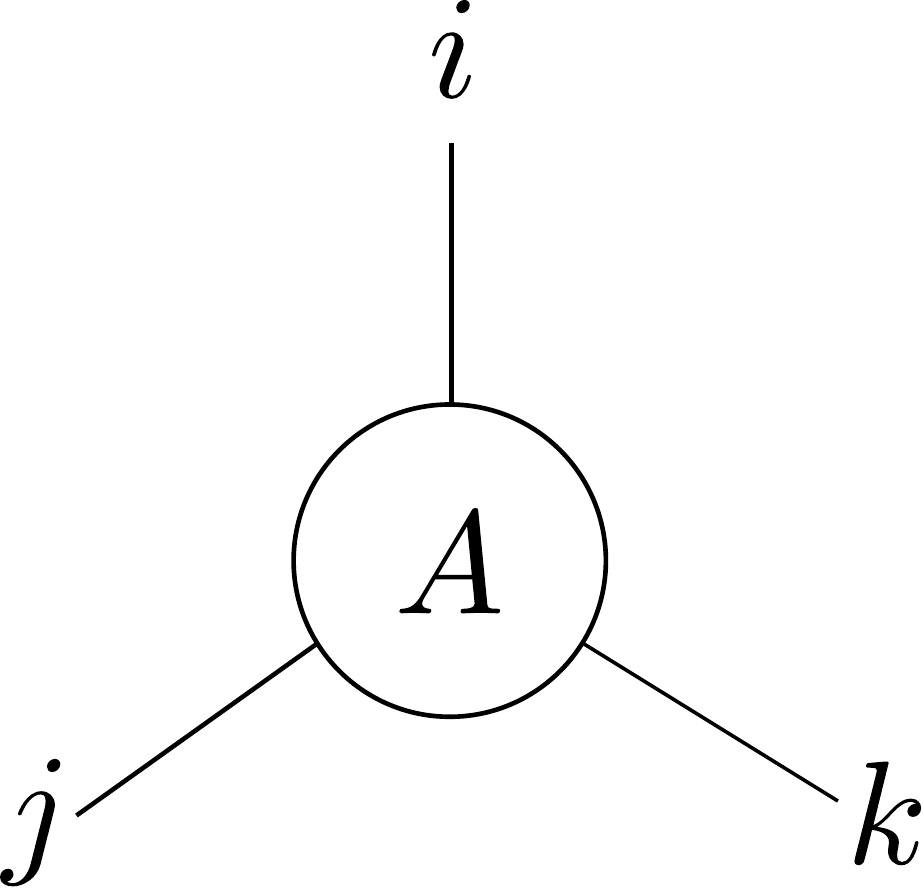} \;.
\end{equation} 
A contraction between different indices of (a single or multiple) tensor(s) is represented by connecting two corresponding legs, e.g.
\begin{equation}
\sum_{ij} A_{ijk}B_{jklm} \; = \;\;\includegraphics[width = 0.44\linewidth,valign=c]{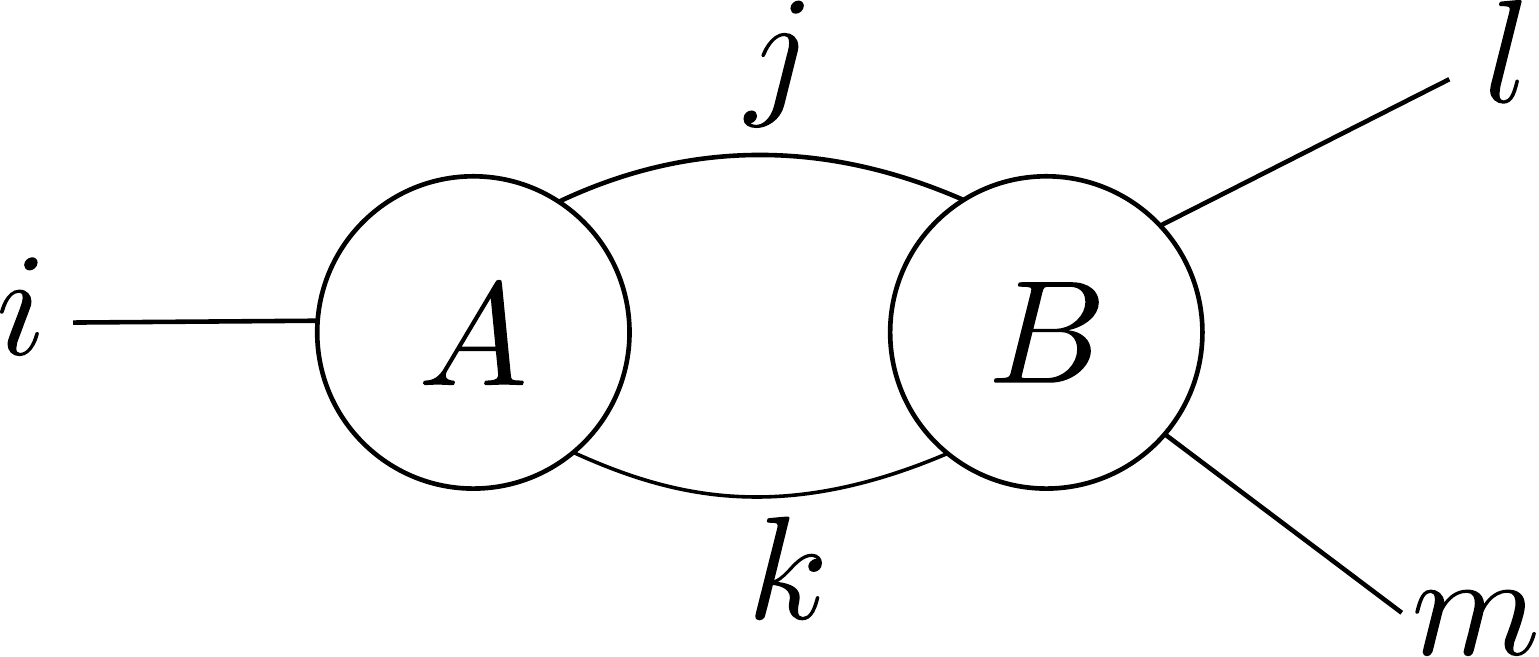}\;.
\end{equation}
Tensors can be blocked or grouped together to form a single tensor. The legs of a given tensor can be combined or split through reshaping. These operations are illustrated as follows,
\begin{equation}
  \includegraphics[width = 0.3\linewidth,valign=c]{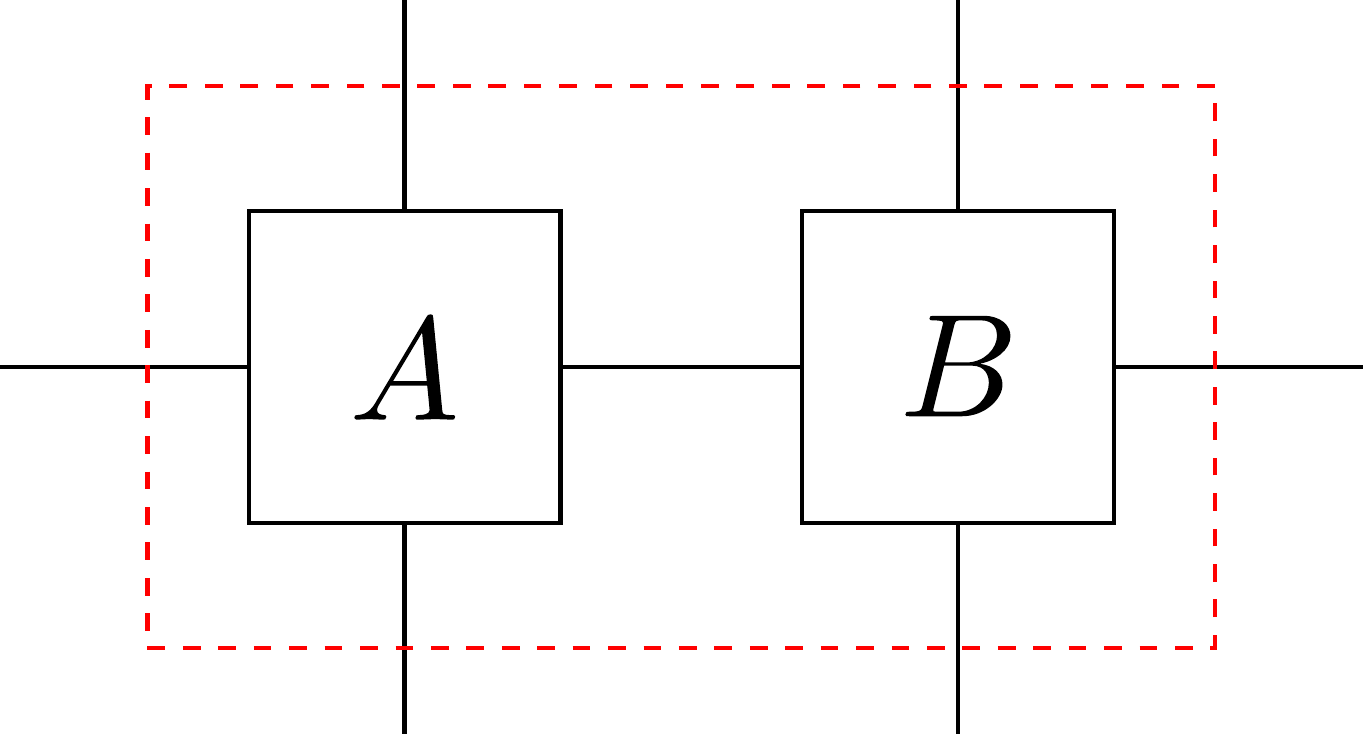}\;\;\equiv\;\; \includegraphics[width = 0.23\linewidth,valign=c]{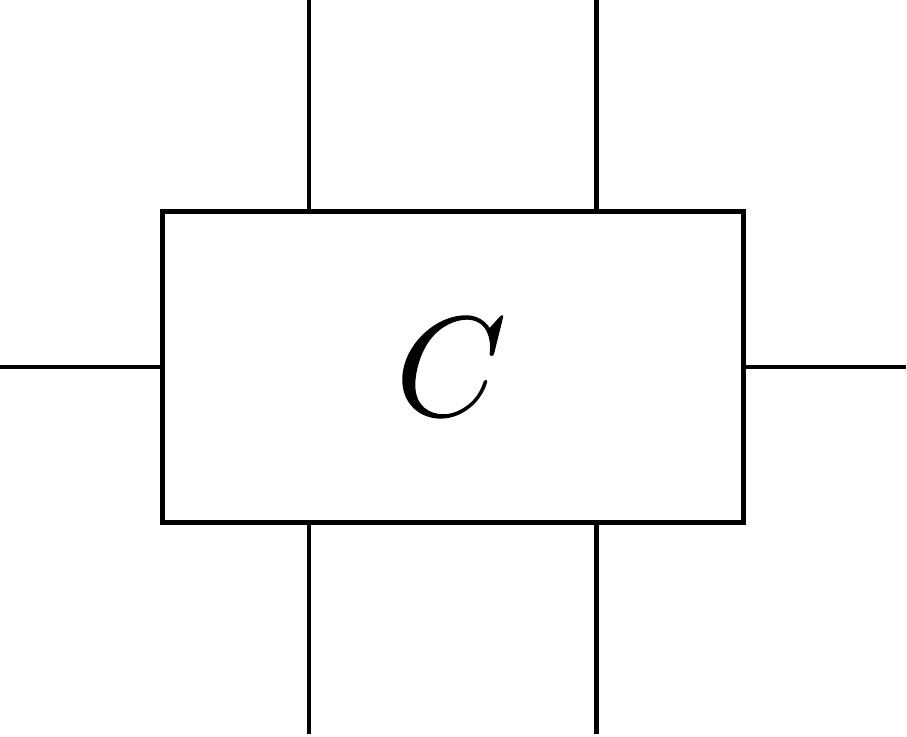}\;\;\equiv\;\;\includegraphics[width = 0.2\linewidth,valign=c]{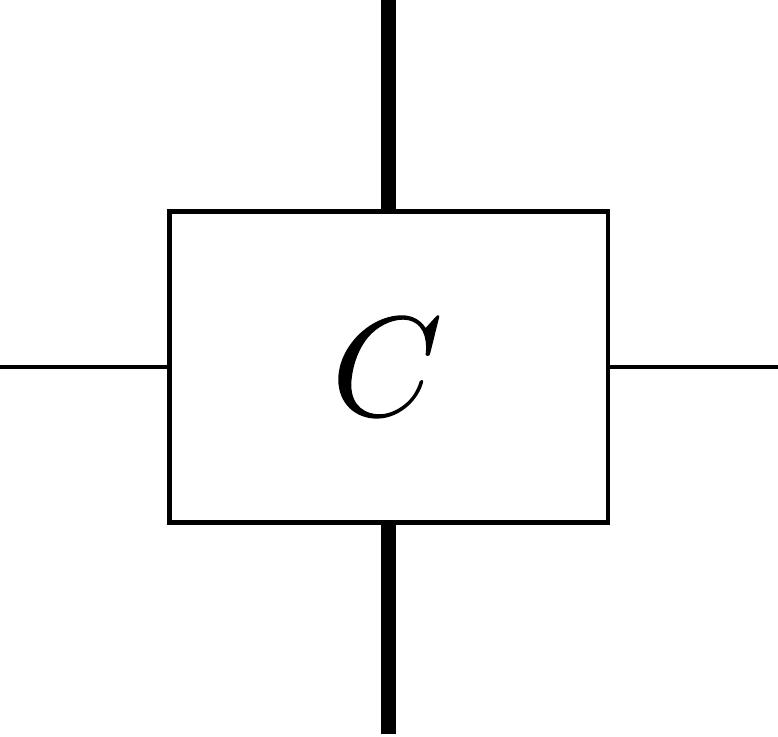}\;.
\end{equation}
The tensor product of two tensors is represented by placing two tensors together, e.g.
\begin{equation}
 \includegraphics[width = 0.22\linewidth,valign=c]{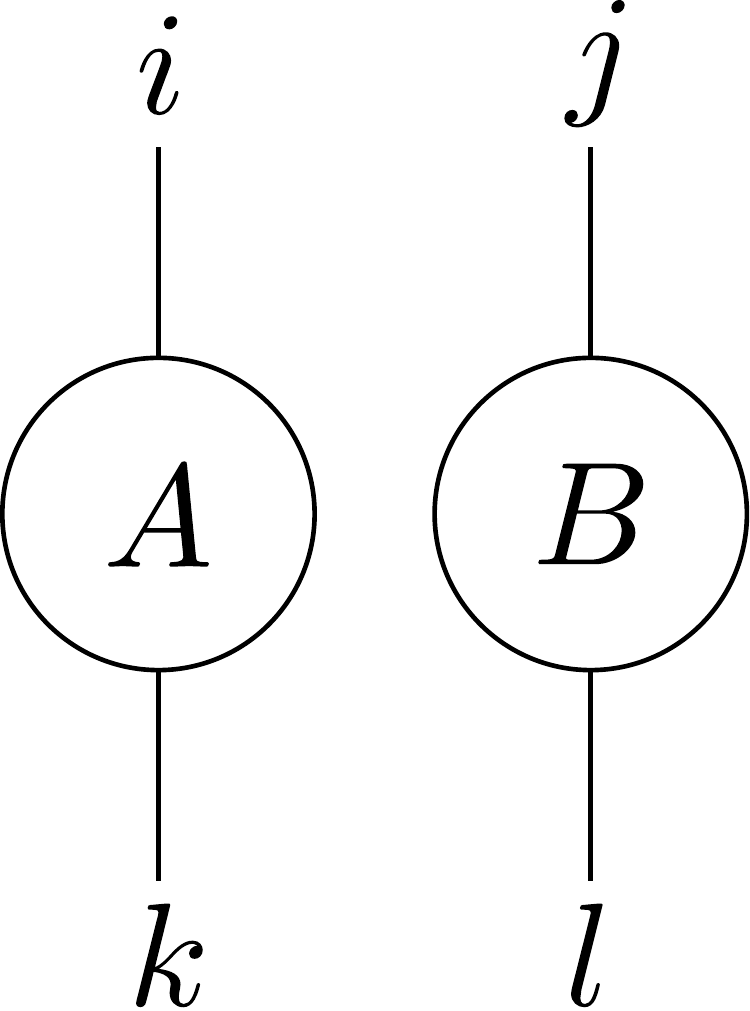}\;\; = \;\;\includegraphics[width = 0.18\linewidth,valign=c]{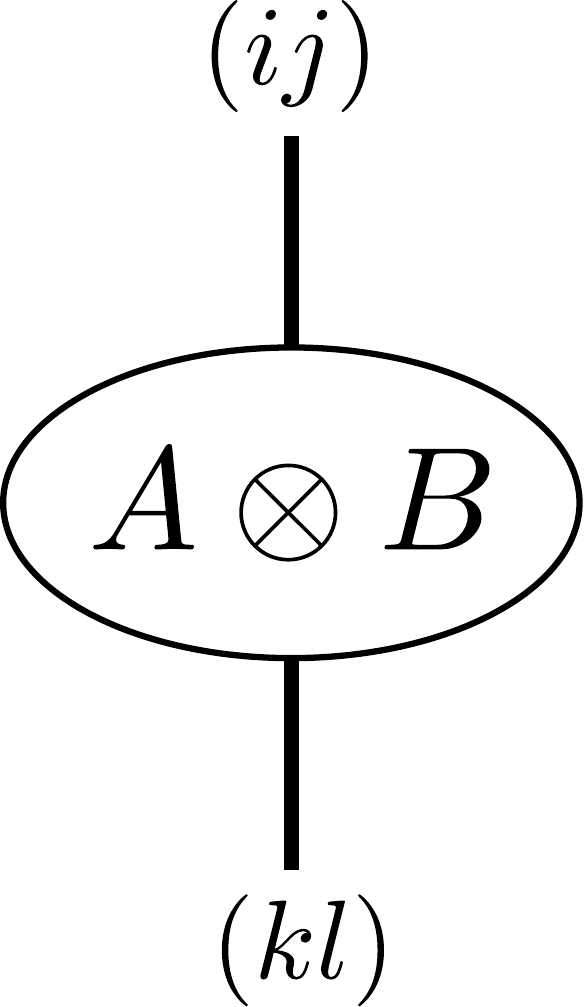}\;.
\end{equation}
The trace operation is a contraction of two legs of the same tensor, e.g.
\begin{equation}
\mathrm{Tr}(A)\; = \;\;\includegraphics[width = 0.2\linewidth,valign=c]{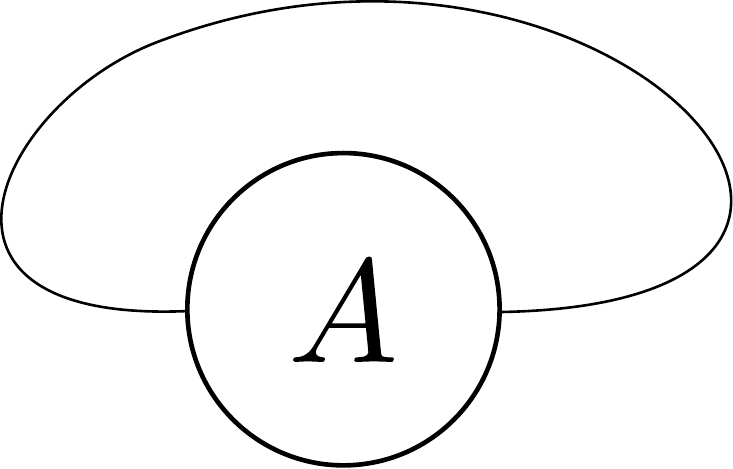}\;.
\end{equation}

A commonly cited problem in quantum many-body physics is the exponential increase of the dimension of the Hilbert space with the system size.  
However, many physically interesting states, such as the ground states of gapped systems, have area-law entanglement and lie in a small region of the Hilbert space, which only scales polynomially with the system size, and hence are expressible in terms of tensor networks.

A classic example is the matrix product state (MPS) in one dimension.  The state of an $N$-site spin chain,
\begin{equation}
    |\psi\rangle = \sum_{i_1 \cdots i_N} \psi_{i_1 \cdots i_N} |i_1 i_2 i_3 \cdots i_N\rangle \; ,
\end{equation}
can be written in the form of an MPS,
\begin{equation}
    |\psi\rangle =\sum_{i_1 \cdots i_N} \mathrm{Tr}\Big(A_{i_1}^{(1)}A_{i_2}^{(2)}A_{i_3}^{(3)}\cdots A_{i_N}^{(N)}\Big)  |i_1 i_2 i_3 \cdots i_N\rangle 
\end{equation}
if we decompose $\psi_{i_1 \cdots i_N}$ as
\begin{equation}
 \includegraphics[width = 0.3\linewidth,valign=c]{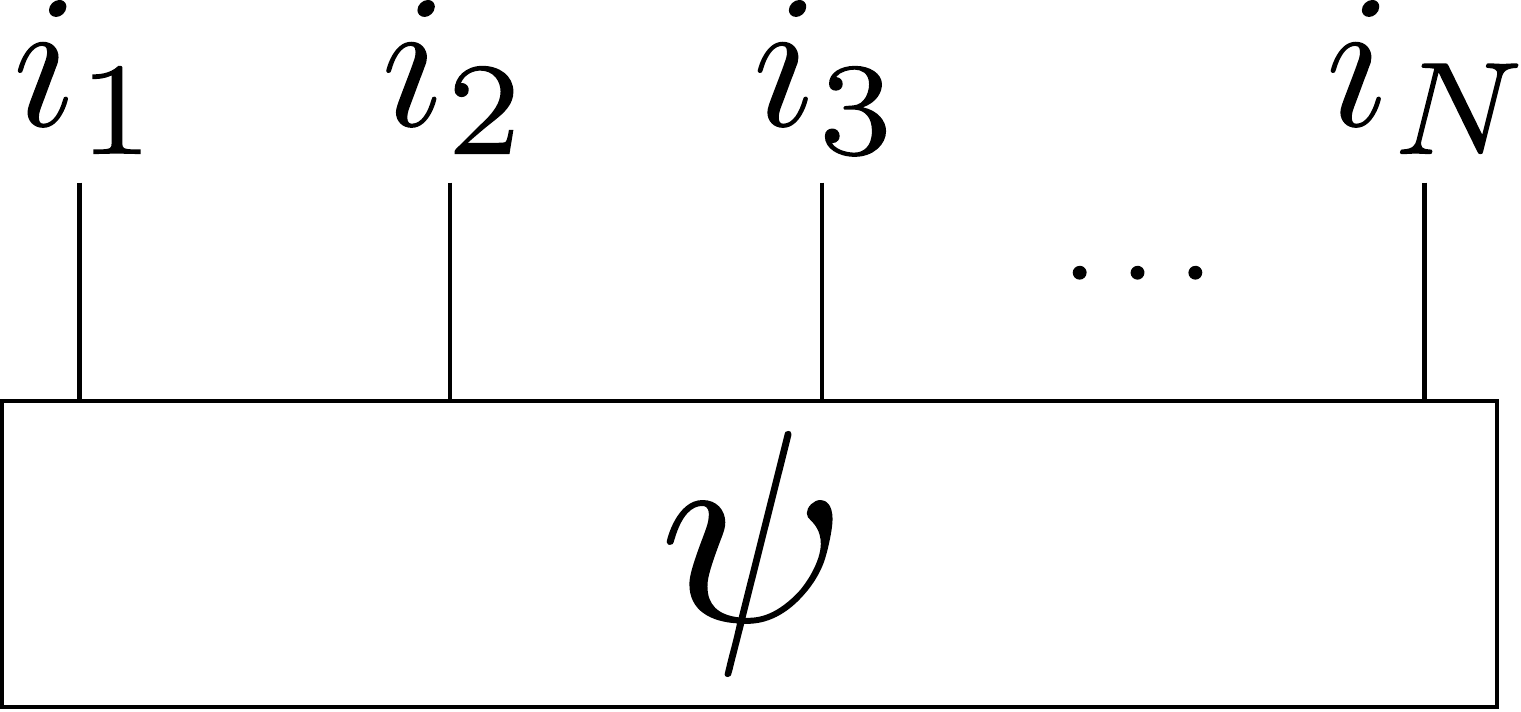}\;= \;\includegraphics[width = 0.5\linewidth,valign=c]{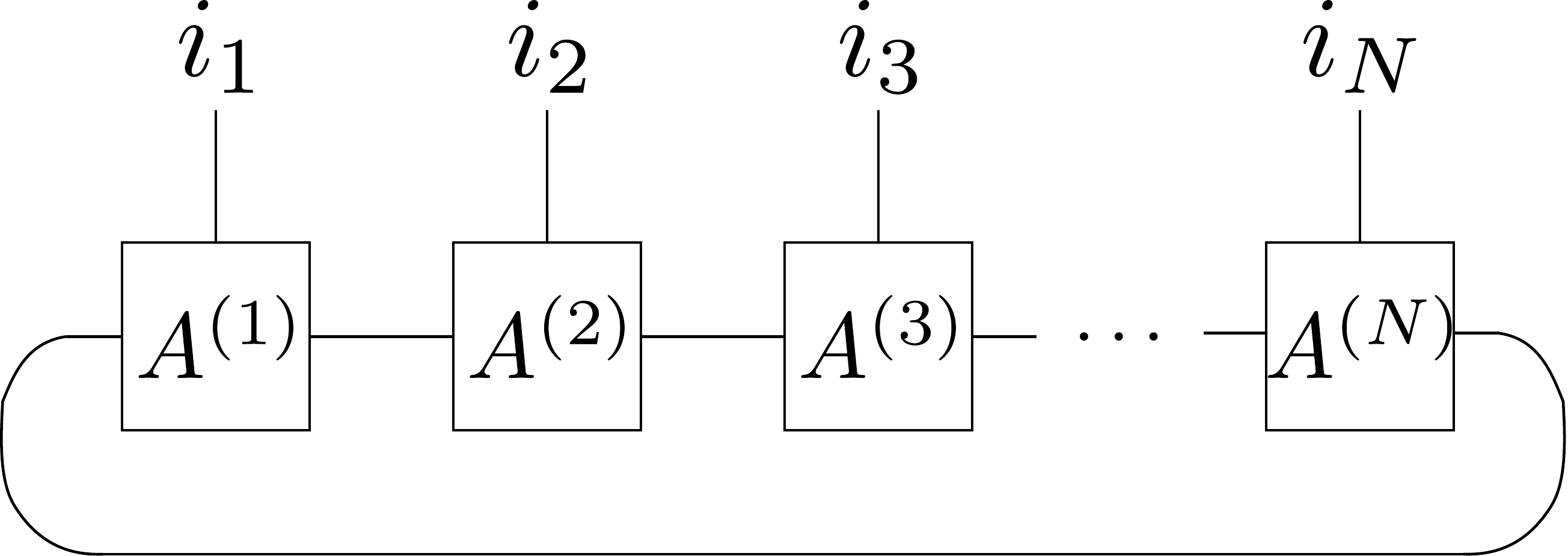}\;.
\end{equation}
Such a decomposition can always be found using, say, a singular value decomposition (SVD). 
This procedure is not always useful since the maximum dimension of the legs of $A^{(n)}$, or the ``bond dimension'', can be exponentially large.  
However, for area-law entangled states, there exist accurate MPS representations with small bond dimensions. Furthermore, in a few cases such as the AKLT model~\cite{AKLT}, exact MPS representations can be found with fixed bond dimensions.
Another example of tensor network states is projected entangled pair states (PEPS). PEPS are $d$-dimensional versions of MPS, with each site represented by a tensor with one ``physical" leg and $2d$ bond legs on a square lattice.  In this paper we will work mostly with unitary quantum circuits (or simply ``quantum circuits"), which is a sequence of unitary quantum gates and can be diagrammatically represented in the tensor network notation.

\section{Non-technical summary of results}\label{sec:nontechnical}

\subsection{Underlying assumptions}\label{sec:assumptions}


Here we give an overview of the main ideas and results. We consider a strongly disordered FMBL Hamiltonian $H$ defined on an $N \times N$ square lattice with periodic boundary conditions. 
Furthermore, we assume that the system is invariant under an on-site symmetry $\mathpzc{v}_g$ with abelian symmetry group $G \ni g$, that is 
\begin{align}
    H = \mathpzc{v}_g^{\otimes N^2} H (\mathpzc{v}_g^\dagger)^{\otimes N^2} . \label{eq:Ham_symmetry}
\end{align}
$\mathpzc{v}_g$ forms a representation of the group, i.e., $\mathpzc{v}_g \mathpzc{v}_h = \mathpzc{v}_{gh}$. For our derivation, we assume that the symmetry group $G$ is abelian. However, non-abelian symmetry groups have been argued to be inconsistent with FMBL even in one dimension~\cite{2016Potter_Vasseur}: The system either spontaneously breaks the symmetry (possibly still keeping an abelian sub-symmetry) or is delocalized. 
Abelian symmetries do not protect degeneracies. We can thus assume that all exact degeneracies have been lifted by a small perturbation. In that case, it can be shown (see Sec.~\ref{sec:2dmblonsite}) that the unitary $U$ diagonalizing the Hamiltonian ($H = U E U^\dg$), fulfills 
\begin{equation}\label{eq:symmetry}
    \mathpzc{v}_g^{\otimes N^2} U = U\Theta_g,
\end{equation}
where $\Theta_g$ is a diagonal matrix where each diagonal element is a complex number of magnitude $1$.

We now consider local unitaries $\tilde U$ of the type described in Sec.~\ref{sec:1DMBL}, i.e., the quantities $\tilde \tau_i^z := \tilde U \sigma_i^z \tilde U^\dg$ have exponentially decaying non-trivial matrix elements, where the corresponding decay lengths $\xi_i$ satisfy the bound $\xi_i \leq c N^\mu$ for some $c > 0, 0 \leq \mu < 1$. Let us focus on the unitary $\tilde U$ which minimizes the quantity $\sum_i \|[H, \tilde \tau_i^z]\|_\mr{op}$. (The $\tilde \tau_i^z$ commute with each other by construction.) For truly randomly disordered systems if the avalanche scenario is wrong, and most likely for systems with strong quasi-periodic disorder in general, the minimum of this figure of merit will be zero, i.e., $\tilde U$ exactly diagonalizes the Hamiltonian. For true random disorder and if the avalanche scenario is correct, $\tilde U$ encodes approximate eigenstates which delocalize under time evolution with $H$ on the time scale given by Eq.~\eqref{eq:time-scale}. In the following, we will analyse the topological properties of these approximate eigenstates. Their topological features will be stable to small (symmetry-preserving) perturbations, but as those are only approximate eigenstates, we have to keep in mind that they would lose their topological properties after times of order Eq.~\eqref{eq:time-scale} due to delocalization. 

\begin{figure}[t]
    \centering
    \includegraphics[width=0.8\linewidth]{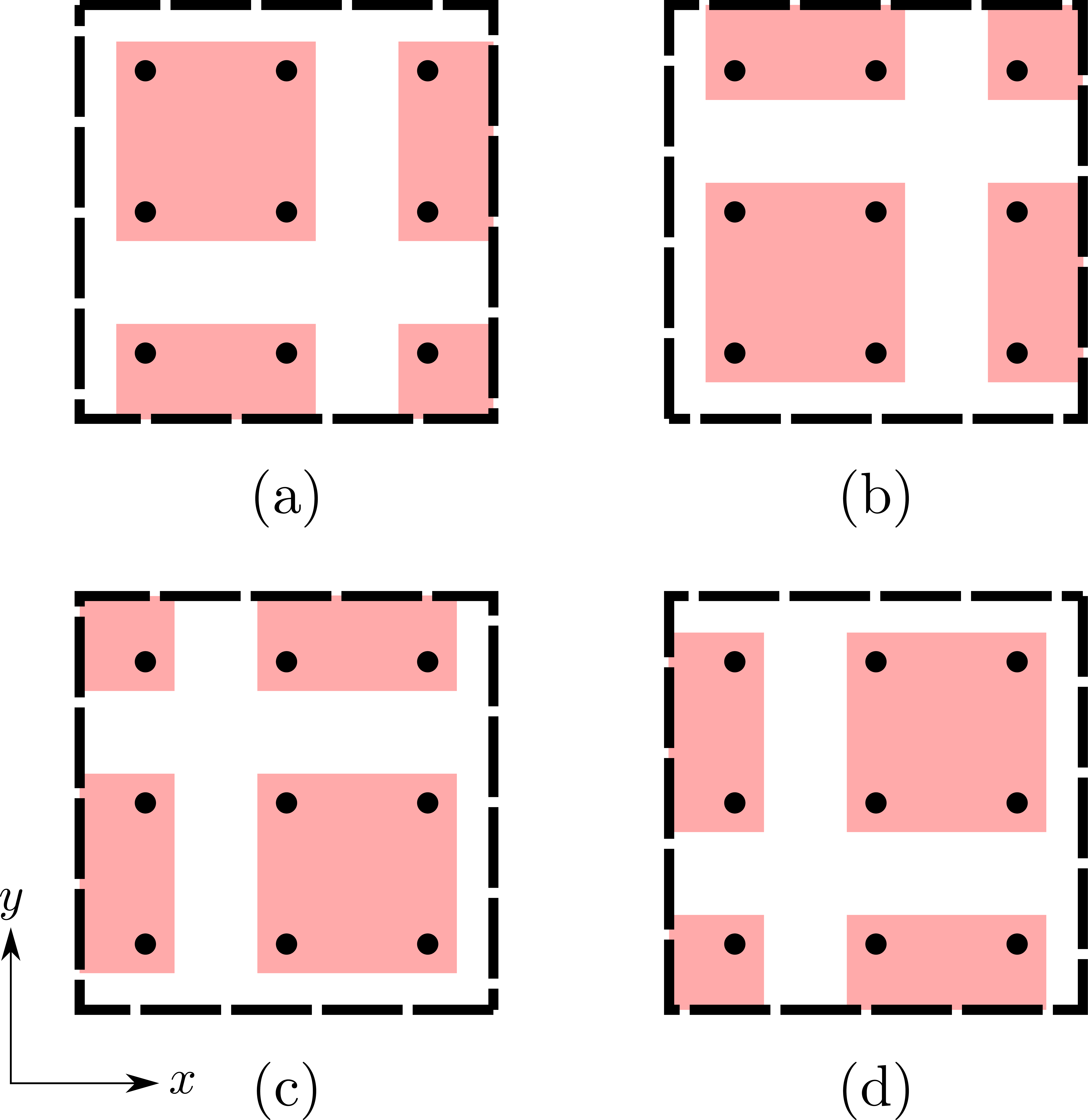} 
    \caption{
    An illustration of the 4-layer quantum circuit approximating the unitary $\tilde U$. The $xy$-plane is parallel to the plane where the sites of the system are located. The unitaries are stacked from bottom to top in the order (a), (b), (c), (d) (parallel to the $z$-axis). (a) represents the top view of the first layer, (b) of the second layer, and so on. A dot represents a group of $\frac{\ell}{2} \times \frac{\ell}{2}$ sites. A red box represents a unitary. The quantum circuit periodically extends beyond the region defined by the dashed lines.
    }
    \label{fig:2dqc}
\end{figure}

Furthermore, we assume that $\tilde U$ can be efficiently approximated by a four-layer quantum circuit $U'$
of the form of Fig.~\ref{fig:2dqc}, where each unitary acts on plaquettes of $\ell\times\ell$ sites. 
For that, we have to require that $\ell = c' N^{\nu}$ with $c' > 0$ and $\mu < \nu < 1$ such that the range of all unitaries is much larger than the longest localization length $\xi_\mr{max}$ in the limit of large $N$~\cite{Wahl2017PRX}. With increasing $N$, the quantum circuit $U'$ thus approximates $\tilde U$ with arbitrary accuracy~\cite{1DSPTMBL}. In order to describe the topological properties of MBL systems within time scales of order Eq.~\eqref{eq:time-scale}, it thus suffices to characterize quantum circuits of the type $U'$. 


Our approach towards the classification of SPT phases differs from the one more commonly found in the literature, where quantum circuits are assumed to have fixed gate length and whose depth is variable, albeit independent of the system size. In contrast, we (i) keep the number of layers constant at four and have a flexible gate length, which (ii) is allowed to grow sublinearly with the system size. The reasons for this modified approach are as follows: (i) MBL systems with true random disorder contain regions of anomalously small disorder. The localization length $\xi_i$ of a LIOM located in the center of such an anomalous region has to be of the order of its size. Since such a ``thermal puddle'' is featureless, the quantum circuit should have of the order of $2^{\xi_i^2}$ parameters in that region to be able to diagonalize the Hamiltonian with any reasonable accuracy~\cite{Wahl2017PRX}. To that end, one could increase the depth of the quantum circuit exponentially with 
$\xi_i^2$, or the length $\ell$ of its gates linearly with $\xi_i$. Hence, a quantum circuit with long gates is the more natural choice for MBL systems. There is no need to increase the depth of the quantum circuit as well, cf. Ref.~\onlinecite{Wahl2017PRX}. (ii) The gate length has to increase with the system size, since the maximum $\xi_i$ does: In the thermodynamic limit, there will be anomalous regions of arbitrarily large size, since there is a finite probability for them to occur. Thus, $\xi_\mr{max} = \max_i (\xi_i)$ diverges in the thermodynamic limit. Therefore, $\ell$ also has to grow with the system size in order to allow for a correct global description of any reasonable accuracy. 

The classification we derive below is based on the question whether such quantum circuits with a diverging gate length can be continuously connected. Consequently, our results also apply to all more restrictive sets of quantum circuits: The central result of our work is that for given gate length $\ell$ the whole set $\mathcal S$ of four-layer quantum circuits with a symmetry decomposes into disconnected sets $\mathcal S_a$ given by the third cohomology class $a$ of the symmetry group, $S = \cup_a S_a$. A quantum circuit contained in $\mathcal S_a$ cannot be continuously connected with a quantum circuit contained in $\mathcal S_b$ for $a \neq b$. Now consider a more restrictive notion of quantum circuits $\mathcal R$ contained in $\mathcal S$, $\mathcal R \subset \mathcal S$. As long as this more restrictive set contains a representative of each cohomology class $a$, the same decomposition has to apply, i.e., $\mathcal R = \cup_a \mathcal R_a$, $\mathcal R_a \neq \{\}$ and $\mathcal R_a \subseteq \mathcal S_a$. The last relation implies likewise that quantum circuits contained in $\mathcal R_a$ and $\mathcal R_b$ cannot be continuously connected for $a \neq b$. An example of such a more restrictive set $\mathcal R$ is the commonly used quantum circuits with a large but fixed number of layers and (small) fixed gate length~\cite{2013Bauer_Nayak}: Those quantum circuits have strict short-range correlations and can thus be approximated with arbitrarily small error by our ansatz if $\ell$ is sufficiently large. (In one dimension, multi-layer quantum circuits can even be written exactly as two-layer long-gate ones.)  Furthermore, those more restrictive quantum circuits have a representative in each cohomology class: Such a representative is the finite-depth, finite-gate-length quantum circuit which maps a product state to a ground state in the corresponding SPT phase. 

The quantum circuit $U'$ 
is the natural generalization of the two-layer quantum circuit with long gates used in one dimension to represent MBL systems~\cite{Wahl2017PRX,Thorsten,1DSPTMBL}: It consists of parallel one-dimensional two-layer quantum circuits, which are themselves coupled with each other in a two-layer quantum circuit structure,
\begin{equation}
    \label{eq:Udef}
    \includegraphics[width=0.9\linewidth,valign=c]{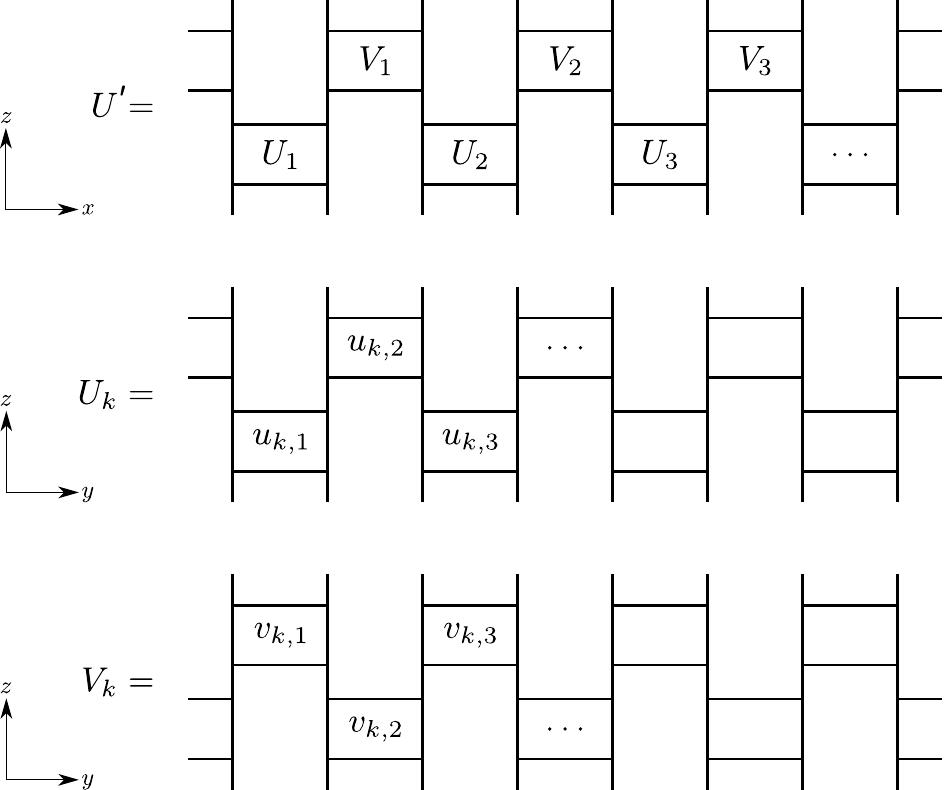}  \; .
\end{equation}
Here, we blocked together sites as in Fig.~\ref{fig:2dqc}, i.e., each tensor leg corresponds to $\frac{\ell}{2} \times \frac{\ell}{2}$ sites. The unitaries of $U_k$ are located in the first two layers of Fig.~\ref{fig:2dqc} (i.e, Fig.~\ref{fig:2dqc}a,b), the unitaries of $V_k$ in the second two layers (Fig.~\ref{fig:2dqc}c,d). 
For the derivation below we assume that one-dimensional unitaries which encode states with strict short-range entanglement can be efficiently approximated by one-dimensional two-layer quantum circuits with long gates, which corresponds to the assumption that one-dimensional MBL systems can be efficiently approximated by such unitaries~\cite{Wahl2017PRX,Thorsten,1DSPTMBL}.

\subsection{Main results}

$U'$ (we will drop the prime symbol from now on) approximately fulfills Eq.~\eqref{eq:symmetry}, as it approximately diagonalizes the Hamiltonian $H$. 
It follows from Eqs.~\eqref{eq:symmetry} and \eqref{eq:Udef} that $\Theta_g$ can likewise be written as a four-layer quantum circuit (see Sec.~\ref{sec:thetaqc} for details), thus making Eq.~\eqref{eq:symmetry} an equality of two short-depth quantum circuits.

Next, we perform manipulations with the quantum circuits. We collapse the quantum circuits of Eq.~\eqref{eq:symmetry} along the $y$-direction, so that \eqref{eq:symmetry} becomes an equality of two one-dimensional quantum circuits, which are stretched out along the $x$-direction. One obtains
\begin{align}
  \includegraphics[width=0.9\linewidth,valign=c]{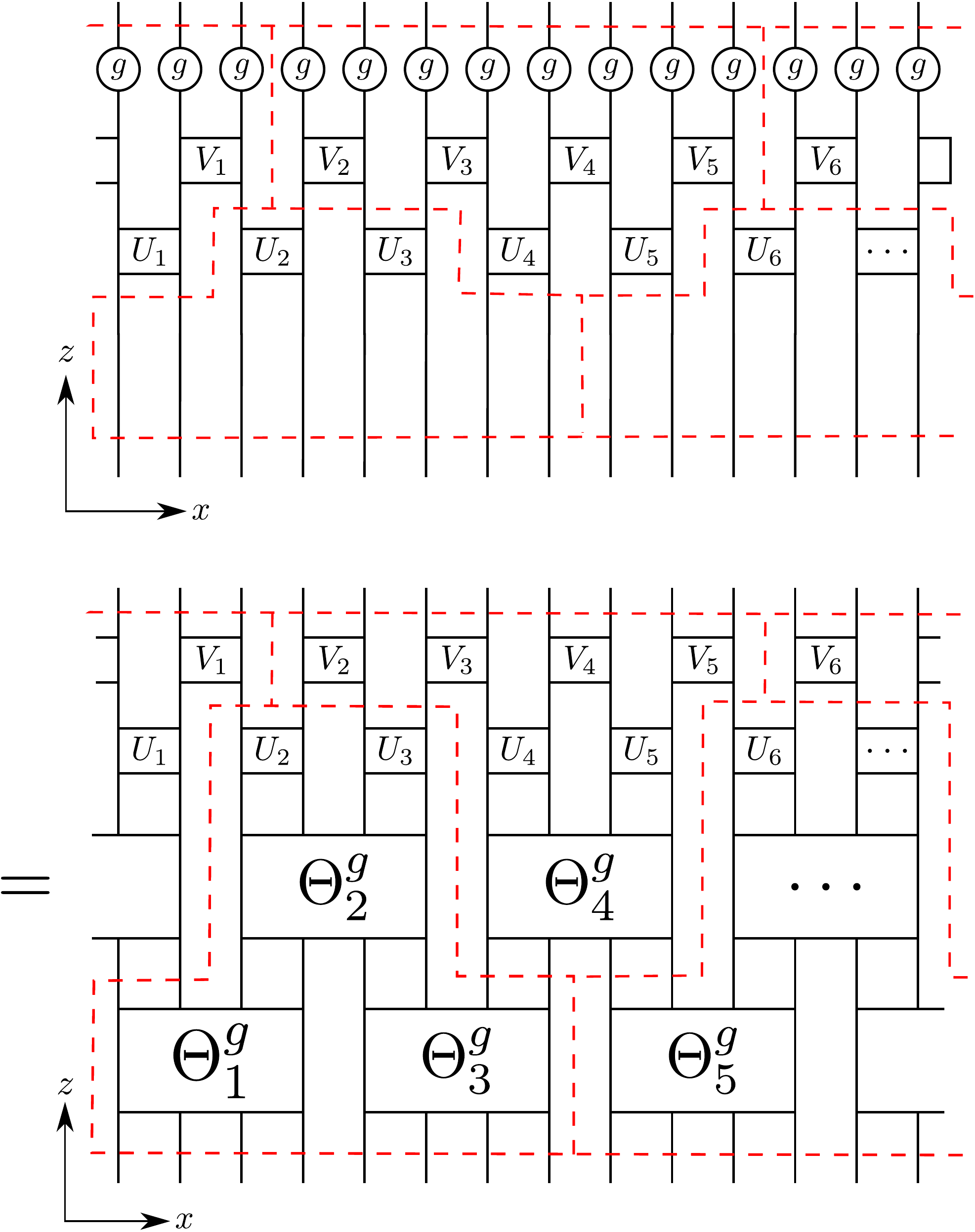}  \; , 
\end{align}
where $g$ represents $\mathpzc{v}_g^{\otimes N \ell/2}$, and the $\Theta^g_j$ are (diagonal) unitaries extended along the $y$-direction. They constitute the quantum circuit representation of $\Theta_g$. This equation is of the general form (see red dashed lines)
\begin{equation}
\begin{tikzpicture}[scale=1.1,baseline=(current  bounding  box.center)]    
\foreach \x in {0,1.6,3.2,4.8}{
\draw[thick](-0.4+\x,-0.6) -- (-0.4+\x,1.4);
\draw[thick](0.4+\x,-0.6) -- (0.4+\x,1.4);
\draw[thick,fill=white] (-0.4+\x,-0.25) rectangle (0.4+\x,0.25);		
}
\foreach \x in {0.8,2.4,4}{
\draw[thick,fill=white] (-0.4+\x,0.55) rectangle (0.4+\x,1.05);		
}
\draw[thick] (-0.8,0.55) -- (-0.4,0.55) -- (-0.4,1.05) -- (-0.8,1.05);
\draw[thick] (5.6,0.55) -- (5.2,0.55) -- (5.2,1.05) -- (5.6,1.05); 
\coordinate[label=right:$U_1''$] (A) at (-0.3,0);
\coordinate[label=right:$U_2''$] (A) at (1.3,0);
\coordinate[label=right:$\ldots$] (A) at (2.9,0);
\coordinate[label=right:$U_n''$] (A) at (4.5,0);
\coordinate[label=right:$V_1''$] (A) at (0.5,0.8);
\coordinate[label=right:$V_2''$] (A) at (2.1,0.8);
\coordinate[label=right:$\ldots$] (A) at (3.7,0.8);
\coordinate[label=right:$V_n''$] (A) at (5.15,0.78);
\coordinate[label=right:$V_n''$] (A) at (-1,0.78);

\begin{scope}[shift={(0,-2.8)}]
\coordinate[label=right:${=}$] (A) at (-2,0.8);
\foreach \x in {0,1.6,3.2,4.8}{
\draw[thick](-0.4+\x,-0.6) -- (-0.4+\x,1.4);
\draw[thick](0.4+\x,-0.6) -- (0.4+\x,1.4);
\draw[thick,fill=white] (-0.4+\x,-0.25) rectangle (0.4+\x,0.25);		
}
\foreach \x in {0.8,2.4,4}{
\draw[thick,fill=white] (-0.4+\x,0.55) rectangle (0.4+\x,1.05);		
}
\draw[thick] (-0.8,0.55) -- (-0.4,0.55) -- (-0.4,1.05) -- (-0.8,1.05);
\draw[thick] (5.6,0.55) -- (5.2,0.55) -- (5.2,1.05) -- (5.6,1.05); 
\coordinate[label=right:$U_1'$] (A) at (-0.3,0);
\coordinate[label=right:$U_2'$] (A) at (1.3,0);
\coordinate[label=right:$\ldots$] (A) at (2.9,0);
\coordinate[label=right:$U_n'$] (A) at (4.5,0);
\coordinate[label=right:$V_1'$] (A) at (0.5,0.8);
\coordinate[label=right:$V_2'$] (A) at (2.1,0.8);
\coordinate[label=right:$\ldots$] (A) at (3.7,0.8);
\coordinate[label=right:$V_n'$] (A) at (5.15,0.78);
\coordinate[label=right:$V_n'$] (A) at (-1,0.78);
\end{scope}

\end{tikzpicture}  \label{eq:qu_circuits}. 
\end{equation}        
As shown in Ref.~\onlinecite{1DSPTMBL}, this equation implies that there have to exist unitaries $W_1, W_2, \ldots, W_{2n}$ such that
\begin{equation} 
\begin{tikzpicture}[scale=1.1,baseline=(current  bounding  box.center)]    

\draw[thick](-0.4,-1.4) -- (-0.4,1.4);
\draw[thick](0.4,-1.4) -- (0.4,1.4);
\draw[thick,fill=white] (-0.4,0.3) rectangle (0.4,0.8);		
\draw[thick,fill=white] (-0.4,-0.8) rectangle (0.4,-0.3);		

\coordinate[label=right:$U_k''$] (A) at (-0.3,0.55);
\coordinate[label=right:${U_k'}^\dg$] (A) at (-0.4,-0.55);

\coordinate[label=right:${=}$] (A) at (1,0);

\begin{scope}[shift={(3.4,0)}]
\draw[thick](-0.7,-1.4) -- (-0.7,1.4);
\draw[thick](0.7,-1.4) -- (0.7,1.4);
\draw[thick,fill=white] (-0.7,0) ellipse (0.55 and 0.55);
\draw[thick,fill=white] (0.7,0) ellipse (0.55 and 0.55);
\coordinate[label=right:$W_{2k-1}$] (A) at (-1.3,0);
\coordinate[label=right:$W_{2k}$] (A) at (0.2,0);

\end{scope}

\end{tikzpicture} \ , \label{eq:gauge1}
\end{equation}        

\begin{equation}
\begin{tikzpicture}[scale=1.1,baseline=(current  bounding  box.center)]    

\draw[thick](-0.4,-1.4) -- (-0.4,1.4);
\draw[thick](0.4,-1.4) -- (0.4,1.4);
\draw[thick,fill=white] (-0.4,0.3) rectangle (0.4,0.8);		
\draw[thick,fill=white] (-0.4,-0.8) rectangle (0.4,-0.3);		

\coordinate[label=right:${V_k''}^\dg$] (A) at (-0.3,0.55);
\coordinate[label=right:$V_k'$] (A) at (-0.3,-0.55);

\coordinate[label=right:${=}$] (A) at (1,0);

\begin{scope}[shift={(3.4,0)}]
\draw[thick](-0.7,-1.4) -- (-0.7,1.4);
\draw[thick](0.7,-1.4) -- (0.7,1.4);
\draw[thick,fill=white] (-0.7,0) ellipse (0.55 and 0.55);
\draw[thick,fill=white] (0.7,0) ellipse (0.55 and 0.55);
\coordinate[label=right:$W_{2k}$] (A) at (-1.2,0);
\coordinate[label=right:$W_{2k+1}$] (A) at (0.1,0);

\end{scope} \label{eq:gauge2}
\end{tikzpicture} \ .
\end{equation}        

Combining the above two equations, we can derive the following useful relation 
\begin{equation}\label{eq:qcpushthru}
    \begin{tikzpicture}[scale=1.1,baseline]
    \draw[thick](-0.8,-1.4) -- (-0.8,1.4);
    \draw[thick](0,-1.4) -- (0,1.4);
    \draw[thick](0.8,-1.4) -- (0.8,1.4);
    \draw[thick,fill=white] (-0.8,-0.8) rectangle (0,-0.3);		
    \draw[thick,fill=white] (0,0.3) rectangle (0.8,0.8);
    \coordinate[label=right:$U_k'$] (A) at (-0.72,-0.55);
    \coordinate[label=right:$V_k'$] (A) at (0.08,0.55);
    \end{tikzpicture}\; \;\; = \;\;\; \begin{tikzpicture}[scale=1.1,baseline]
    \draw[thick](-0.8,-1.4) -- (-0.8,1.4);
    \draw[thick](0,-1.4) -- (0,1.4);
    \draw[thick](0.8,-1.4) -- (0.8,1.4);
    \draw[thick,fill=white] (-0.8,-0.8) rectangle (0,-0.3);		
    \draw[thick,fill=white] (0,0.3) rectangle (0.8,0.8);
    \draw[thick,fill=white] (-0.8,0.5) ellipse (0.55 and 0.55);
    \draw[thick,fill=white] (0.8,-0.5) ellipse (0.55 and 0.55);
    \coordinate[label=right:$U_k''$] (A) at (-0.72,-0.55);
    \coordinate[label=right:$V_k''$] (A) at (0.08,0.55);
    \coordinate[label=right:$W_{2k-1}^\dg$] (A) at (-1.42,0.5);
    \coordinate[label=right:$W_{2k+1}$] (A) at (0.19,-0.5);
    \end{tikzpicture} \   .
\end{equation}
As the quantum circuits in Eq.~\eqref{eq:qu_circuits} depend on the group elements $g$, so do the unitaries $W_j$. Sequential application of the symmetry operation $\mathpzc{v}_g^{\otimes N^2}$ and $\mathpzc{v}_h^{\otimes N^2}$ and comparison to $\mathpzc{v}_{gh}^{\otimes N^2}$ in Eq.~\eqref{eq:symmetry} then yields the relation~\cite{1DSPTMBL}
%
%
\begin{equation}
    \label{eq:Wrep}
    \includegraphics[width=0.13\linewidth,valign=c]{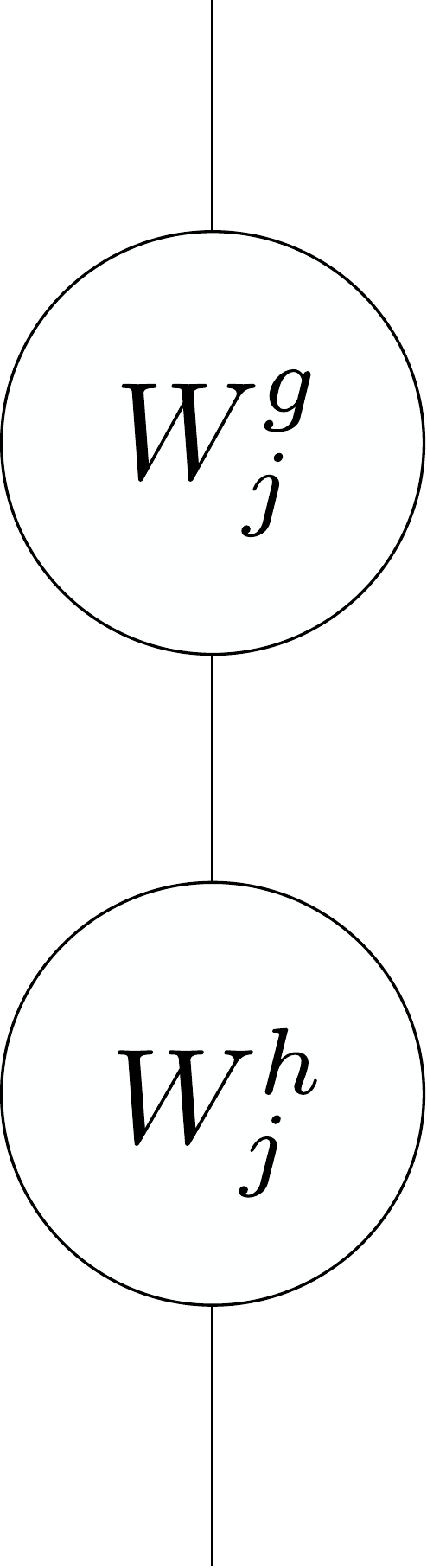}\;\;=\;\;\beta(g,h)\;\,\includegraphics[width=0.13\linewidth,valign=c]{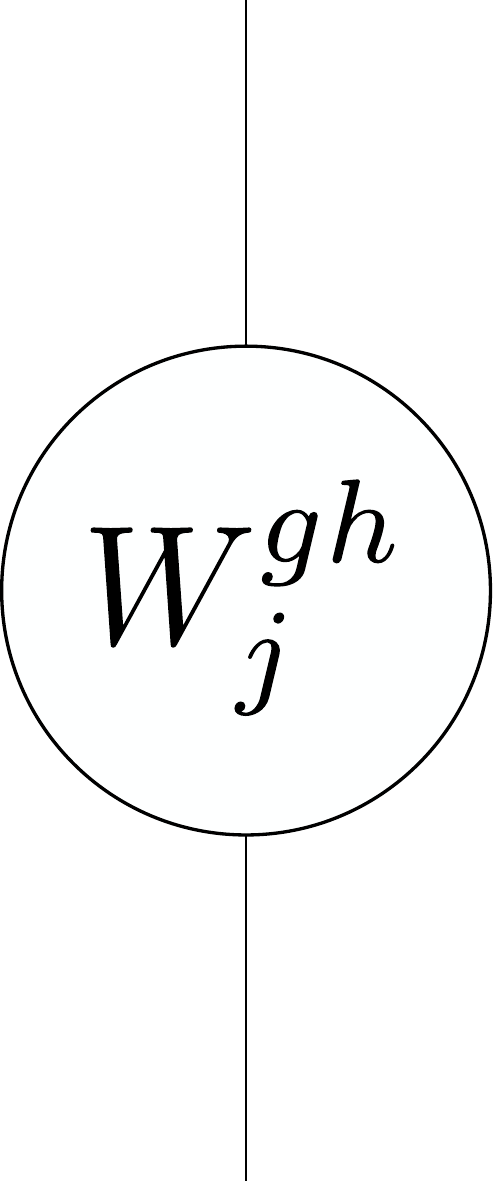} \, 
\end{equation}
with $|\beta(g,h)| = 1$.  That is, $W_j^g$ is a projective representation of the symmetry group $G$. 
In our two-dimensional case, each $W_j^g$ is a tensor that extends along the $y$-direction. Eqs.~\eqref{eq:gauge1} and~\eqref{eq:gauge2} imply that it has strict short-range correlations along $y$-direction. Hence, it can be efficiently approximated by a quantum circuit, 
\begin{align}
    \label{eq:Wqc}
    \includegraphics[width=0.1\linewidth,valign=c]{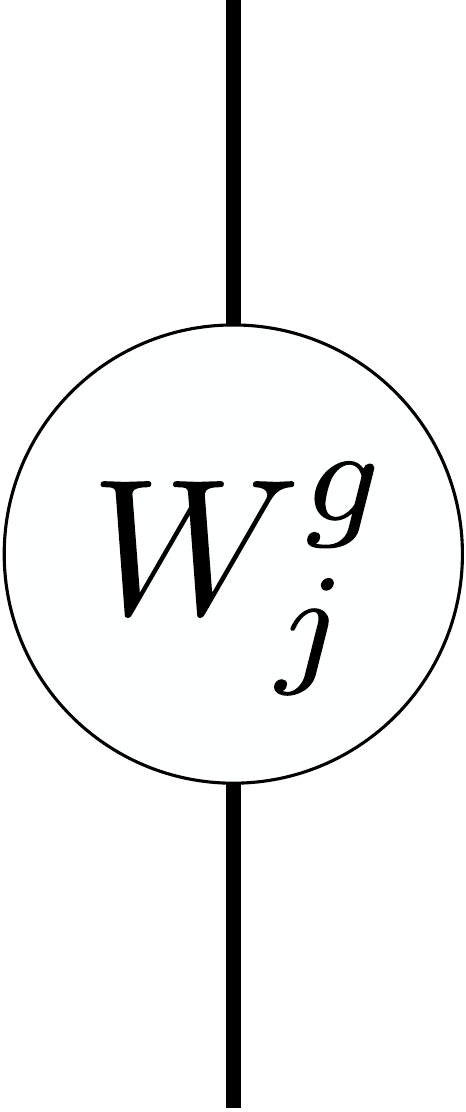} \,=
    \includegraphics[width=0.66\linewidth,valign=c]{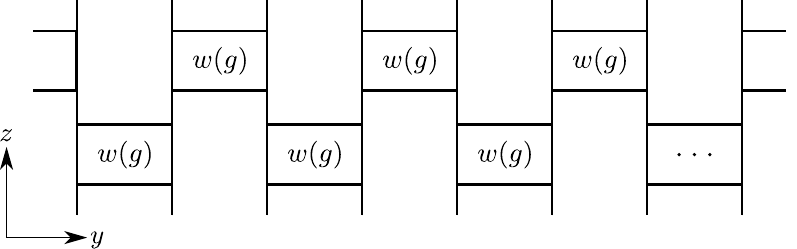}
    \; ,
\end{align}
where the $j$ subscript and indices corresponding to the position along the $y$-direction have been suppressed on the right hand side. 
In the technical derivation, we will suppress the indices of constituting unitaries (e.g., of $U_k$ and $V_k$) when there are no ambiguities, but we emphasize that the quantum circuits 
are typically not translationally invariant. As an example,  the left hand side of Eq.~\eqref{eq:qu_circuits} would be written with all the upper layer tensors labeled $V$ and all the lower layer tensors labeled $U$.  

In Sec.~\ref{sec:qcreps} we prove the following lemma (for quantum circuits of the type Eq.~\eqref{eq:Wqc}): \textit{Two-layer quantum circuit projective representations of a group $G$ have a topological index given by an element of the third cohomology group $H^3(G,U(1))$ of $G$.}  Together with the existence of the $W_j^g$ acting on the boundary, the lemma implies that two-dimensional SPT MBL phases are labeled by the elements of the third cohomology group $H^3(G,U(1))$.  Since the cohomology group is discrete, different cohomology classes, and therefore different SPT MBL phases, cannot be continuously connected.

To complete the argument, one has to show that the cohomology class (i.e. the topological index) is the same independently of the $x$-coordinate ($k$) of $W_{2k-1}^g$; we do this in section~\ref{sec:wxwTopTrivial} by proving that $W_j^g \otimes W_{j+1}^g$ is topologically trivial, i.e. that it is a quantum circuit representation that corresponds to the identity element of $H^3(G,U(1))$.

\subsection{Intuitive overview of the proof of the lemma}

Here we give an intuitive overview of the ideas behind the proof of the above lemma.  
Following Refs.~\onlinecite{MPOgaugingandedge, chen2011twodim}, we review the ``pentagon equation'', which applies to the tensor network symmetry operator that acts on an edge of a two-dimensional symmetric tensor network state, 
such as matrix product operators (MPOs) acting on a PEPS. 
The pentagon equation shows that those symmetry operators can be classified by the elements of the third cohomology group, implying that the overall symmetric states have those elements as topological indices. 
In the technical part, we demonstrate that $W_j(g)$ satisfies the pentagon equation and consequently two-dimensional MBL SPT phases can be labeled by the elements of the third cohomology group.

\begin{figure}
    \centering
    \includegraphics[width =1\linewidth]{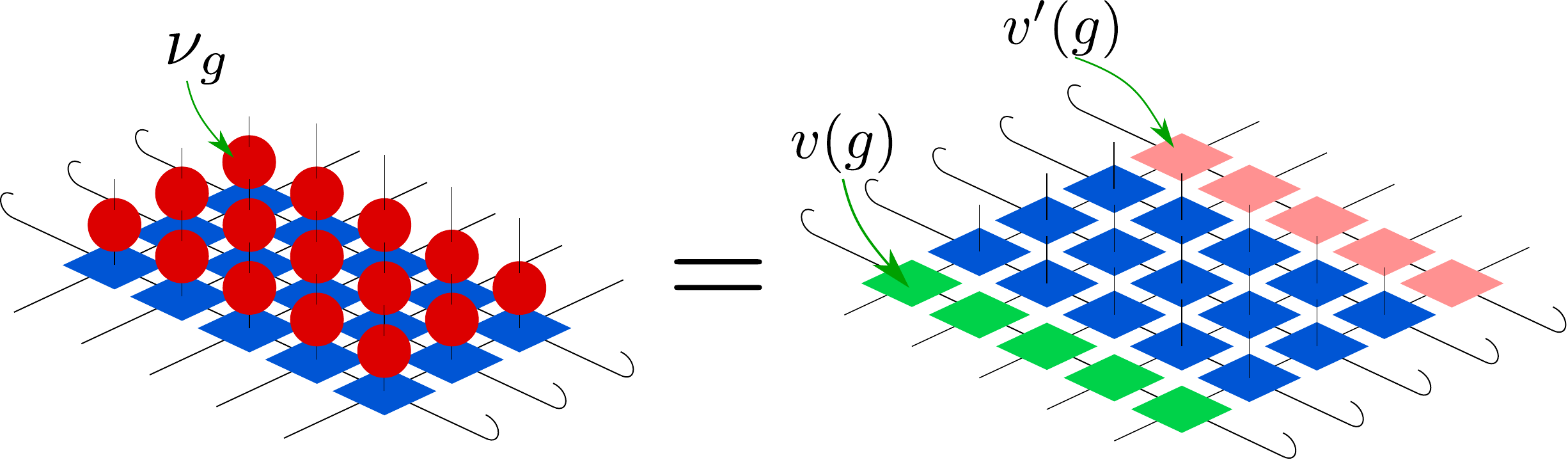}
    \includegraphics[width =0.8\linewidth]{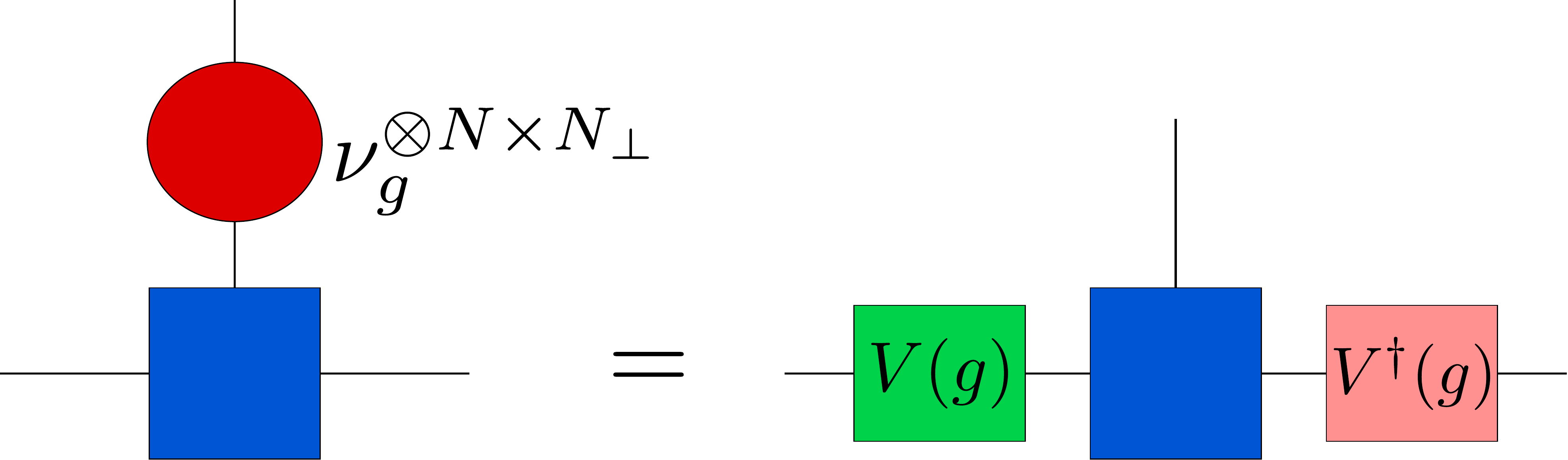}
    \caption{Top equation: On-site symmetry operators $\mathpzc{v}_g$ (red balls on the left hand side of the equation) can be `pushed through' to become $V(g)$ and $V^\dagger(g)$ (MPOs indicated as rows of blocks on the right hand side---note that $v(g)$ and $v'(g)$ refer to individual MPO tensors) acting on the open legs along the edge of the partially contracted PEPS whose tensors are indicated by blue boxes. 
    The bent lines indicate periodic boundary conditions along one direction. 
    Bottom equation: A schematic side-view of the top equation, where and $V(g)$ and $V^\dagger(g)$ are MPOs.  Note that if this is interpreted as a one dimensional equation, with the blue square being an MPS tensor, this is simply the one-dimensional result that symmetries can be "pushed through" to the virtual indices \cite{perez2008}.  }
    \label{fig:PEPSsymmetry}
\end{figure}

Specifically, these operators appear in translationally invariant PEPS invariant under the symmetry if only a patch of PEPS tensors is contracted (rather than the full PEPS). In Fig.~\ref{fig:PEPSsymmetry} we show a PEPS which has been fully contracted along one direction, but only partially along the orthogonal direction, i.e., there are dangling bonds of the PEPS, see Fig.~\ref{fig:PEPSsymmetry}. If the symmetry operation $\mathpzc{v}_g^{\otimes (N N_\perp)}$ is applied on that patch ($N_\perp$ corresponds to the incomplete orthogonal contraction), this is equivalent to applying certain MPOs $V(g)$ and $V'(g)$ along the open boundaries of the PEPS.  $V(g)$ and $V'(g)$ correspond to ${W_j^g}^\dg$ and $W_{j'}^g$ in the MBL case, respectively. 

The $V(g)$ operator is in general not a group representation in the usual sense, since given two symmetry operations $V(g)$ and $V(h)$, their composition would correspond to an MPO with a larger bond dimension, whose tensors thus are different from those of $V(gh)$.  Rather, we need a ``combining" operator~\cite{MPOgaugingandedge} $X_{L,R}(g,h)$ satisfying
\begin{equation} \label{eq:combineop}
    \includegraphics[width = 0.25\linewidth,valign=c]{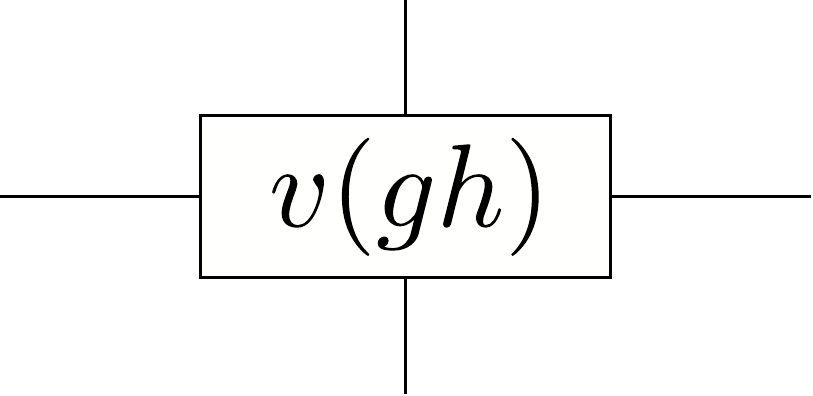} \;=\; \includegraphics[width = 0.43\linewidth,valign=c]{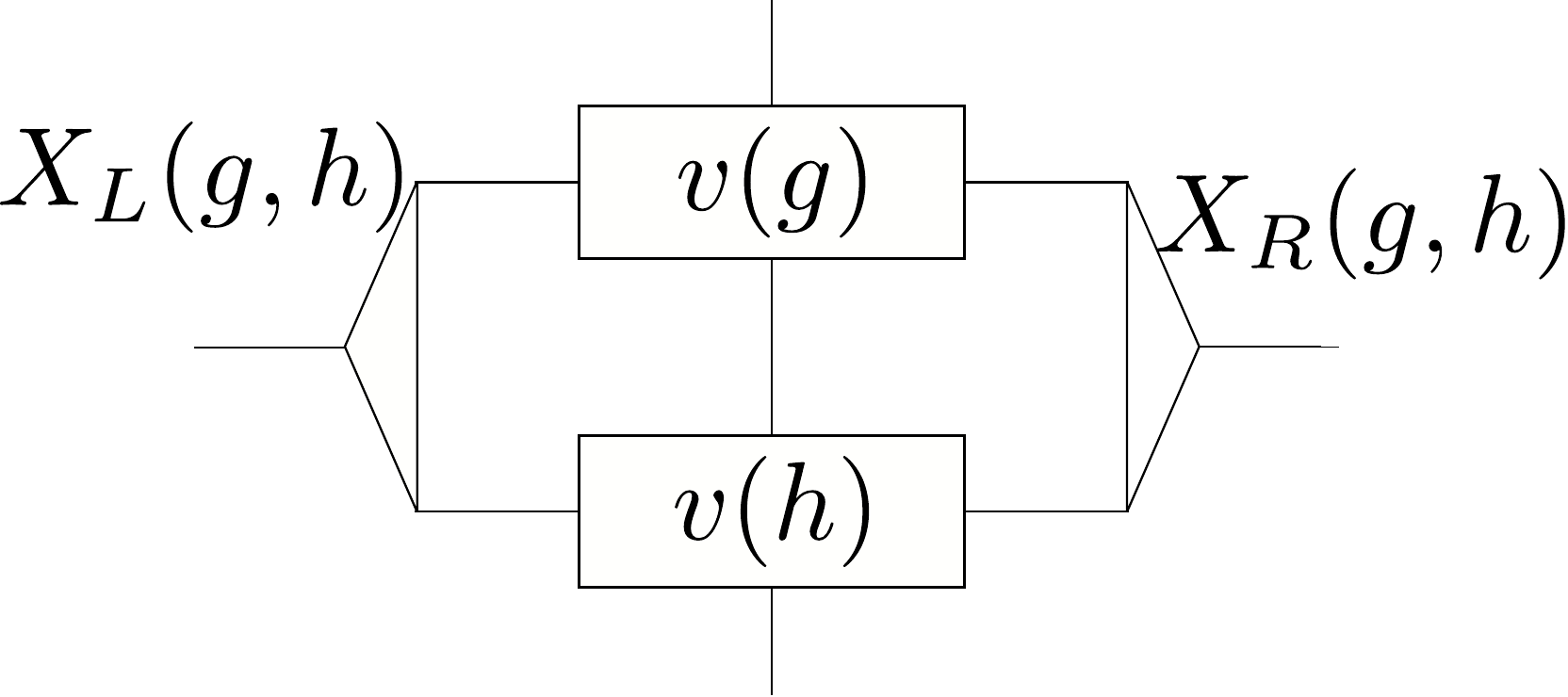} \ ,
\end{equation}
where $v(g)$ are the constituting tensors of the MPO $V(g)$ (see Fig.~\ref{fig:PEPSsymmetry}). 

This equation is invariant under the transformation $X_L(g,h) \rightarrow X_L(g,h)/\chi(g,h)$ and $X_R(g,h) \rightarrow X_R(g,h) \chi(g,h)$. \textit{A priori} $\chi(g,h)$ could be any complex number. However, we have to exclude $\chi(g,h) = 0$ (and $\chi(g,h) = \infty$) such that $X_L(g,h)$ and $X_R(g,h)$ remain well-defined. This is topologically equivalent to constraining to $|\chi(g,h)| = 1$, i.e., no rescaling of $X_L(g,h)$ and $X_R(g,h)$ is allowed. 
For the quantum circuit case we focus on in this paper, $\chi(g,h)$ will appear as a result of a gauge degree of freedom in the quantum circuit unitaries.

For three group elements we then have
\begin{align}
     \includegraphics[width = 0.27\linewidth,valign=c]{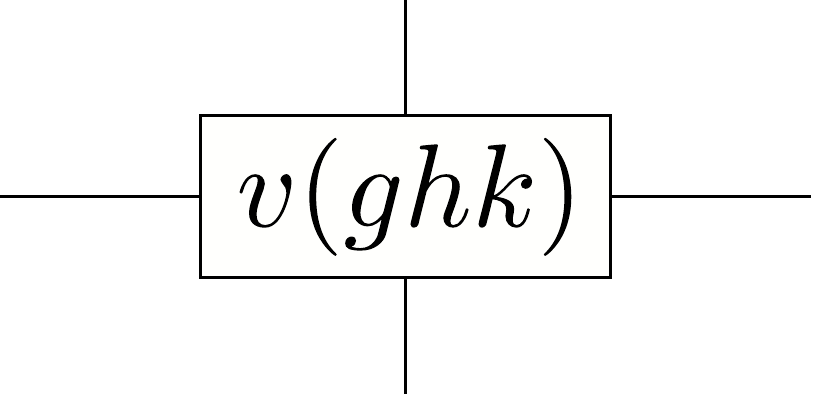} \;\;=\;\;  \includegraphics[width = 0.53\linewidth,valign=c]{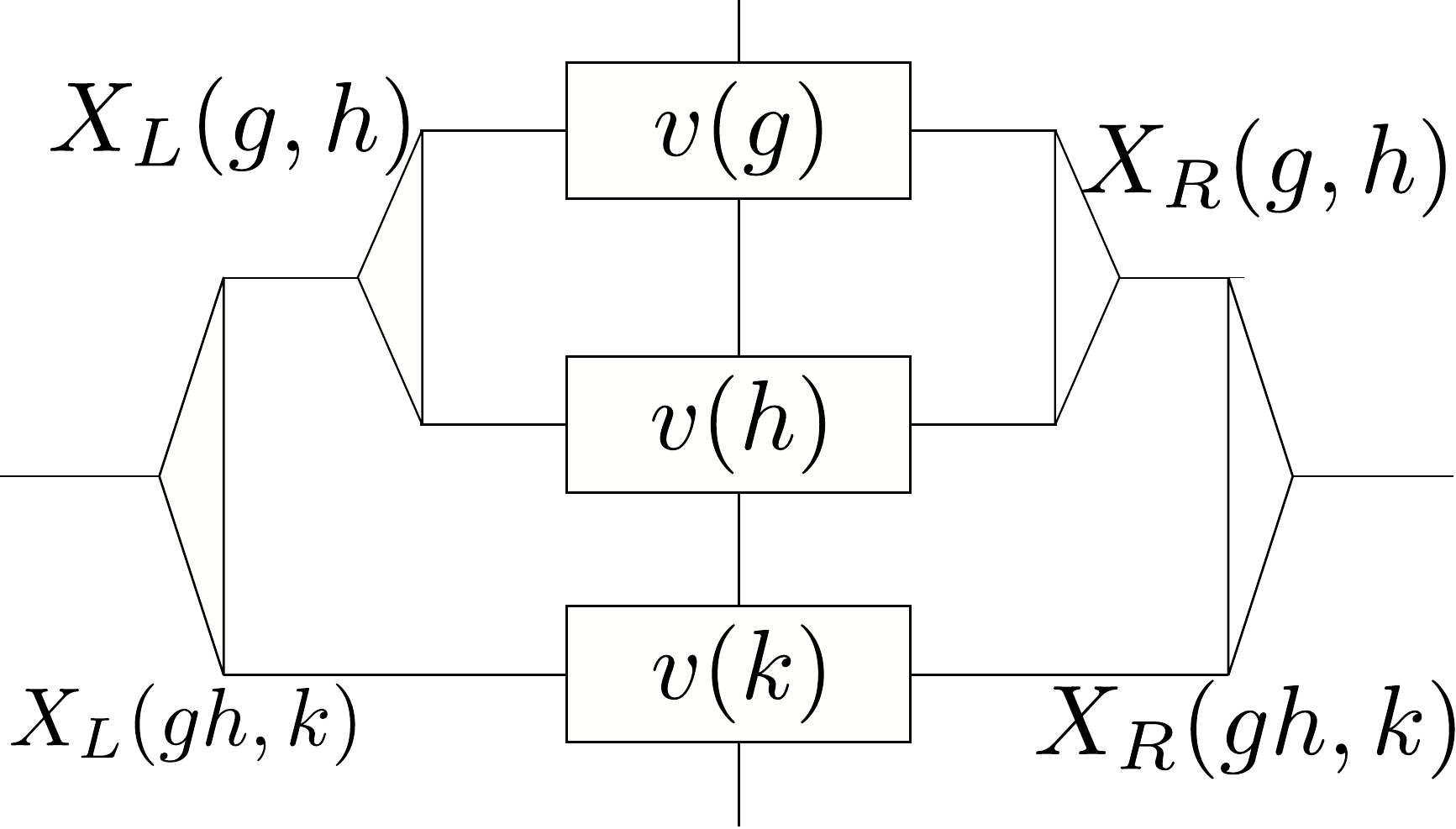}\nonumber\\=\;\; \includegraphics[width = 0.53\linewidth,valign=c]{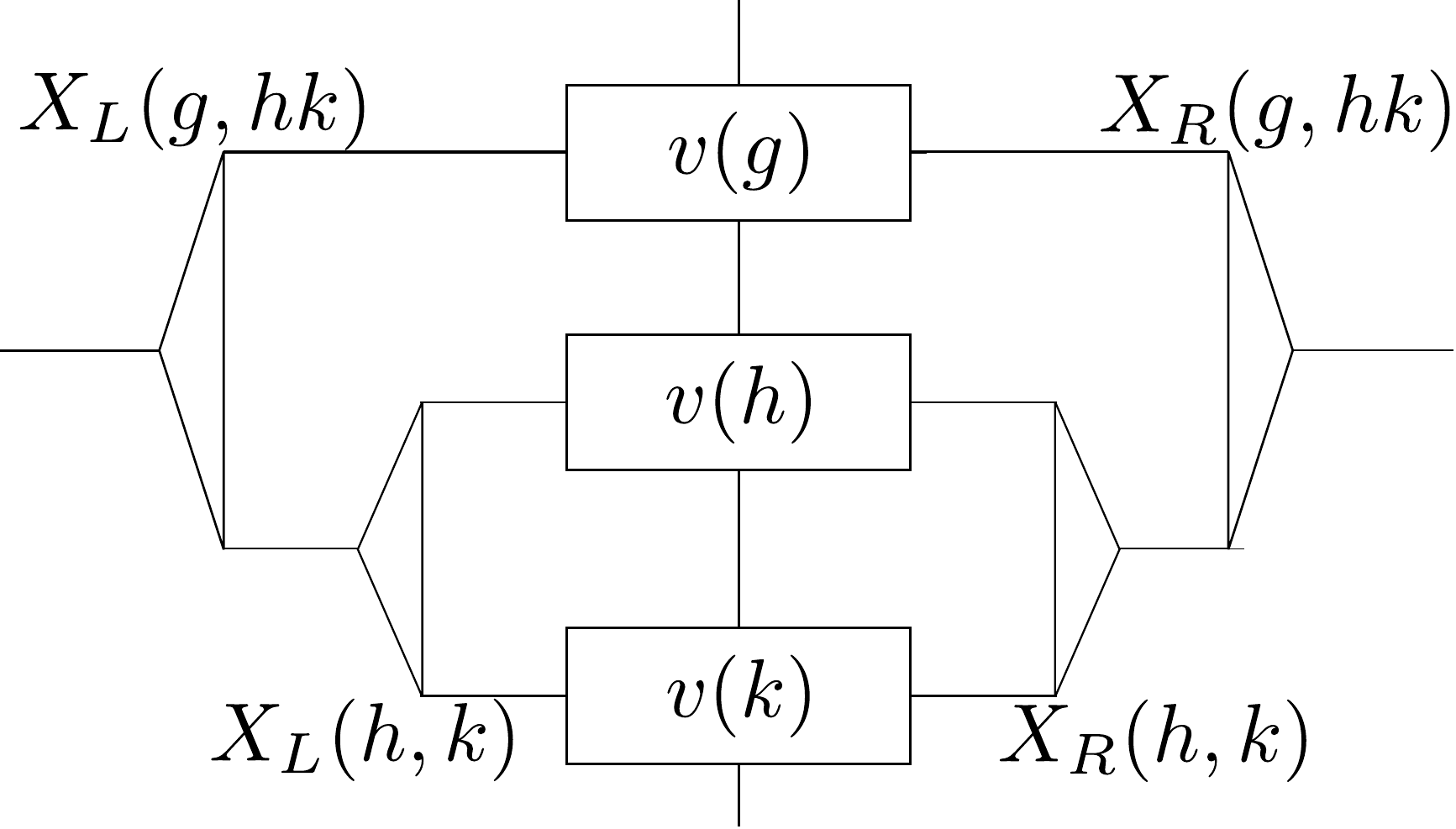} \ .
\end{align} 

If one considers operating on the left edge and right edge separately, one may deduce~\cite{MPOgaugingandedge} (if $V(g)$ is injective~\cite{PerezGarcia2007}) that
\begin{equation} \label{eq:xxxx}
    \includegraphics[width = 0.19\linewidth,valign=c]{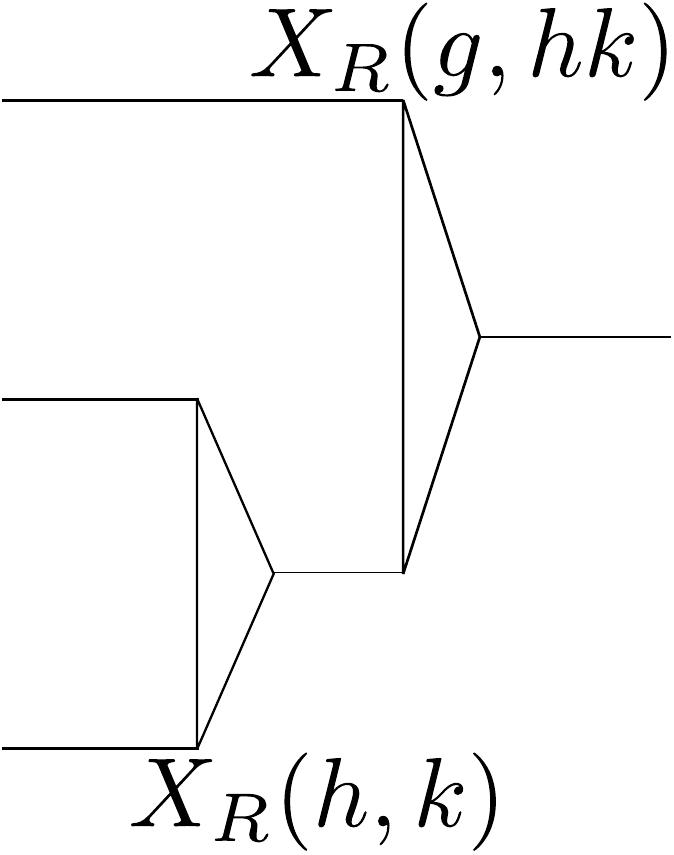}\;\;=\;\;\alpha(g,h,k)\;\;\includegraphics[width = 0.19\linewidth,valign=c]{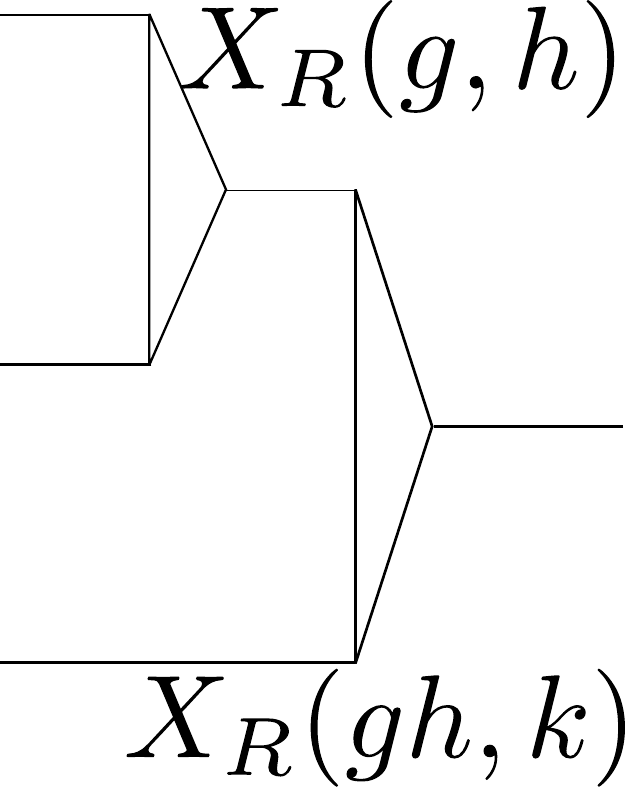}\;,
\end{equation}
as well as a similar equation for the $X_L(g,h)$ with a factor $1/\alpha(g,h,k)$. Since no rescaling of $X_L(g,h)$ and $X_R(g,h)$ is allowed, $|\alpha(g,h,k)| = 1$. 

In the quantum circuit case, a similar equation as Eq.~\eqref{eq:xxxx} holds, because the quantum circuits are only short-range correlated and hence the left and right boundary operators can be separated. 

We note that the $X_R(g,h)$ (or equivalently the $X_L(g,h)$) in Eq.~\eqref{eq:xxxx} in some sense form a ``representation" of the group $G\ni g,h$, but with not one but two group elements associated to each operator.  This kind of representation is sometimes called a \textit{gerbal} representation and has been studied in the mathematics literature\cite{gerbal}. 

We can use the gauge degree of freedom of $X_R(g,h)$ to show that $ \alpha(g,h,k)$ is only defined up to a  3-coboundary
\begin{equation}
    \alpha(g,h,k)\rightarrow \alpha'(g,h,k) = \alpha(g,h,k)\frac{\chi(g,hk)\chi(h,k)}{\chi(g,h)\chi(gh,k)}. \label{eq:coboundary}
\end{equation}

Using Eq.~\eqref{eq:xxxx}, we can perform the following sequence of manipulations on the combination of $V(g)$, $V(h)$, $V(k)$ and $V(l)$ leading to the same result in two different ways (cf. pentagon equation in topological quantum field theories~\cite{topQC}) 
\begin{equation}\label{eq:penta}
    \includegraphics[width=0.77\linewidth,valign=c]{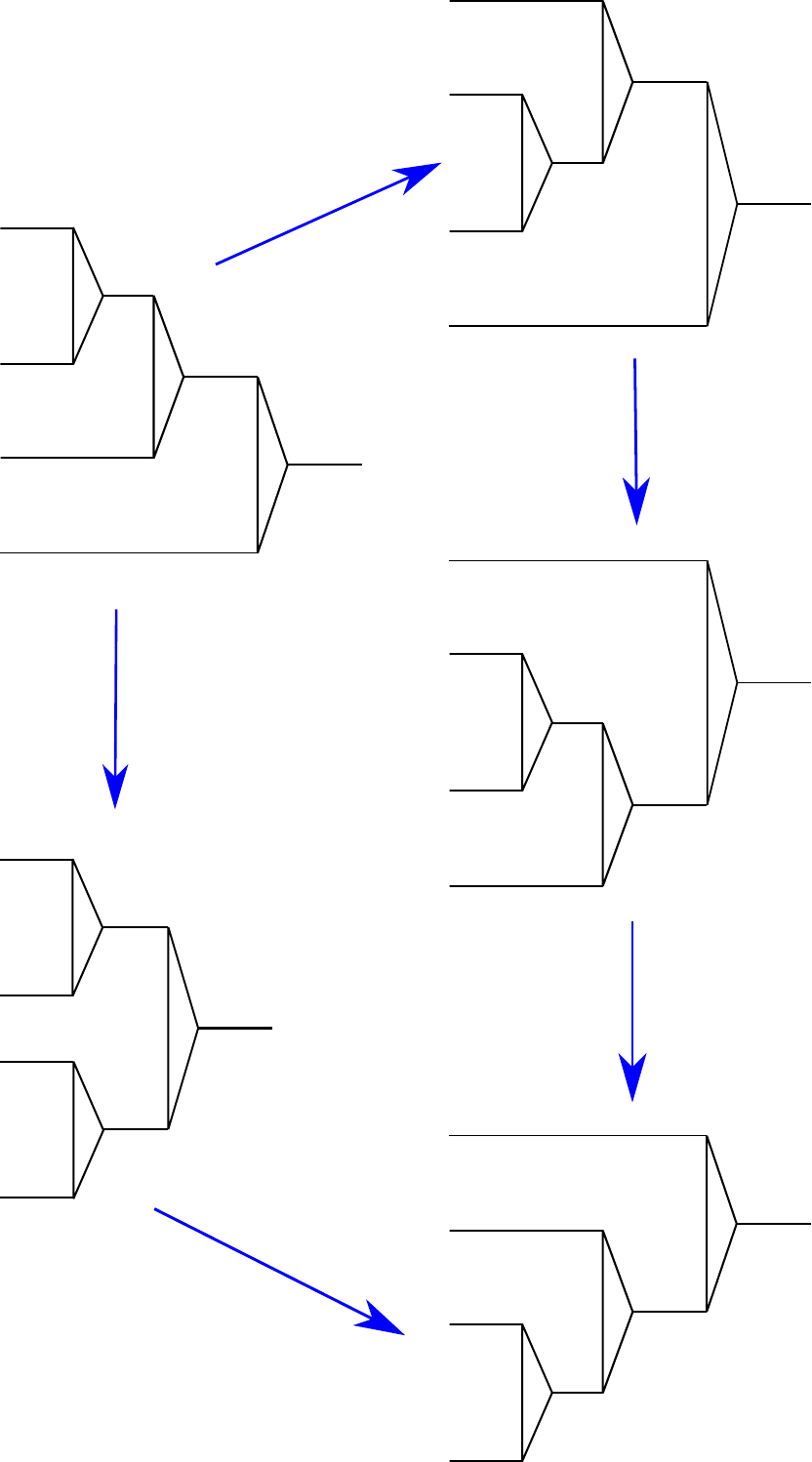} \;. 
\end{equation}
This implies that the incurred phases $\alpha(g,h,k)$ must fulfill the following consistency relation
\begin{equation}
    \frac{\alpha(g,h,k)\alpha(g,hk,l)\alpha(h,k,l)}{\alpha(gh,k,l)\alpha(g,h,kl)} = 1,
\end{equation}
which is known as a 3-cocycle. 
Recall that the cohomology group $H^n(G,U(1))$ consists of the equivalence classes of $n$-cocycles that differ by only an $n$-coboundary (Eq.~\eqref{eq:coboundary} in our case).  So, we have essentially shown that a projective representation in the form of an MPO acting on the edge of a two-dimensional tensor network corresponds to an element of the third cohomology group of the symmetry group.  

For the case where $V(g)$ is an injective~\cite{PerezGarcia2007} MPO, the above calculation is the complete argument\cite{MPOgaugingandedge, chen2011twodim}. 
In the context of two-dimensional SPT MBL, $V(g)$ has to be replaced by the quantum circuit $W^g_j$, 
and it is not obvious how to define a combining operation in terms of $X_{L,R}(g,h)$ tensors such as those in Eq.~\eqref{eq:combineop}. 
In Sec.~~\ref{sec:qcreps}, we construct a suitable combining operation and show that it satisfies the corresponding pentagon equation and hence the 3-cocycle condition. Thus, SPT MBL phases in two dimensions are also labeled by an element of the third cohomology group. Moreover, we explicitly demonstrate below that all eigenstates of the MBL system must correspond to the \textit{same} element of the third cohomology group, just as in one dimension~\cite{1DSPTMBL}. Finally, we show that the obtained topological labels are stable to small perturbations and can only change if perturbations are made strong enough that the system becomes delocalized along the way. 



\section{Classification of two-dimensional SPT MBL phases with quantum circuits}\label{sec:main}

We will show that two-dimensional MBL SPT phases are labeled by the elements of the third cohomology group of the symmetry group $G$. 
Due to a mathematical result (proven in Sec.~\ref{sec:qcreps}), this reduces to the problem of finding a projective representation of $G$ in terms of quantum circuits.  This follows from  projecting the two-dimensional problem into one dimension and then applying the results of the calculations for the classification of SPT phases in one-dimensional MBL systems, as done in Ref.~\onlinecite{1DSPTMBL}. Note that we do not show that MBL Hamiltonians corresponding to the same third cohomology class can be continuously connected (without violating FMBL), i.e., we do not demonstrate completeness of our classification. 

Consider a two-dimensional spin system on an $N \times N$ lattice. We shall work with an FMBL Hamiltonian invariant under an on-site abelian symmetry. As elaborated on above, we represent the unitary which diagonalizes the Hamiltonian by 
 a four-layer quantum circuit with gates acting on plaquettes of $\ell \times \ell$ sites, cf. Fig.~\ref{fig:2dqc}, and we choose $\ell \propto N^\nu$ with $\nu < 1$ to carry out our classification.

\subsection{2D MBL systems with an on-site symmetry}\label{sec:2dmblonsite}

We assume the strongly disordered FMBL Hamiltonian $H$ to be invariant under a local unitary symmetry operator $\mathpzc{v}_g$, for $g\in G$. That is, $H$ commutes with the symmetry operator, 
\begin{align} 
H = \mathpzc{v}_g^{\otimes N^2} H (\mathpzc{v}_g^\dg)^{\otimes N^2}. \label{eq:Ham_sym}
\end{align}
 
Let $U$ be the unitary matrix that diagonalizes the Hamiltonian, and $E$ the diagonal matrix of energies, i.e. $H = U E U^\dg$.  By the same line of reasoning as in Ref.~\onlinecite{Thorsten}, one can derive the action of the symmetry on $U$. Eq.~\eqref{eq:Ham_sym} implies that
\begin{align}
E = U^\dg \mathpzc{v}_g^{\otimes N^2} U E U^\dg (\mathpzc{v}_g^\dg)^{\otimes N^2} U. \label{eq:insert_diagonalisation}
\end{align}
As the symmetry group is abelian, $E$ cannot have any symmetry-enforced degeneracies.  Assuming $E$ to be non-degenerate, Eq.~\eqref{eq:insert_diagonalisation} implies
\begin{align} 
\Theta_g = U^\dagger \mathpzc{v}_g^{\otimes N^2} U, 
\label{eq:theta}
\end{align}
with $\Theta_g$ being a diagonal matrix whose diagonal elements have magnitude $1$. Accidental degeneracies can be removed and are treated explicitly in Section~\ref{sec:robustness}. 

Note that the eigenstates $|\psi_{l_1 \cdots l_{N^2}}\rangle$ can be obtained by fixing the lower indices of the unitary $U$ to the corresponding l-bit labels $l_1, l_2, \ldots, l_{N^2} = \pm 1$,
\begin{equation}
    \includegraphics[width=0.7\linewidth,valign=c]{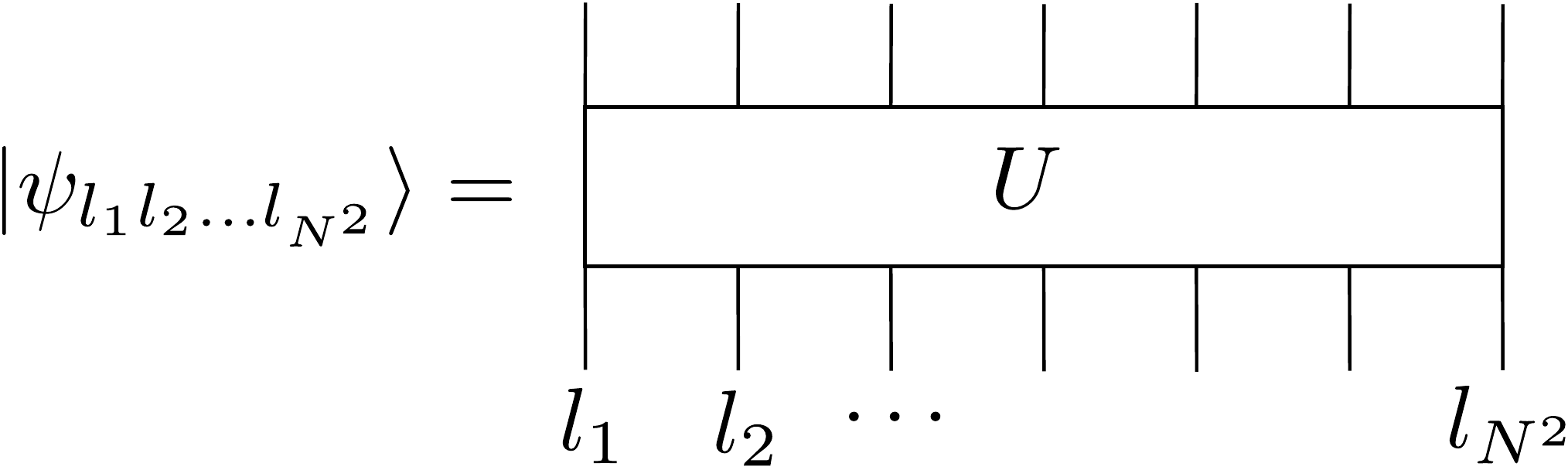}\, .
    \label{eq:psi_lbit}
\end{equation}

\subsubsection{Quantum circuit representation of the $\Theta_g$ matrix}\label{sec:thetaqc}

Next we will show that the tensor $\Theta_g$ can be written as a four-layer quantum circuit as in Fig.~\ref{fig:2dqc} (recall that \textit{a priori} only $U$ is assumed to have that property).  The derivation is the two-dimensional version of the one-dimensional case in Ref.~\onlinecite{Thorsten}.

Let us set up a coordinate system where $\mb{k}\in \mathbb{Z}^2$ labels a block of $\ell \times \ell$ sites, or equivalently, a $u$-tensor in the lowest layer of $U$ (red squares in Fig.~\ref{fig:2dqc}(a)).  Let $\mb l_\mb{k}$ denote the $l$-bit indices associated with the legs at $\mb{k}$. 
Making the definition $Z_{g,k} = V_k^\dag (\mathpzc{v}_g^{\otimes \ell^2}) V_k$, we write the diagonal elements of $\Theta_g$ as (note that we use the convention that multiplication order left to right in algebraic notation corresponds to top to bottom in diagrammatic notation)
\begin{equation}\label{eq:diagtheta}
    \includegraphics[width=0.85\linewidth,valign=c]{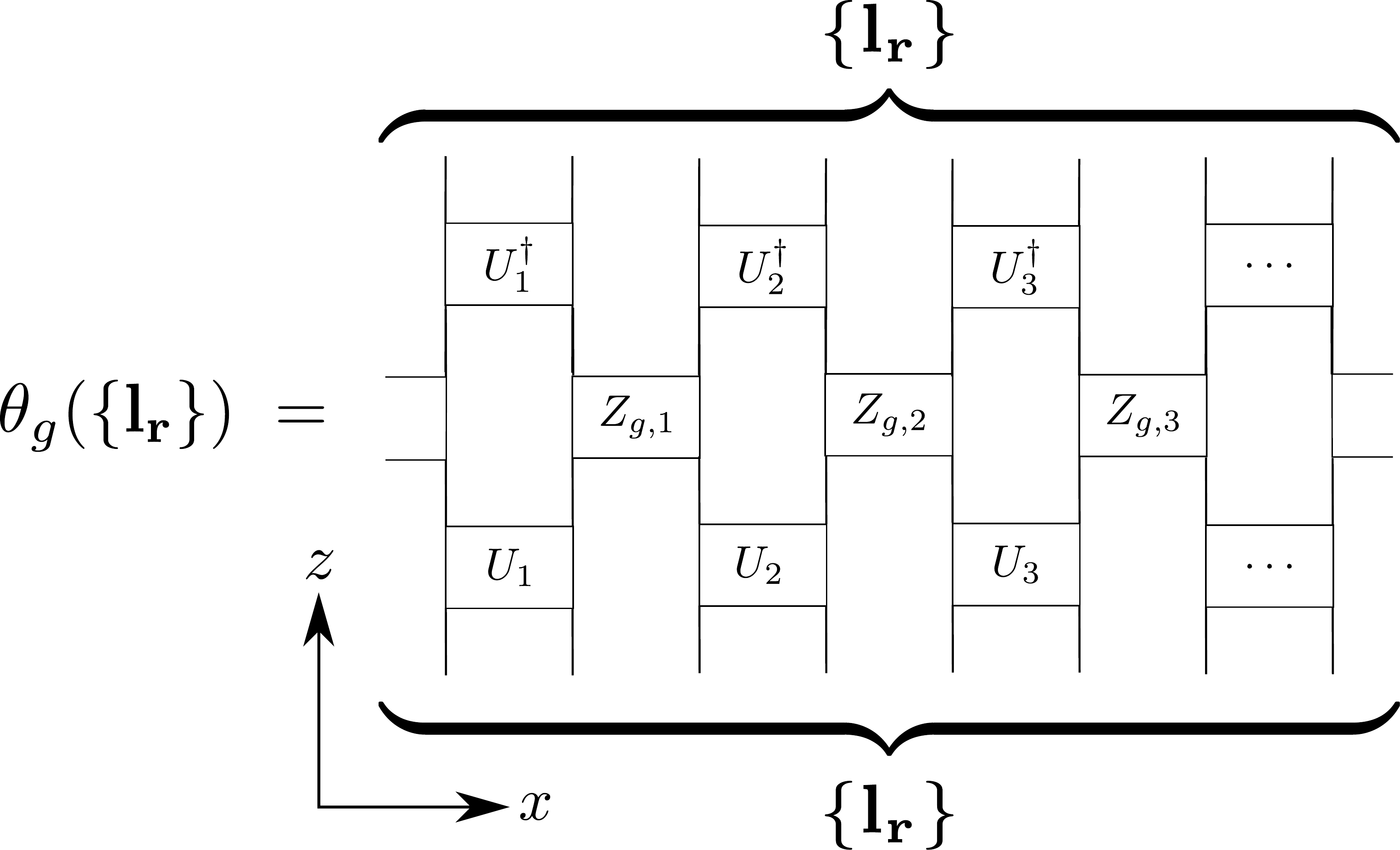}\, .
\end{equation} 
Note that \eqref{eq:diagtheta} is the projected view onto the $xz$-plane of a two-dimensional seven-layer quantum circuit where the locations of the unitaries in the individual layers are as illustrated in panels (a,b,c,d,c,b,a) of Fig.~\ref{fig:2dqc}, respectively. (The uppermost layer Fig.~\ref{fig:2dqc}(d) can be combined with $\mathpzc{v}_g^{\otimes N^2}$ and its adjoint.)  

\begin{figure}
    \centering
    \includegraphics[width = 0.6\linewidth]{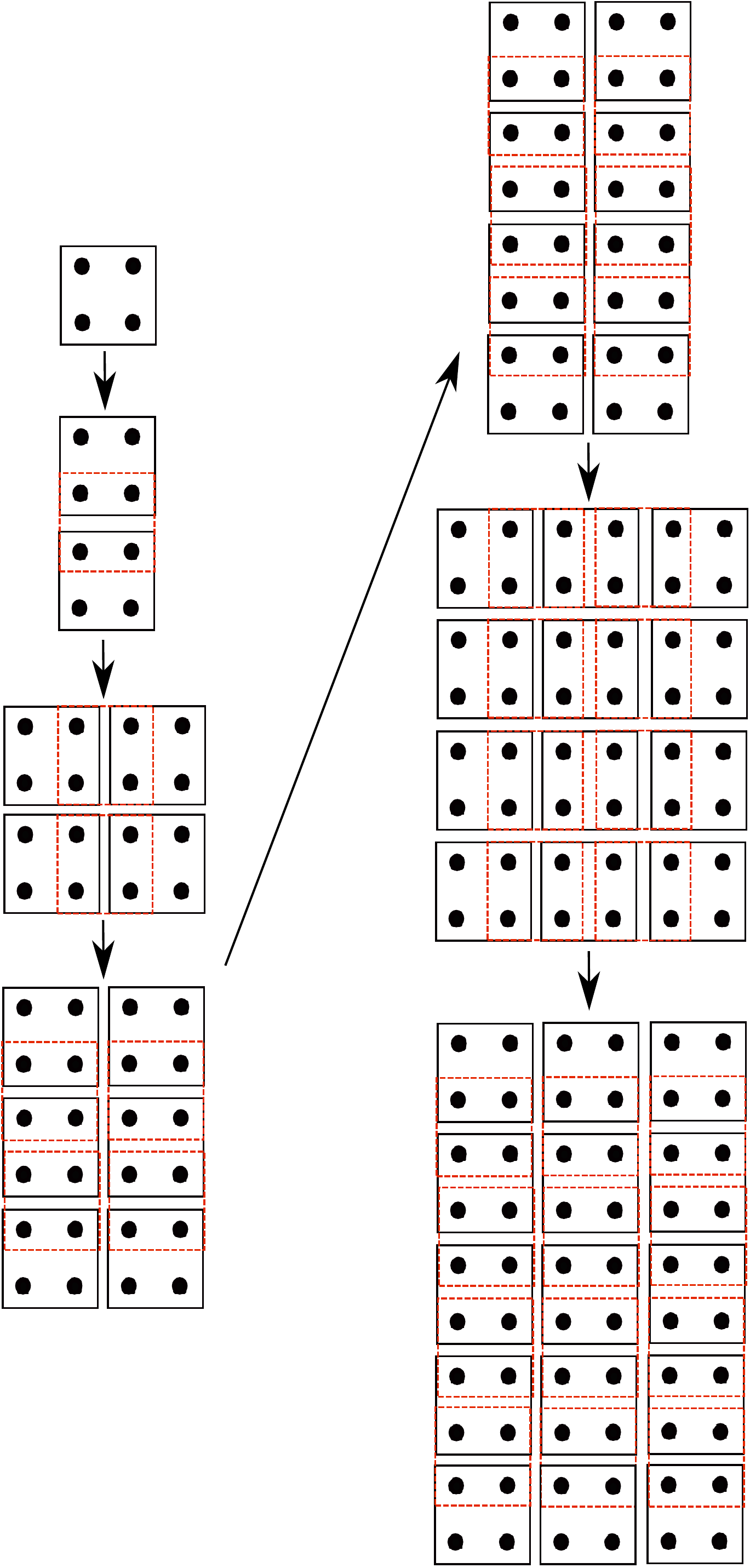}
    \caption{Layers of the lower half of the causal cone ordered from top to bottom as denoted by arrows. The unitaries of the respective upper layer are indicated by red dashed lines. Each dot corresponds to $\frac{\ell}{2} \times \frac{\ell}{2}$ sites.}
    \label{fig:causal}
\end{figure}

\begin{figure}
    \includegraphics[width=0.8\linewidth]{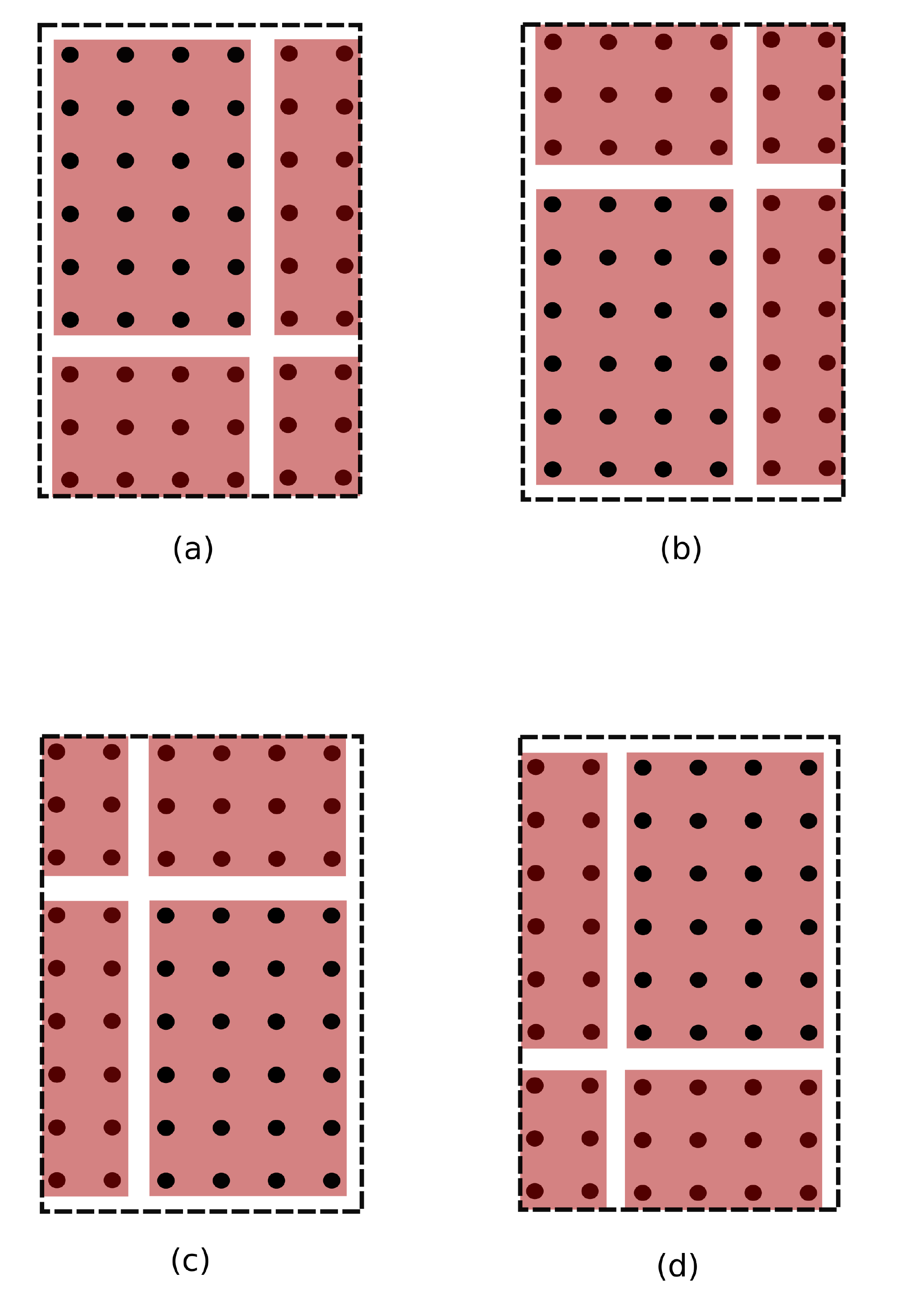}
    \caption{Layers of the $\Theta_g$ quantum circuit within the region indicated by dashed lines. The unitaries are stacked from bottom to top in the order (a), (b), (c), (d). The quantum circuit periodically extends beyond the regions indicated by dashed lines.}
    \label{fig:4x6}
\end{figure}

Consider for some $\mb{k}$, the product of numbers $\theta^*_g(\mathbf{l}_\mb{k},\{\mathbf{l}_\mathbf{r}|\, \forall \, \mb{r}\neq \mb{k}\}) \theta_g(\mathbf{l}'_\mb{k},\{\mathbf{l}_\mathbf{r}|\, \forall \, \mb{r}\neq \mb{k}\})$, which can be written diagrammatically (with the same convention as in Eq.~\eqref{eq:diagtheta} and with implicit subscripts) as 
\begin{align}\label{eq:causalcone}
    \theta^*_g(\mathbf{l}_\mb{k},\{\mathbf{l}_\mathbf{r}|\, \forall \, \mb{r}\neq \mb{k}\}) \theta_g(\mathbf{l}'_\mb{k},\{\mathbf{l}_\mathbf{r}| \, \forall \, \mb{r}\neq \mb{k}\}) \;= \nonumber\\ \includegraphics[width=0.9\linewidth,valign=c]{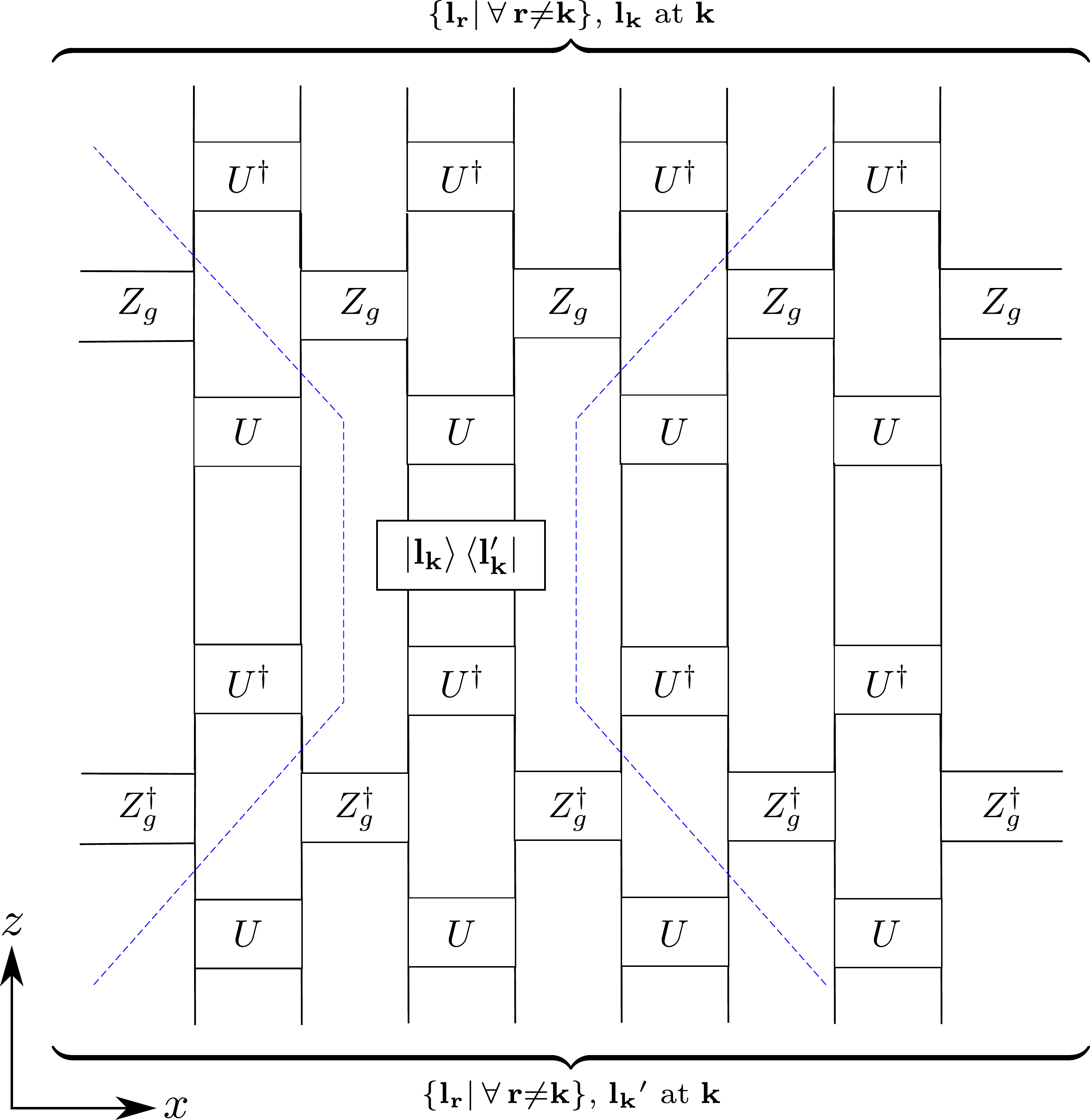}
    \, ,
\end{align}
where we have used the fact that Eq.~\eqref{eq:diagtheta} is diagonal, and where the operator $\ket{\mathbf{l}_\mathbf{k}} \bra{\mathbf{l}_\mathbf{k}'}$ acts non-trivially only on the block of sites labeled by $\mathbf{k}$.
All the unitaries outside the causal cone (blue dashed line) cancel. The causal cone also has a finite extension along $y$-direction and its lower half is shown in detail in Fig.~\ref{fig:causal}. Consequently, the product becomes a phase that depends \textit{only} on the degrees of freedom that lie within the causal cone in Eq.~\eqref{eq:causalcone},
\begin{align}\label{eq:algebraiclightcone}
    \theta_g^*\begin{pmatrix}
    & & \vdots & & \\
    & \mb{l}_{\mb{k}-\mb{\hat{x}}+2 \mb{\hat{y}}} &  \mb{l}_{\mb{k}+2 \mb{\hat{y}}} & \mb{l}_{\mb{k}+\mb{\hat{x}}+2 \mb{\hat{y}}}& \\
        & \mb{l}_{\mb{k}-\mb{\hat{x}}+\mb{\hat{y}}} &  \mb{l}_{\mb{k}+\mb{\hat{y}}} & \mb{l}_{\mb{k}+\mb{\hat{x}}+\mb{\hat{y}}}& \\
    \cdots&  \mb{l}_{\mb{k}-\mb{\hat{x}}} &  \mb{l}_{\mb{k}} & \mb{l}_{\mb{k}+\mb{\hat{x}}} &\cdots\\
     &  \mb{l}_{\mb{k}-\mb{\hat{x}}-\mb{\hat{y}}} &  \mb{l}_{\mb{k}-\mb{\hat{y}}} & \mb{l}_{\mb{k}+\mb{\hat{x}}-\mb{\hat{y}}} & \\
         &  \mb{l}_{\mb{k}-\mb{\hat{x}}-2 \mb{\hat{y}}} &  \mb{l}_{\mb{k}-2 \mb{\hat{y}}} & \mb{l}_{\mb{k}+\mb{\hat{x}}-2 \mb{\hat{y}}} & \\
     & & \vdots & & 
    \end{pmatrix} \times \nonumber\\
    \theta_g\begin{pmatrix}
    & & \vdots & & \\
     & \mb{l}_{\mb{k}-\mb{\hat{x}}+2 \mb{\hat{y}}} &  \mb{l}_{\mb{k}+2\mb{\hat{y}}} & \mb{l}_{\mb{k}+\mb{\hat{x}}+2\mb{\hat{y}}}& \\
    & \mb{l}_{\mb{k}-\mb{\hat{x}}+\mb{\hat{y}}} &  \mb{l}_{\mb{k}+\mb{\hat{y}}} & \mb{l}_{\mb{k}+\mb{\hat{x}}+\mb{\hat{y}}}& \\
    \cdots&  \mb{l}_{\mb{k}-\mb{\hat{x}}} &  \mb{l}_{\mb{k}}' & \mb{l}_{\mb{k}+\mb{\hat{x}}} &\cdots\\
     &  \mb{l}_{\mb{k}-\mb{\hat{x}}-\mb{\hat{y}}} &  \mb{l}_{\mb{k}-\mb{\hat{y}}} & \mb{l}_{\mb{k}+\mb{\hat{x}}-\mb{\hat{y}}} & \\
        &  \mb{l}_{\mb{k}-\mb{\hat{x}}-2\mb{\hat{y}}} &  \mb{l}_{\mb{k}-2\mb{\hat{y}}} & \mb{l}_{\mb{k}+\mb{\hat{x}}-2\mb{\hat{y}}} & \\
     & & \vdots & & 
    \end{pmatrix}\nonumber\\
    = \exp\left[-ip_\mb{k}^g\begin{pmatrix}
      \mb{l}_{\mb{k}-\mb{\hat{x}}+2\mb{\hat{y}}} &  \mb{l}_{\mb{k}+2\mb{\hat{y}}} & \mb{l}_{\mb{k}+\mb{\hat{x}}+2\mb{\hat{y}}}& & \\     \mb{l}_{\mb{k}-\mb{\hat{x}}+\mb{\hat{y}}} &  \mb{l}_{\mb{k}+\mb{\hat{y}}} & \mb{l}_{\mb{k}+\mb{\hat{x}}+\mb{\hat{y}}}& & \\
      \mb{l}_{\mb{k}-\mb{\hat{x}}} &  \mb{l}_{\mb{k}} & \mb{l}_{\mb{k}+\mb{\hat{x}}} & , 
      & \mb{l}_{\mb{k}}'\\
       \mb{l}_{\mb{k}-\mb{\hat{x}}-\mb{\hat{y}}} &  \mb{l}_{\mb{k}-\mb{\hat{y}}} &  \mb{l}_{\mb{k}+\mb{\hat{x}}-\mb{\hat{y}}} & & \\
           \mb{l}_{\mb{k}-\mb{\hat{x}}-2\mb{\hat{y}}} &  \mb{l}_{\mb{k}-2\mb{\hat{y}}} &  \mb{l}_{\mb{k}+\mb{\hat{x}}-2\mb{\hat{y}}} & &
    \end{pmatrix}\right]
     \, ,
 \end{align}
 for some functions $p^g_\mb{k}\in \mathbb{R}$. Note that the arguments of $\theta_g$ and $p_\mb{k}^g$ were written out in a two-dimensional array, such that the dependence on the l-bit indices within the causal cone of $\ket{\mathbf{l}_\mathbf{k}} \bra{\mathbf{l}_\mathbf{k}'}$ in \eqref{eq:algebraiclightcone} is apparent. Let us introduce $f_g(\{\mathbf{l}_\mb{k}\})$ defined by $\theta_g(\{\mathbf{l}_\mb{k}\}) = \exp[if_g(\{\mathbf{l}_\mb{k}\})]$, so that we have
\begin{align}
    f_g\begin{pmatrix}
    & & \vdots & & \\
        & \mb{l}_{\mb{k}-\mb{\hat{x}}+2 \mb{\hat{y}}} &  \mb{l}_{\mb{k}+2 \mb{\hat{y}}} & \mb{l}_{\mb{k}+\mb{\hat{x}}+2 \mb{\hat{y}}}& \\
    & \mb{l}_{\mb{k}-\mb{\hat{x}}+\mb{\hat{y}}} &  \mb{l}_{\mb{k}+\mb{\hat{y}}} & \mb{l}_{\mb{k}+\mb{\hat{x}}+\mb{\hat{y}}}& \\
    \cdots&  \mb{l}_{\mb{k}-\mb{\hat{x}}} &  \mb{l}_{\mb{k}} & \mb{l}_{\mb{k}+\mb{\hat{x}}} &\cdots\\
     &  \mb{l}_{\mb{k}-\mb{\hat{x}}-\mb{\hat{y}}} &  \mb{l}_{\mb{k}-\mb{\hat{y}}} & \mb{l}_{\mb{k}+\mb{\hat{x}}-\mb{\hat{y}}} & \\
          &  \mb{l}_{\mb{k}-\mb{\hat{x}}-2\mb{\hat{y}}} &  \mb{l}_{\mb{k}-2\mb{\hat{y}}} & \mb{l}_{\mb{k}+\mb{\hat{x}}-2\mb{\hat{y}}} & \\
     & & \vdots & & 
    \end{pmatrix} - \nonumber\\
    f_g\begin{pmatrix}
    & & \vdots & & \\
        & \mb{l}_{\mb{k}-\mb{\hat{x}}+2\mb{\hat{y}}} &  \mb{l}_{\mb{k}+2\mb{\hat{y}}} & \mb{l}_{\mb{k}+\mb{\hat{x}}+2\mb{\hat{y}}}& \\
    & \mb{l}_{\mb{k}-\mb{\hat{x}}+\mb{\hat{y}}} &  \mb{l}_{\mb{k}+\mb{\hat{y}}} & \mb{l}_{\mb{k}+\mb{\hat{x}}+\mb{\hat{y}}}& \\
    \cdots&  \mb{l}_{\mb{k}-\mb{\hat{x}}} &  \mb{l}_{\mb{k}}' & \mb{l}_{\mb{k}+\mb{\hat{x}}} &\cdots\\
     &  \mb{l}_{\mb{k}-\mb{\hat{x}}-\mb{\hat{y}}} &  \mb{l}_{\mb{k}-\mb{\hat{y}}} & \mb{l}_{\mb{k}+\mb{\hat{x}}-\mb{\hat{y}}} & \\
          &  \mb{l}_{\mb{k}-\mb{\hat{x}}-2\mb{\hat{y}}} &  \mb{l}_{\mb{k}-2\mb{\hat{y}}} & \mb{l}_{\mb{k}+\mb{\hat{x}}-2\mb{\hat{y}}} & \\
     & & \vdots & & 
    \end{pmatrix}\nonumber\\
    = p_\mb{k}^g\begin{pmatrix}
         \mb{l}_{\mb{k}-\mb{\hat{x}}+2\mb{\hat{y}}} &  \mb{l}_{\mb{k}+2\mb{\hat{y}}} & \mb{l}_{\mb{k}+\mb{\hat{x}}+2\mb{\hat{y}}}& & \\
     \mb{l}_{\mb{k}-\mb{\hat{x}}+\mb{\hat{y}}} &  \mb{l}_{\mb{k}+\mb{\hat{y}}} & \mb{l}_{\mb{k}+\mb{\hat{x}}+\mb{\hat{y}}}& & \\
      \mb{l}_{\mb{k}-\mb{\hat{x}}} &  \mb{l}_{\mb{k}} & \mb{l}_{\mb{k}+\mb{\hat{x}}} & , 
      & \mb{l}_{\mb{k}}'\\
       \mb{l}_{\mb{k}-\mb{\hat{x}}-\mb{\hat{y}}} &  \mb{l}_{\mb{k}-\mb{\hat{y}}} &  \mb{l}_{\mb{k}+\mb{\hat{x}}-\mb{\hat{y}}} & & \\
           \mb{l}_{\mb{k}-\mb{\hat{x}}-2\mb{\hat{y}}} &  \mb{l}_{\mb{k}-2\mb{\hat{y}}} &  \mb{l}_{\mb{k}+\mb{\hat{x}}-2\mb{\hat{y}}} & &
    \end{pmatrix}  \;\mathrm{mod}\;2\pi,
\end{align}
and
\begin{align}
    f_g\begin{pmatrix}
    & & \vdots & & \\
        & \mb{l}_{\mb{k}-\mb{\hat{x}} +\mb{\hat{y}}} &  \mb{l}_{\mb{k} + \mb{\hat{y}}}& \mb{l}_{\mb{k}+\mb{\hat{x}} +\mb{\hat{y}} }& \\
    & \mb{l}_{\mb{k}-\mb{\hat{x}}} &  \mb{l}_{\mb{k}}' & \mb{l}_{\mb{k}+\mb{\hat{x}}}& \\
    \cdots& \mb{l}_{\mb{k}-\mb{\hat{x}}-\mb{\hat{y}}} &  \mb{l}_{\mb{k}-\mb{\hat{y}}} & \mb{l}_{\mb{k}+\mb{\hat{x}}-\mb{\hat{y}}} &\cdots\\
     &  \mb{l}_{\mb{k}-\mb{\hat{x}}-2\mb{\hat{y}}} &  \mb{l}_{\mb{k}-2\mb{\hat{y}}} & \mb{l}_{\mb{k}+\mb{\hat{x}}-2\mb{\hat{y}}} & \\
      &  \mb{l}_{\mb{k}-\mb{\hat{x}}-3\mb{\hat{y}}} &  \mb{l}_{\mb{k}-3\mb{\hat{y}}} & \mb{l}_{\mb{k}+\mb{\hat{x}}-3\mb{\hat{y}}} & \\
     & & \vdots & & 
    \end{pmatrix} - \nonumber\\
    f_g\begin{pmatrix}
    & & \vdots & & \\
        & \mb{l}_{\mb{k}-\mb{\hat{x}} +\mb{\hat{y}}} &  \mb{l}_{\mb{k} + \mb{\hat{y}}} & \mb{l}_{\mb{k}+\mb{\hat{x}} + \mb{\hat{y}}}& \\
    & \mb{l}_{\mb{k}-\mb{\hat{x}}} &  \mb{l}_{\mb{k}}' & \mb{l}_{\mb{k}+\mb{\hat{x}}}& \\
    \cdots& 
    \mb{l}_{\mb{k}-\mb{\hat{x}}-\mb{\hat{y}}} &  \mb{l}_{\mb{k}-\mb{\hat{y}}}' & \mb{l}_{\mb{k}+\mb{\hat{x}}-\mb{\hat{y}}} 
    &\cdots\\
     &  \mb{l}_{\mb{k}-\mb{\hat{x}}-2\mb{\hat{y}}} &  \mb{l}_{\mb{k}-2\mb{\hat{y}}} & \mb{l}_{\mb{k}+\mb{\hat{x}}-2\mb{\hat{y}}} & \\
     &  \mb{l}_{\mb{k}-\mb{\hat{x}}-3\mb{\hat{y}}} &  \mb{l}_{\mb{k}-3\mb{\hat{y}}} & \mb{l}_{\mb{k}+\mb{\hat{x}}-3\mb{\hat{y}}} & \\
     & & \vdots & & 
    \end{pmatrix}\nonumber\\
    = p_\mb{k-\hat{y}}^g\begin{pmatrix}
       \mb{l}_{\mb{k}-\mb{\hat{x}} + \mb{\hat{y}}} &  \mb{l}_{\mb{k} +\mb{\hat{y}}} & \mb{l}_{\mb{k}+\mb{\hat{x}} +\mb{\hat{y}}} & & \\
      \mb{l}_{\mb{k}-\mb{\hat{x}}} &  \mb{l}_{\mb{k}}' & \mb{l}_{\mb{k}+\mb{\hat{x}}} & & \\
      \mb{l}_{\mb{k}-\mb{\hat{x}}-\mb{\hat{y}}} &  \mb{l}_{\mb{k}-\mb{\hat{y}}} &  \mb{l}_{\mb{k}+\mb{\hat{x}}-\mb{\hat{y}}}
      & , 
      & \mb{l}_{\mb{k}-\mb{\hat{y}}}'\\
       \mb{l}_{\mb{k}-\mb{\hat{x}}-2\mb{\hat{y}}} &  \mb{l}_{\mb{k}-2\mb{\hat{y}}} &  \mb{l}_{\mb{k}+\mb{\hat{x}}-2\mb{\hat{y}}} & &
       \\
       \mb{l}_{\mb{k}-\mb{\hat{x}}-3\mb{\hat{y}}} &  \mb{l}_{\mb{k}-3\mb{\hat{y}}} &  \mb{l}_{\mb{k}+\mb{\hat{x}}-3\mb{\hat{y}}} & &
    \end{pmatrix}  \;\mathrm{mod}\;2\pi,
\end{align}
where in the second equation we act with $|\mb l_{\mb k- \hat{\mb y}}\rangle \langle \mb l'_{\mb k - \hat{\mb y}}|$ on the block of sites at $\mathbf{k}-\hat{\mathbf{y}}$ instead of $\mathbf{k}$. We sweep column-by-column through the lattice and write down analogous equations corresponding to cases where that operator acts on other blocks.  
As an example, at an intermediate step, we have, at some point $\mb{r}$
\begin{align}
    f_g\begin{pmatrix}
    & & \vdots & & \\
   & \mb{l}_{\mb{r}-\mb{\hat{x}}+2\mb{\hat{y}}}' &  \mb{l}_{\mb{r}+2\mb{\hat{y}}}' & \mb{l}_{\mb{r}+\mb{\hat{x}}+2\mb{\hat{y}}}& \\
    & \mb{l}_{\mb{r}-\mb{\hat{x}}+\mb{\hat{y}}}' &  \mb{l}_{\mb{r}+\mb{\hat{y}}}' & \mb{l}_{\mb{r}+\mb{\hat{x}}+\mb{\hat{y}}}& \\
    \cdots&  \mb{l}_{\mb{r}-\mb{\hat{x}}}' &  \mb{l}_{\mb{r}} & \mb{l}_{\mb{r}+\mb{\hat{x}}} &\cdots\\
     &  \mb{l}_{\mb{r}-\mb{\hat{x}}-\mb{\hat{y}}}' &  \mb{l}_{\mb{r}-\mb{\hat{y}}} & \mb{l}_{\mb{r}+\mb{\hat{x}}-\mb{\hat{y}}} & \\
     &  \mb{l}_{\mb{r}-\mb{\hat{x}}-2\mb{\hat{y}}}' &  \mb{l}_{\mb{r}-2\mb{\hat{y}}} & \mb{l}_{\mb{r}+\mb{\hat{x}}-2\mb{\hat{y}}} & \\
     & & \vdots & & 
    \end{pmatrix} - \nonumber\\
    f_g\begin{pmatrix}
    & & \vdots & & \\
    & \mb{l}_{\mb{r}-\mb{\hat{x}}+2\mb{\hat{y}}}' &  \mb{l}_{\mb{r}+2\mb{\hat{y}}}' & \mb{l}_{\mb{r}+\mb{\hat{x}}+2\mb{\hat{y}}}& \\
    & \mb{l}_{\mb{r}-\mb{\hat{x}}+\mb{\hat{y}}}' &  \mb{l}_{\mb{r}+\mb{\hat{y}}}' & \mb{l}_{\mb{r}+\mb{\hat{x}}+\mb{\hat{y}}}& \\
    \cdots&  \mb{l}_{\mb{r}-\mb{\hat{x}}}' &  \mb{l}_{\mb{r}}' & \mb{l}_{\mb{r}+\mb{\hat{x}}} &\cdots\\
     &  \mb{l}_{\mb{r}-\mb{\hat{x}}-\mb{\hat{y}}}' &  \mb{l}_{\mb{r}-\mb{\hat{y}}} & \mb{l}_{\mb{r}+\mb{\hat{x}}-\mb{\hat{y}}} & \\
     &  \mb{l}_{\mb{r}-\mb{\hat{x}}-2\mb{\hat{y}}}' &  \mb{l}_{\mb{r}-2\mb{\hat{y}}} & \mb{l}_{\mb{r}+\mb{\hat{x}}-2\mb{\hat{y}}} & \\ 
     & & \vdots & & 
    \end{pmatrix}\nonumber\\
    = p_\mb{r}^g\begin{pmatrix}
        \mb{l}_{\mb{r}-\mb{\hat{x}}+2\mb{\hat{y}}}' &  \mb{l}_{\mb{r}+2\mb{\hat{y}}}' & \mb{l}_{\mb{r}+\mb{\hat{x}}+2\mb{\hat{y}}}& & \\
     \mb{l}_{\mb{r}-\mb{\hat{x}}+\mb{\hat{y}}}' &  \mb{l}_{\mb{r}+\mb{\hat{y}}}' & \mb{l}_{\mb{r}+\mb{\hat{x}}+\mb{\hat{y}}}& & \\
      \mb{l}_{\mb{r}-\mb{\hat{x}}}' &  \mb{l}_{\mb{r}} & \mb{l}_{\mb{r}+\mb{\hat{x}}} & , 
      & \mb{l}_{\mb{r}}'\\
       \mb{l}_{\mb{r}-\mb{\hat{x}}-\mb{\hat{y}}}' &  \mb{l}_{\mb{r}-\mb{\hat{y}}} &  \mb{l}_{\mb{r}+\mb{\hat{x}}-\mb{\hat{y}}} & &
       \\
       \mb{l}_{\mb{r}-\mb{\hat{x}}-2\mb{\hat{y}}}' &  \mb{l}_{\mb{r}-2\mb{\hat{y}}} &  \mb{l}_{\mb{r}+\mb{\hat{x}}-2\mb{\hat{y}}} & &
    \end{pmatrix} \;\mathrm{mod}\;2\pi
\end{align}
Adding up all of these equations leads to 
\begin{align}
        f_g(\{\mb{l}_\mb{k}\}) - f_g(\{\mb{l}_\mb{k}'\}) = \notag 
        \\
        \sum_{\mb{r}} p_\mb{r}^g
        \begin{pmatrix}
      \mb{l}_{\mb{r}-\mb{\hat{x}}+2\mb{\hat{y}}}' &  \mb{l}_{\mb{r}+2\mb{\hat{y}}}' & \mb{l}_{\mb{r}+\mb{\hat{x}}+2\mb{\hat{y}}}& & \\  
     \mb{l}_{\mb{r}-\mb{\hat{x}}+\mb{\hat{y}}}' &  \mb{l}_{\mb{r}+\mb{\hat{y}}}' & \mb{l}_{\mb{r}+\mb{\hat{x}}+\mb{\hat{y}}}& & \\
      \mb{l}_{\mb{r}-\mb{\hat{x}}}' &  \mb{l}_{\mb{r}} & \mb{l}_{\mb{r}+\mb{\hat{x}}} & , 
      & \mb{l}_{\mb{r}}'\\
       \mb{l}_{\mb{r}-\mb{\hat{x}}-\mb{\hat{y}}}' &  \mb{l}_{\mb{r}-\mb{\hat{y}}} &  \mb{l}_{\mb{r}+\mb{\hat{x}}-\mb{\hat{y}}} & & \\
         \mb{l}_{\mb{r}-\mb{\hat{x}}-2\mb{\hat{y}}}' &  \mb{l}_{\mb{r}-2\mb{\hat{y}}} &  \mb{l}_{\mb{r}+\mb{\hat{x}}-2\mb{\hat{y}}} & &
    \end{pmatrix} \notag \\ + \;\mathrm{boundary\;terms} \;\;\mathrm{mod}\;2\pi \, .
\end{align}
Now we set all the primed indices to zero, i.e. let $\mb{l_k'} = (\mathbf{0},\mathbf{0},\mathbf{0},\mathbf{0})$ for all $\mb{k}$.  This implies that there exist functions of five $\mb{l_k}$ indices $q_\mb{r}^g$ such that we can write
\begin{equation}\label{eq:5indices}
        f_g(\{\mb{l}_\mb{k}\})  =   \sum\limits_{\mb{r}} q_\mb{r}^g\begin{pmatrix}
     & \mb{l}_{\mb{r}+\mb{\hat{x}}+2\mb{\hat{y}}} \\
 & \mb{l}_{\mb{r}+\mb{\hat{x}}+\mb{\hat{y}}} \\
    \mb{l}_{\mb{r}} & \mb{l}_{\mb{r}+\mb{\hat{x}}}\\
     \mb{l}_{\mb{r}-\mb{\hat{y}}} &  \mb{l}_{\mb{r}+\mb{\hat{x}}-\mb{\hat{y}}}\\
     \mb{l}_{\mb{r}-2\mb{\hat{y}}} &  \mb{l}_{\mb{r}+\mb{\hat{x}}-2\mb{\hat{y}}}
    \end{pmatrix} \ .
\end{equation} 
But we could have just as well applied the above argument sweeping row-by-row, which leads to
\begin{equation}\label{eq:5indicesv2}
        f_g(\{\mb{l}_\mb{k}\})  =   \sum\limits_{\mb{r}} q_\mb{r}^g\begin{pmatrix}
  &\mb{l}_{\mb{r}}& 
    \mb{l}_{\mb{r}+\mb{\hat{x}}}  \\
\mb{l}_{\mb r - \mb{\hat x} - \mb{\hat y}}     & \mb{l}_{\mb r - \mb{\hat y}} 
    & \mb{l}_{\mb r + \mb{\hat x} - \mb{\hat y}} \\
        \mb{l}_{\mb{r} - \mb{\hat x} - 2 \mb{\hat y}}     &
        \mb{l}_{\mb{r} - 2\mb{\hat{y}}} &
        \mb{l}_{\mb{r}+ \mb{\hat x} - 2 \mb{\hat{y}}} 
    \end{pmatrix} \ .
\end{equation} 
Comparing the last two equations shows that there must exist functions $s_\mb{r}^g$ of six $\mb{l_k}$ indices such that we can write
\begin{equation}\label{eq:4indices}
        f_g(\{\mb{l}_\mb{k}\})  =   \sum\limits_{\mb{r}} s_\mb{r}^g\begin{pmatrix}
    \mb{l}_{\mb{r}} & \mb{l}_{\mb{r}+\mb{\hat{x}}}\\
     \mb{l}_{\mb{r}-\mb{\hat{y}}} &  \mb{l}_{\mb{r}+\mb{\hat{x}}-\mb{\hat{y}}} \\
     \mb{l}_{\mb{r}-2\mb{\hat{y}}} &  \mb{l}_{\mb{r}+\mb{\hat{x}}-2\mb{\hat{y}}}     
    \end{pmatrix} \; .
\end{equation}
Therefore $\Theta_g$ can be expressed as a four-layer quantum circuit whose unitary matrices $\theta_{g, \mb k}$ are all diagonal and can be arranged as shown in Fig.~\ref{fig:4x6}. Those unitaries act on plaquettes of $2 \ell \times 3 \ell$ sites. 

%

\subsection{Reduction to one dimensional problem} \label{sec:red1dim}

We have $\mathpzc{v}_g^{\otimes N^2}U = U\Theta_g$, where the left hand side (LHS) is a four-layer quantum circuit like Fig.~\ref{fig:2dqc}, and the right hand side (RHS) is an eight-layer quantum circuit.  We then reduce the two-dimensional quantum circuit to a one-dimensional one by blocking unitaries along the $y$-direction.  We then obtain, along the $x$-direction
\begin{align}
    \label{eq:1d1}
  \includegraphics[width=0.9\linewidth,valign=c]{2019_07_draw4_v2.eps}  \; , 
\end{align}
where the encircled $g$ denotes the tensor product $\mathpzc{v}_g^{\otimes N \ell/2}$, i.e., a stripe along the $y$-direction. Each $U_k$, $V_k$ corresponds to a quantum circuit along the  $y$-direction as in Eq.~\eqref{eq:Udef}.
Each $\Theta^g_k$ in Eq.~\eqref{eq:1d1} is a quantum circuit of diagonal matrices $\theta_{g}$ acting on plaquettes of $2 \ell \times 3 \ell$ sites each
\begin{equation}\label{eq:thetadecomp}
  \Theta_k^g = \;\includegraphics[width=0.7\linewidth,valign=c]{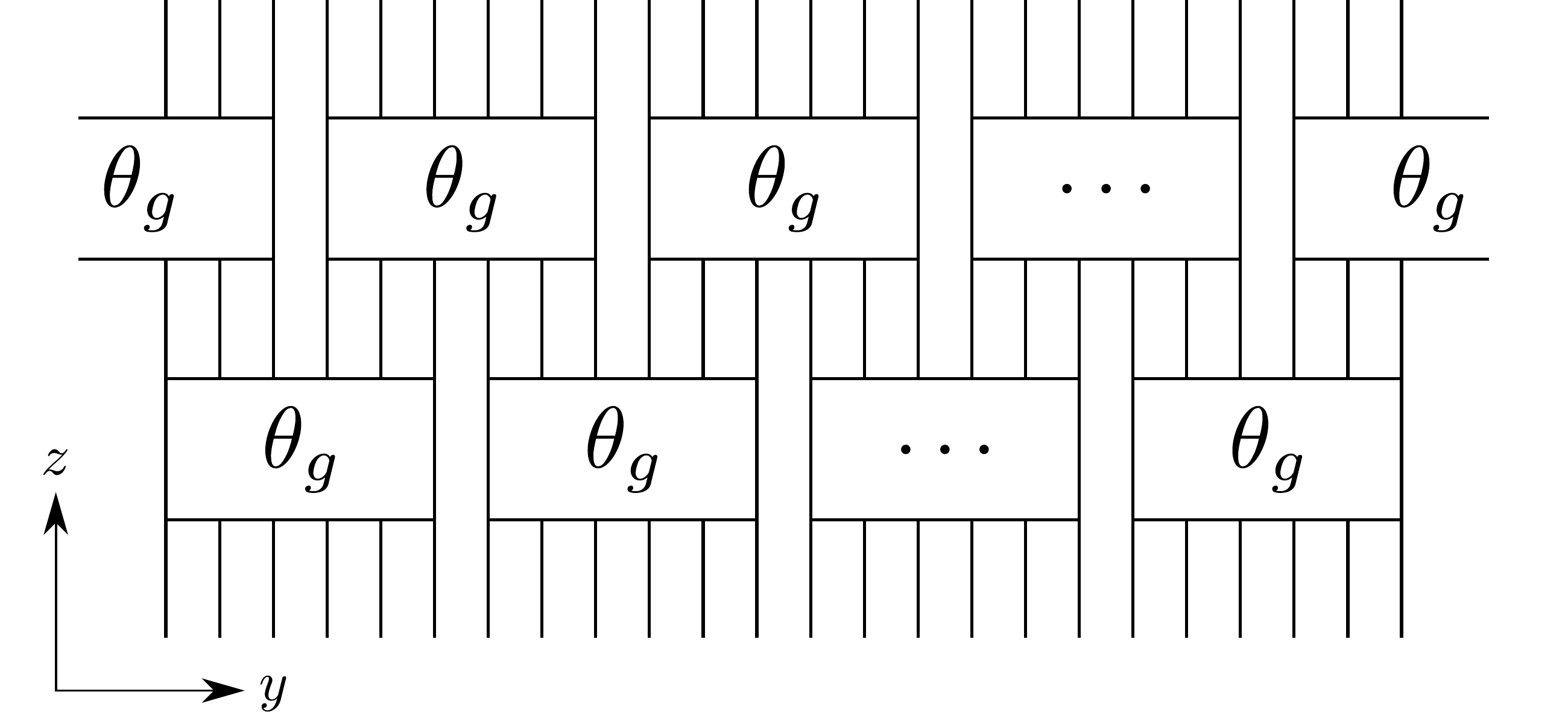}.
\end{equation}
Note that in Eq.~\eqref{eq:1d1} we have used indices on the $U$, $V$, and $\Theta$-tensors to emphasize the non-translation invariance, while we have employed the index-free notation on the RHS of  Eq.~\eqref{eq:thetadecomp}.

Eq.~\eqref{eq:1d1} also appears in the exact same form in the one-dimensional classification of SPT MBL phases~\cite{1DSPTMBL}.  Using the blocking indicated by dashed lines in Eq.~\eqref{eq:1d1} reveals that it is an equation relating two one-dimensional two-layer quantum circuits. Hence, we can use the results below Eq.~\eqref{eq:qu_circuits} and deduce the existence of gauge tensors $W_k$, which transform unitaries of both sides of the equation into each other. These unitaries depend on the group element $g$, and we refer to them as $W_k^g$. 
The result (see Ref.~\onlinecite{1DSPTMBL} for details) is that the $W_k^g$ have to fulfill 
\begin{align}
    \label{eq:wxw1}
   \includegraphics[width = 0.45\linewidth,valign=c]{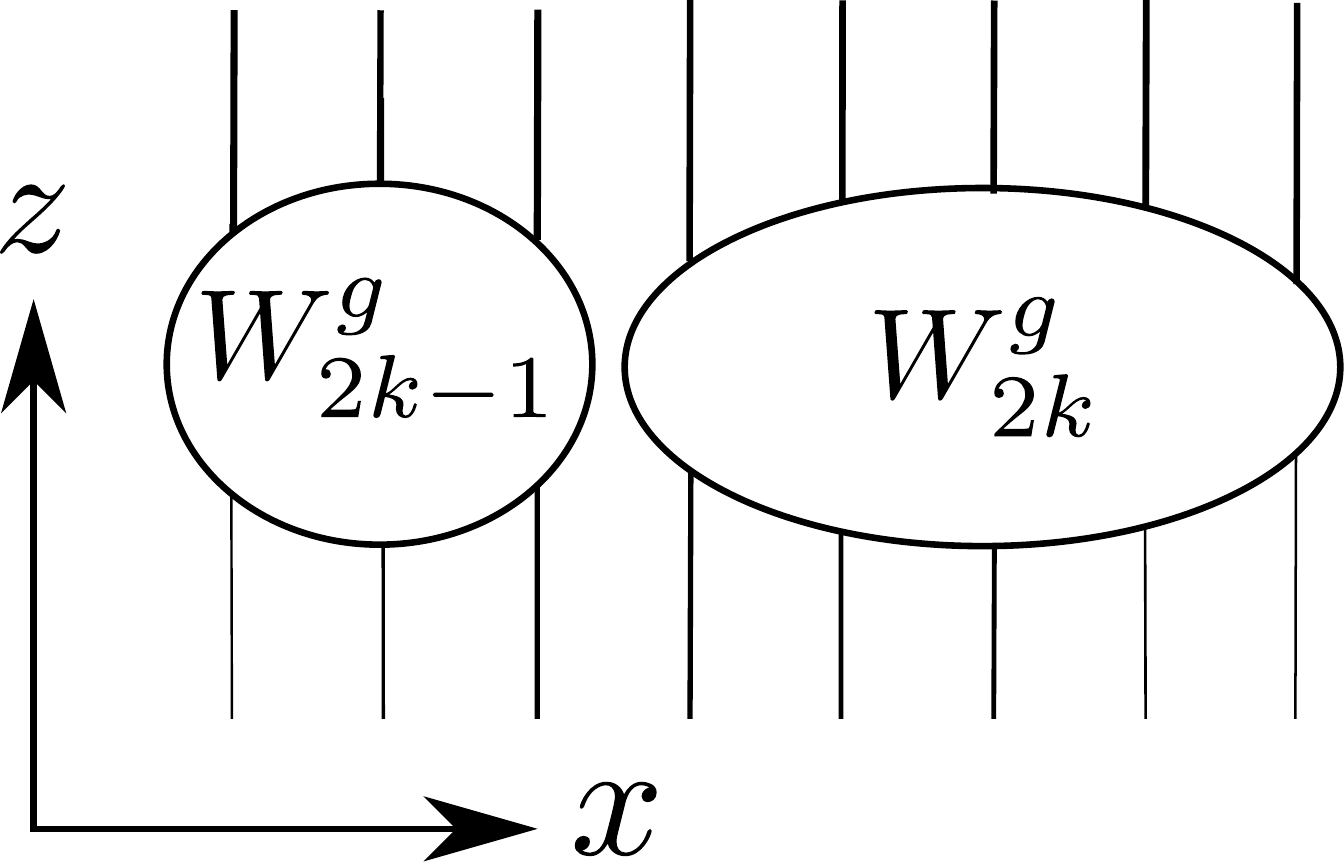} \;\;=\nonumber\\  \includegraphics[width = 0.7\linewidth,valign=c]{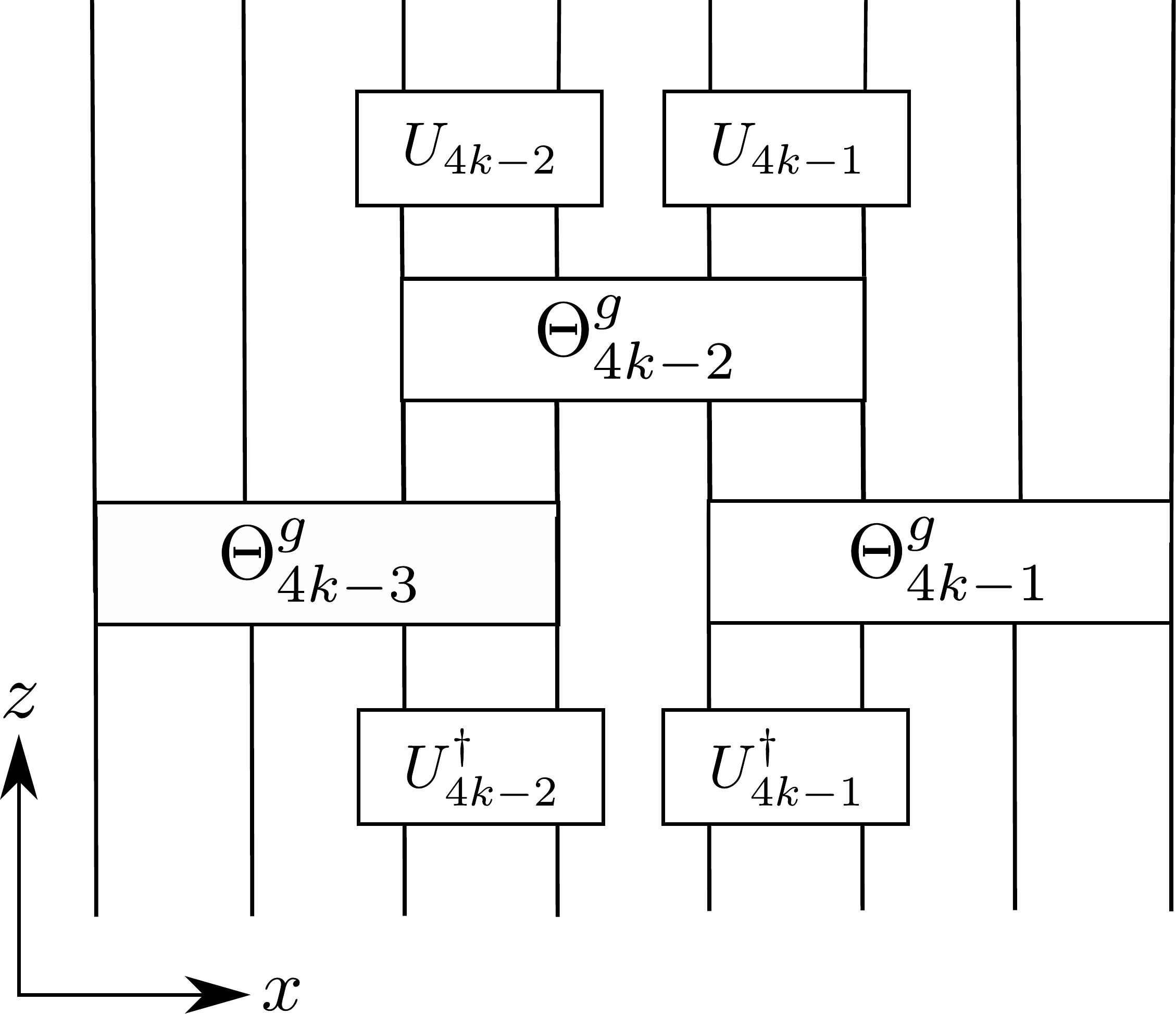}  \; \;,
\end{align}
where the differing numbers of legs of $W_k^g$ for $k$ even and odd are due to the blocking scheme used in that calculation, and also
\begin{align}
    \label{eq:wxw2}
   \includegraphics[width = 0.45\linewidth,valign=c]{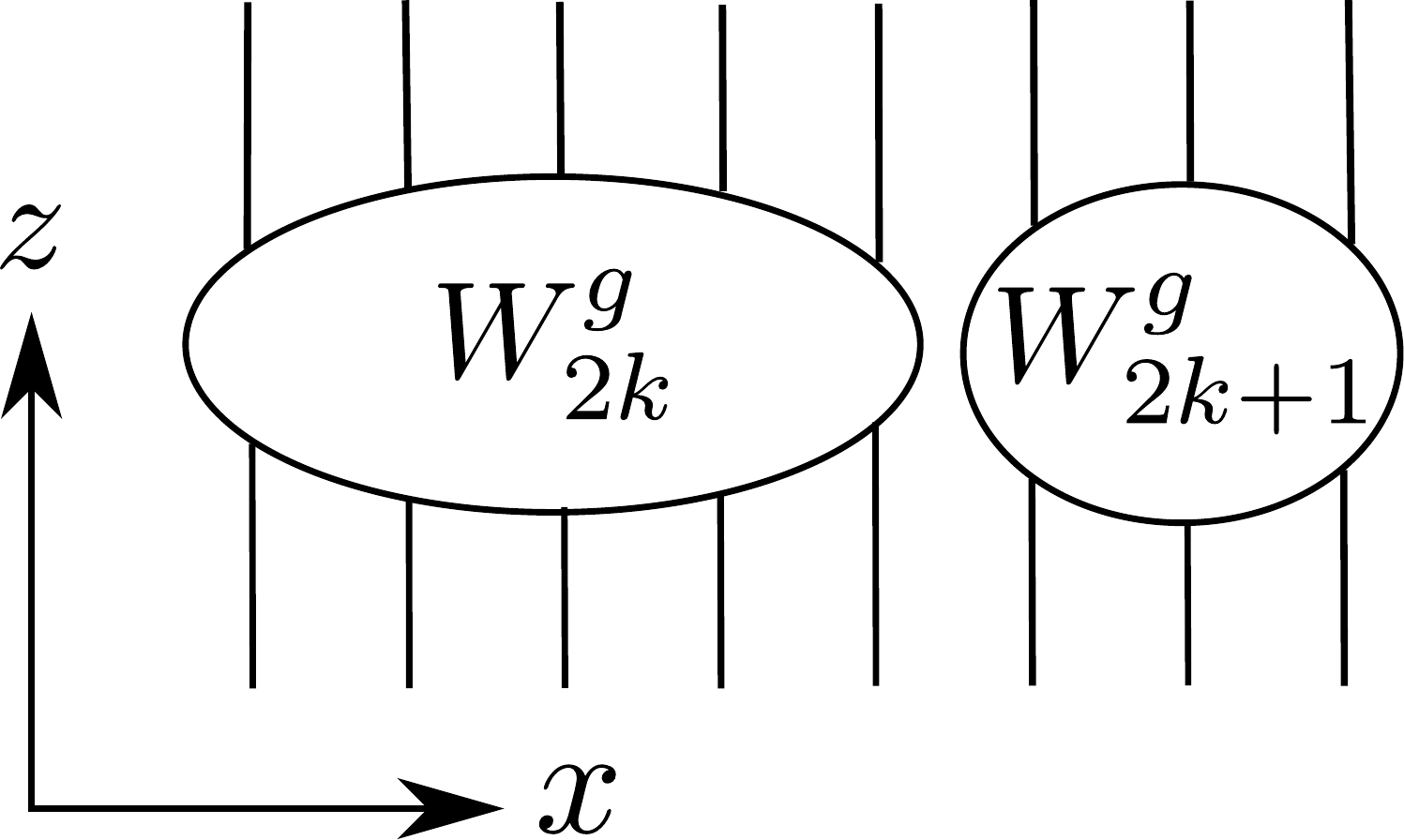} \;\;=\nonumber\\  \includegraphics[width = 0.7\linewidth,valign=c]{2019_07_ww4_v2} \; \; .
\end{align}
Since all unitaries on the right hand side of Eqs.~\eqref{eq:wxw1} and~\eqref{eq:wxw2} are quantum circuits along the  $y$-direction, the $W_k^g$ must also be strictly short-range correlated along that direction. Thus, they can at least be efficiently approximated by two-layer one-dimensional quantum circuits along the  $y$-direction (cf. assumptions made in the beginning of Section~\ref{sec:nontechnical}). 

It can also be shown (again see Ref.~\onlinecite{1DSPTMBL} for details) that the $W_k^g$ form a projective representation of $G$, i.e. for all $k$, it is the case that $W_k^g W_k^h = \beta_k(g,h)W_k^{gh}$ for some $\beta_k(g,h)\in U(1)$.  For the two dimensional classification of SPT MBL phases, we will use this result combined with the lemma below.

\subsection{Quantum circuit representations and the third cohomology group}\label{sec:qcreps}

We now prove our main statement:  
The quantum circuit projective representations of a given group $G$ are labeled by the elements of the third cohomology group $H^3(G,U(1))$. That is, quantum circuits corresponding to different third cohomology classes cannot be continuously connected with each other while preserving the fact that they projectively represent the symmetry group. 
We note that this statement is related to the result from Ref.~\onlinecite{MPOgaugingandedge} that injective matrix product operator (MPO) representations of $G$ likewise correspond to the elements of $H^3(G,U(1))$.  Let us also mention that (for a group $G$ and G-module $F$), the third cohomology group $H^3(G,F)$ has a representation theoretic interpretation: similar to how elements of $H^2(G,F)$ correspond to projective representations of $G$, elements of $H^3(G,F)$ correspond to gerbal representations\cite{gerbal} of $G$; see Appendix~\ref{app:endix} for details.

Each $g\in G$ is associated a quantum circuit, which we denote as
\begin{equation}\label{eq:qcugvg}
\begin{tikzpicture}[scale=1.1,baseline=(current  bounding  box.center)]    
\foreach \x in {0,1.6,3.2,4.8}{
\draw[thick](-0.4+\x,-0.6) -- (-0.4+\x,1.4);
\draw[thick](0.4+\x,-0.6) -- (0.4+\x,1.4);
\draw[thick,fill=white] (-0.4+\x,-0.25) rectangle (0.4+\x,0.25);		
}
\foreach \x in {0.8,2.4,4}{
\draw[thick,fill=white] (-0.4+\x,0.55) rectangle (0.4+\x,1.05);		
}
\draw[thick] (-0.8,0.55) -- (-0.4,0.55) -- (-0.4,1.05) -- (-0.8,1.05);
\draw[thick] (5.6,0.55) -- (5.2,0.55) -- (5.2,1.05) -- (5.6,1.05); 
\coordinate[label=right:$u_1^g$] (A) at (-0.3,0);
\coordinate[label=right:$u_2^g$] (A) at (1.3,0);
\coordinate[label=right:$\ldots$] (A) at (2.9,0);
\coordinate[label=right:$u_n^g$] (A) at (4.5,0);
\coordinate[label=right:$v_1^g$] (A) at (0.5,0.8);
\coordinate[label=right:$v_2^g$] (A) at (2.1,0.8);
\coordinate[label=right:$\ldots$] (A) at (3.7,0.8);
\coordinate[label=right:$v_n^g$] (A) at (5.15,0.78);
\coordinate[label=right:$v_n^g$] (A) at (-1,0.78);
\end{tikzpicture}
\end{equation}
To reduce clutter let us adopt a shorthand notation where we label all the $u_k^g$ and $v_k^g$ tensors as simply $g$.  That is, we would write Eq.~\eqref{eq:qcugvg} as
\begin{equation}\label{eq:qcugvg1}
\begin{tikzpicture}[scale=1.1,baseline=(current  bounding  box.center)]    
\foreach \x in {0,1.6,3.2,4.8}{
\draw[thick](-0.4+\x,-0.6) -- (-0.4+\x,1.4);
\draw[thick](0.4+\x,-0.6) -- (0.4+\x,1.4);
\draw[thick,fill=white] (-0.4+\x,-0.25) rectangle (0.4+\x,0.25);		
}
\foreach \x in {0.8,2.4,4}{
\draw[thick,fill=white] (-0.4+\x,0.55) rectangle (0.4+\x,1.05);		
}
\draw[thick] (-0.8,0.55) -- (-0.4,0.55) -- (-0.4,1.05) -- (-0.8,1.05);
\draw[thick] (5.6,0.55) -- (5.2,0.55) -- (5.2,1.05) -- (5.6,1.05); 
\coordinate[label=right:$g$] (A) at (-0.3,0);
\coordinate[label=right:$g$] (A) at (1.3,0);
\coordinate[label=right:$\ldots$] (A) at (2.9,0);
\coordinate[label=right:$g$] (A) at (4.5,0);
\coordinate[label=right:$g$] (A) at (0.5,0.8);
\coordinate[label=right:$g$] (A) at (2.1,0.8);
\coordinate[label=right:$\ldots$] (A) at (3.7,0.8);
\coordinate[label=right:$g$] (A) at (5.15,0.78);
\coordinate[label=right:$g$] (A) at (-1,0.78);
\end{tikzpicture}.
\end{equation}
Let the quantum circuits associated with $g$, $h\in G$ be a projective representation of $G$, i.e. 
\begin{equation}\label{eq:qcrep}
  \cdots  \includegraphics[width=0.25\linewidth,valign=c]{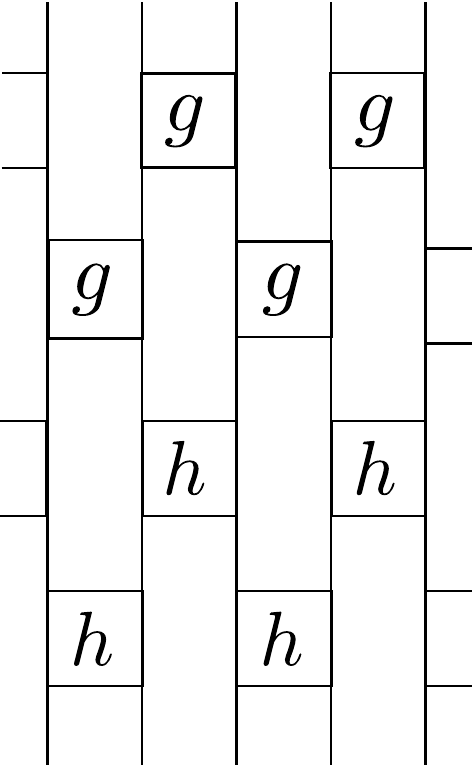}\cdots\;= \;\beta(g,h)\Bigg(\cdots\includegraphics[width=0.25\linewidth,valign=c]{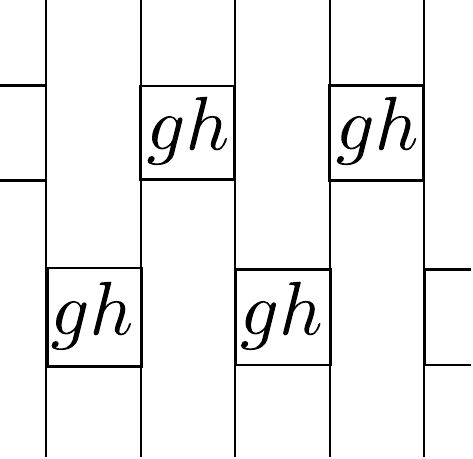}\cdots\Bigg),
\end{equation}
for some $\beta(g,h)\in U(1)$.  In the following steps of the calculation, the factors of $\beta(g,h)$ will only lead to factors like $\frac{\beta(g,h)\beta(gh,k)}{\beta(g,hk)\beta(h,k)}$, which are equal to $1$, due to the 2-cocycle condition for projective representations.  So we omit all the factors of $\beta(g,h)$ hereafter.

Consider blocking unitaries in \eqref{eq:qcugvg1} as follows, 
\begin{align}
    \label{blocking}
    \cdots\includegraphics[width=0.6\linewidth,valign=c]{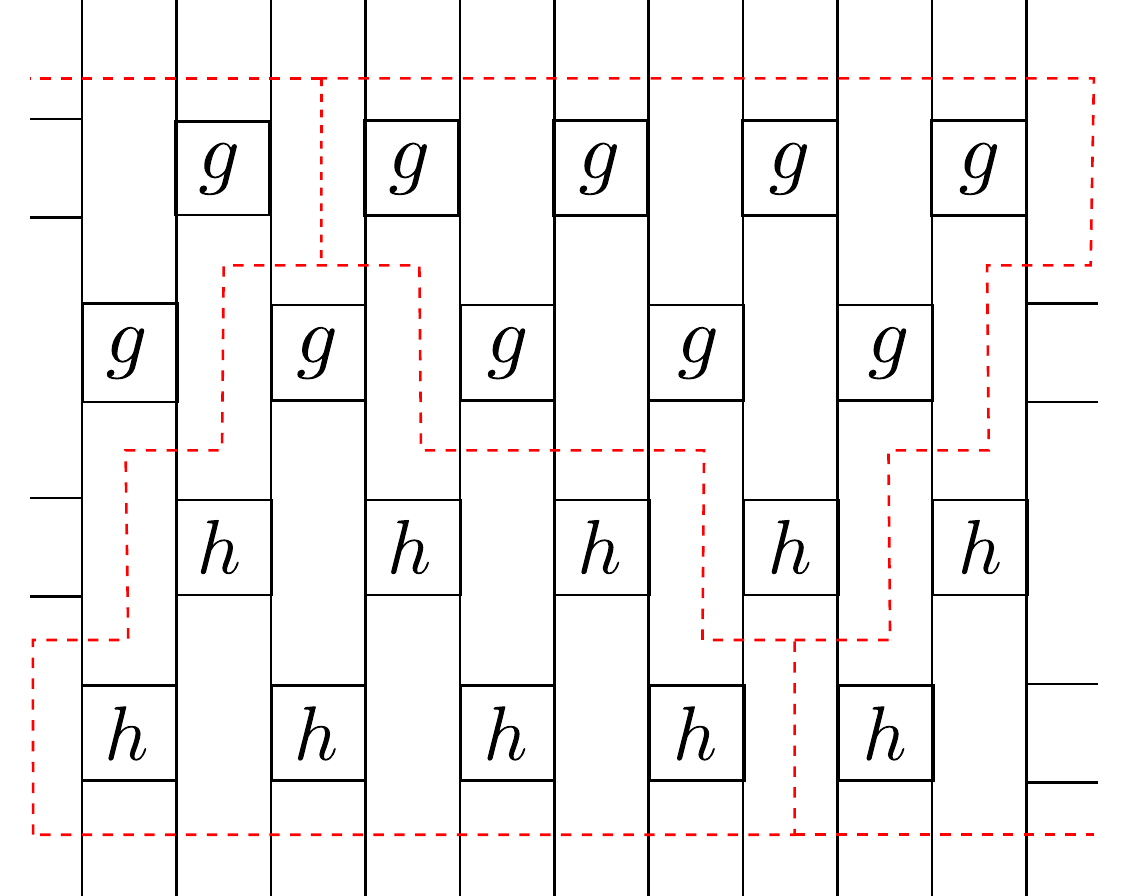}\cdots \; \\ =\; \cdots\includegraphics[width=0.6\linewidth,valign=c]{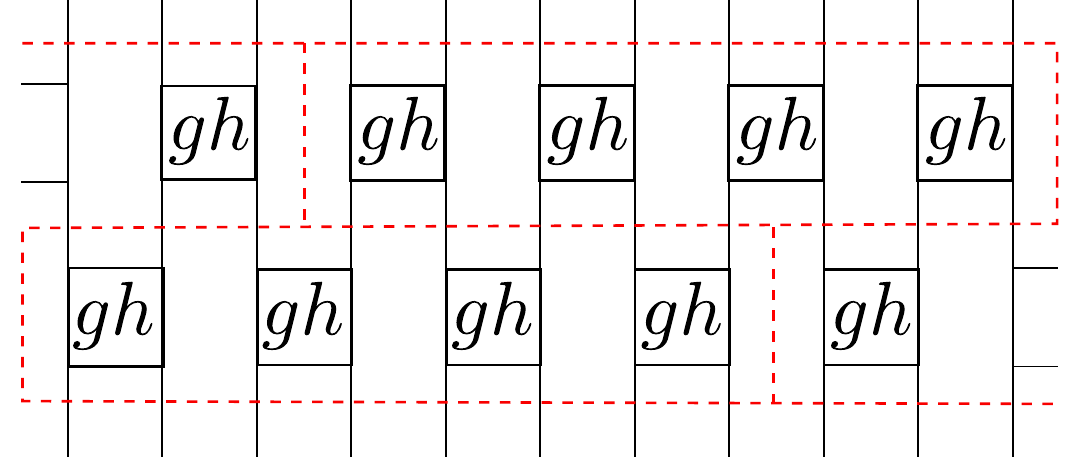}\cdots \; .
\end{align}
In the language of Eq.~\eqref{eq:qu_circuits}, let the blocked tensors on the left hand side be $U'$, $V'$ and the ones on the right hand side $U''$, $V''$.  We deduce the existence of the $W_k$ tensors (which are functions of two group elements here) and plug in Eq.~\eqref{eq:qcpushthru} to obtain
\begin{widetext}
\begin{align}
    \label{blocking1}
    \includegraphics[width=0.59\linewidth,valign=c]{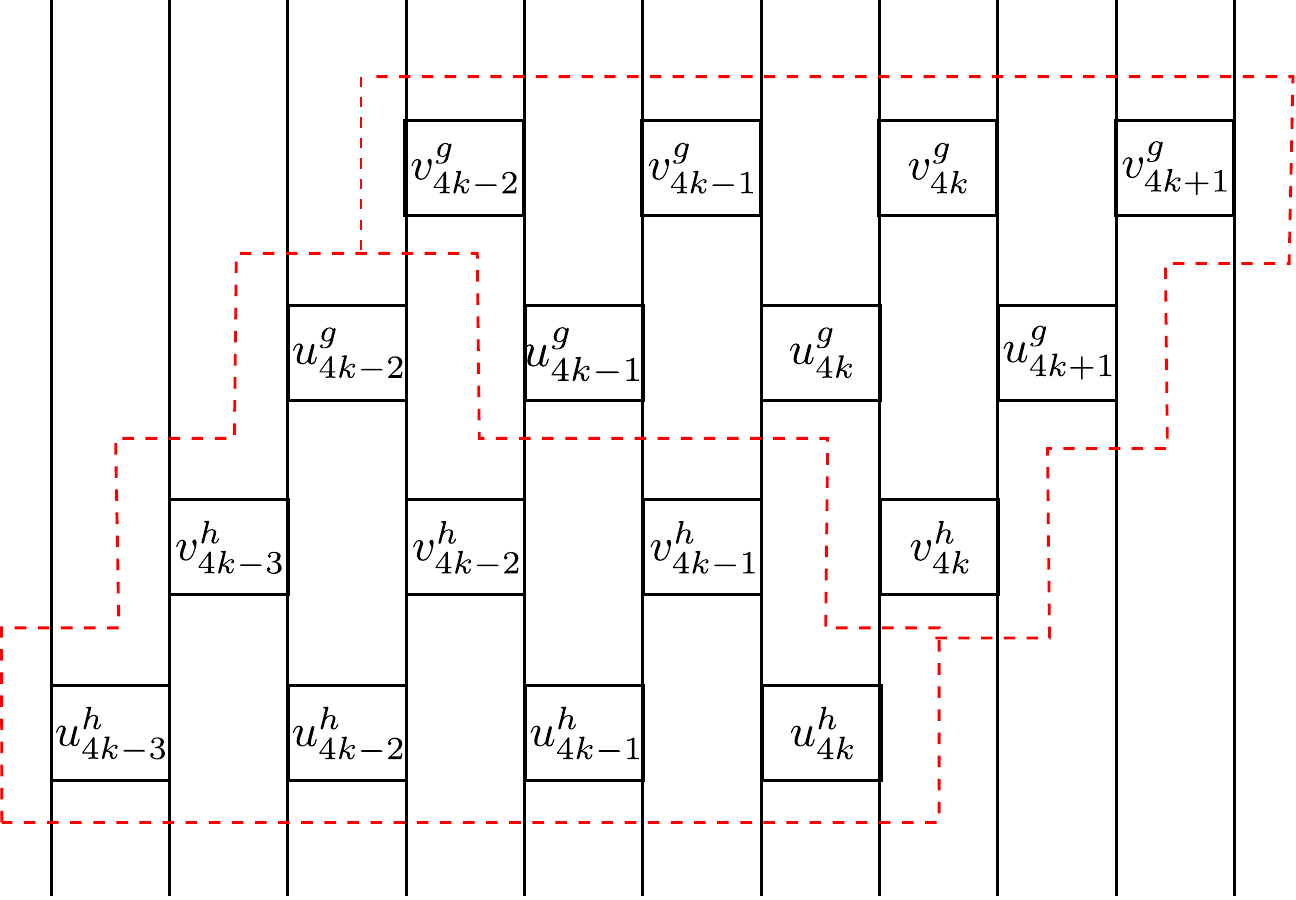} \nonumber \\ =\;\; \includegraphics[width=0.59\linewidth,valign=c]{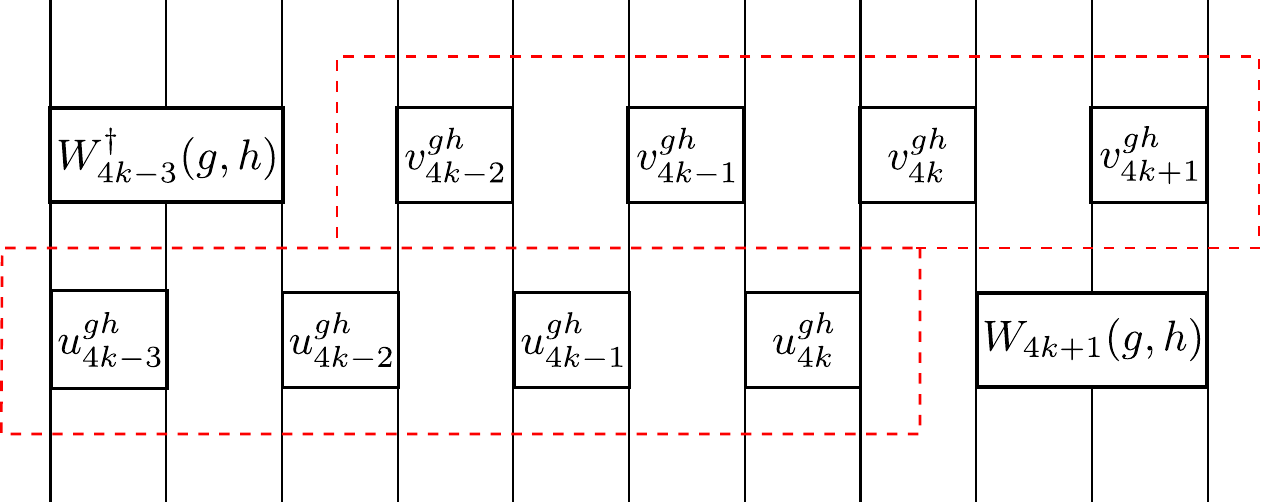} ,
\end{align}
\end{widetext}
where we have temporarily reverted from the abbreviated notation for clarity.  Let us now return to the abbreviated notation, and denote $W_{4k-3}(g,h)$ as $W_L(g,h)$ and $W_{4k+1}(g,h)$ as $W_R(g,h)$.  Eq.~\eqref{blocking1} can be rearranged into
\begin{align}
    \label{blocking2}
    \includegraphics[width=0.63\linewidth,valign=c]{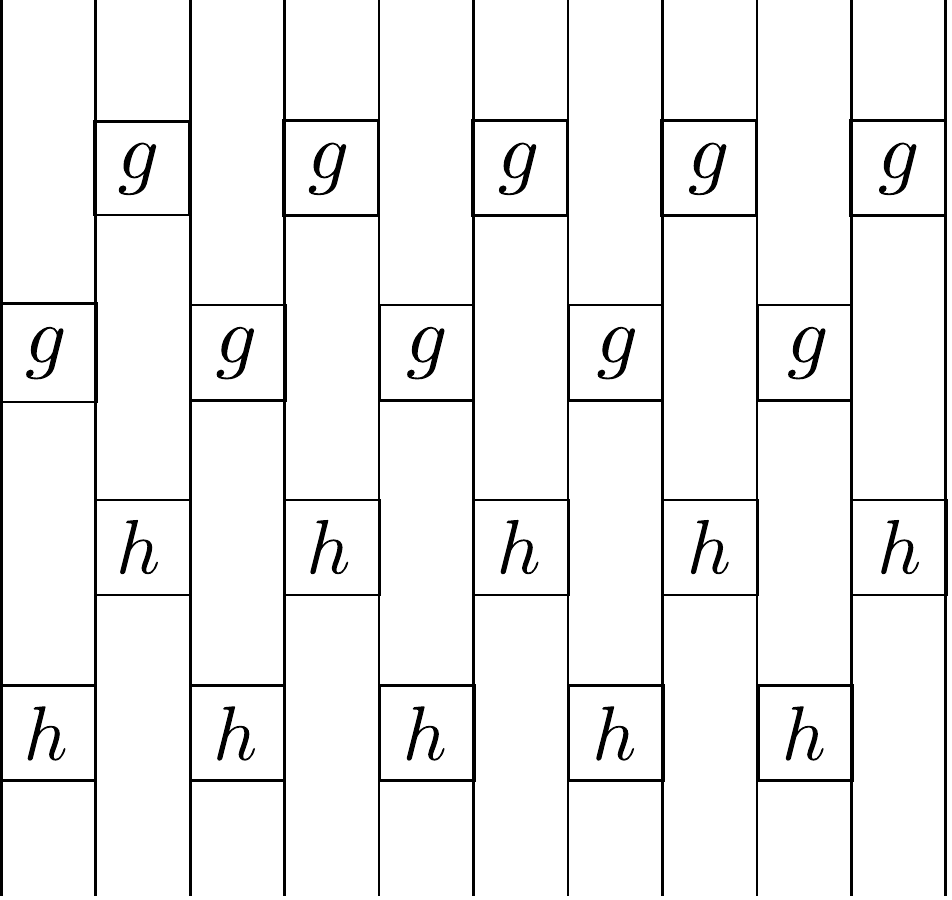}\nonumber\\ =\; \;\includegraphics[width=0.75\linewidth,valign=c]{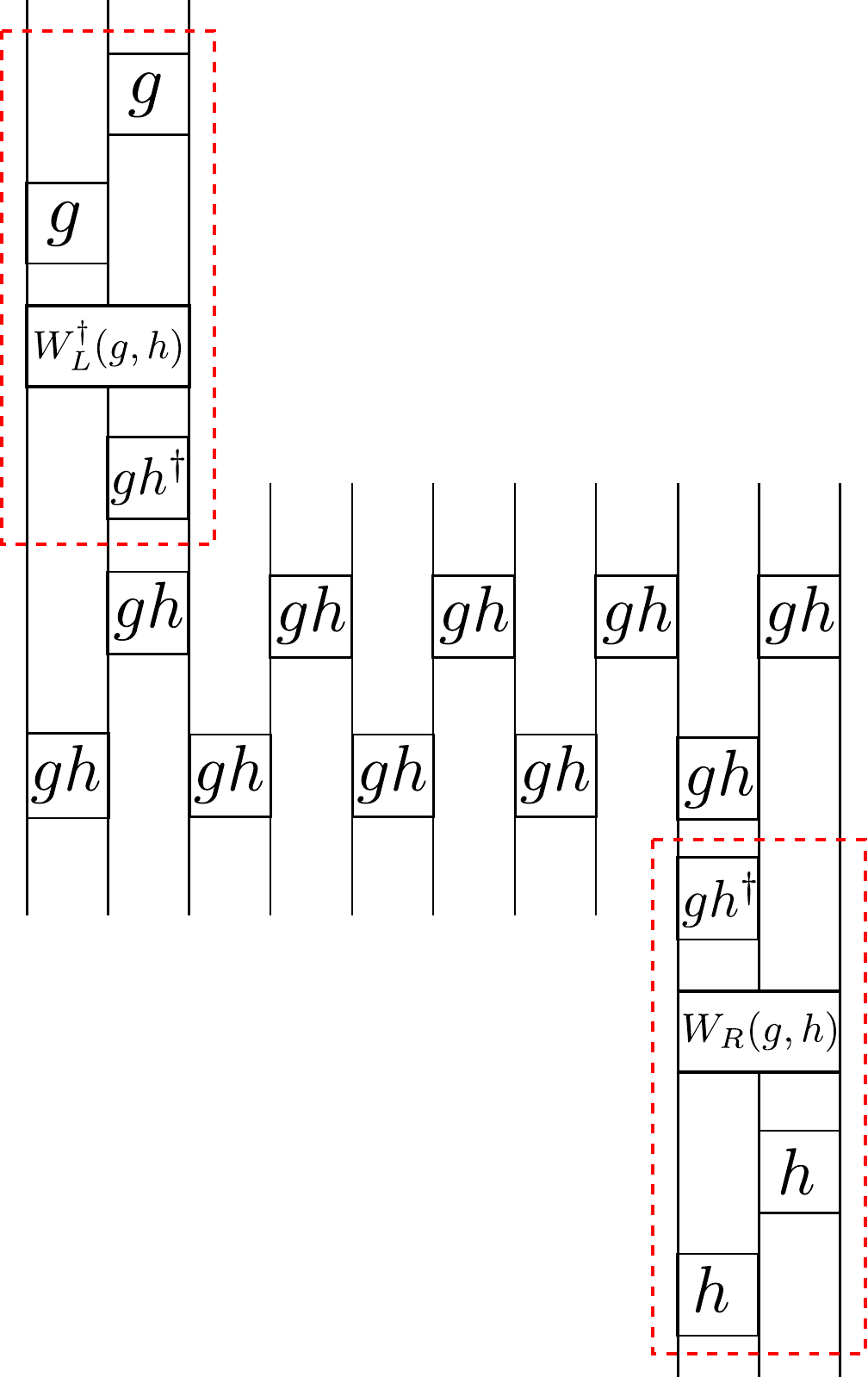}\; ,
\end{align}
whence it becomes clear that we can define new tensors $W(g,h)$ and $W'(g,h)$ such that we have
\begin{align}
    \label{eq:WWprime}
   \includegraphics[width=0.6\linewidth,valign=c]{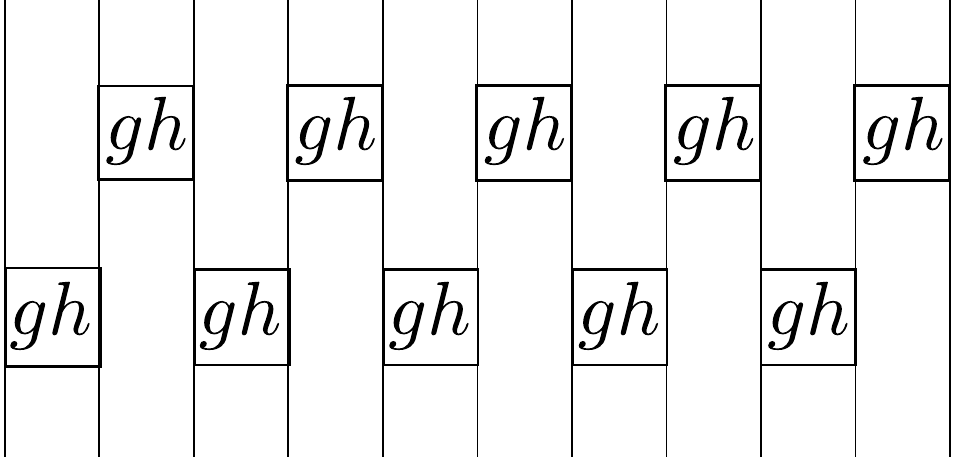} \;  \nonumber \\ =\;\includegraphics[width=0.6\linewidth,valign=c]{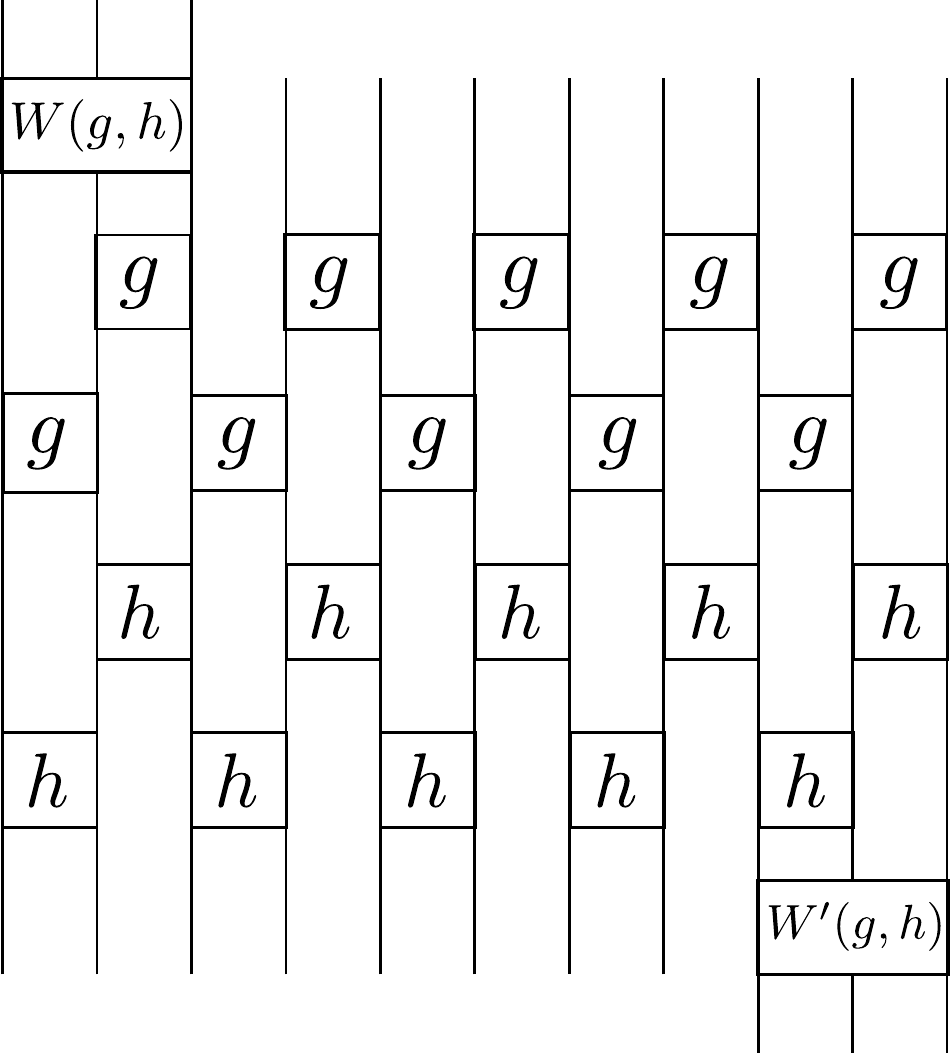} \; .
\end{align}
$W(g,h)$ and $W'(g,h)$ act as gerbal representation operators: 
Informally, $W(g,h)$ and $W'(g,h)$ ``convert" a combination of a section of the $g$ quantum circuit and the $h$ quantum circuit into a section of the $gh$ quantum circuit, playing a role analogous to that of $X_L(g,h)$ and $X_R(g,h)$ in Eq.~\eqref{eq:combineop}, respectively. 
%

To show that the quantum circuit projective representations of $G$ satisfy the pentagon equation \eqref{eq:penta}, 
we must find an associated function of three group elements $\alpha(g,h,k)\in U(1)$ which is a 3-cocycle invariant up to multiplication by a 3-coboundary. Consider three group elements $g,h,k \in G$:
\begin{align}
    \label{eq:ghk}
   \includegraphics[width=0.7\linewidth,valign=c]{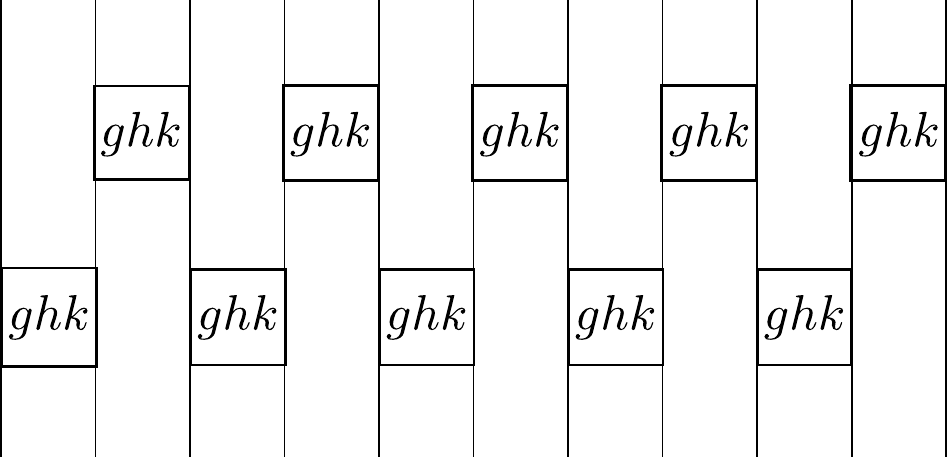} \; \nonumber\\ =\;\includegraphics[width=0.7\linewidth,valign=c]{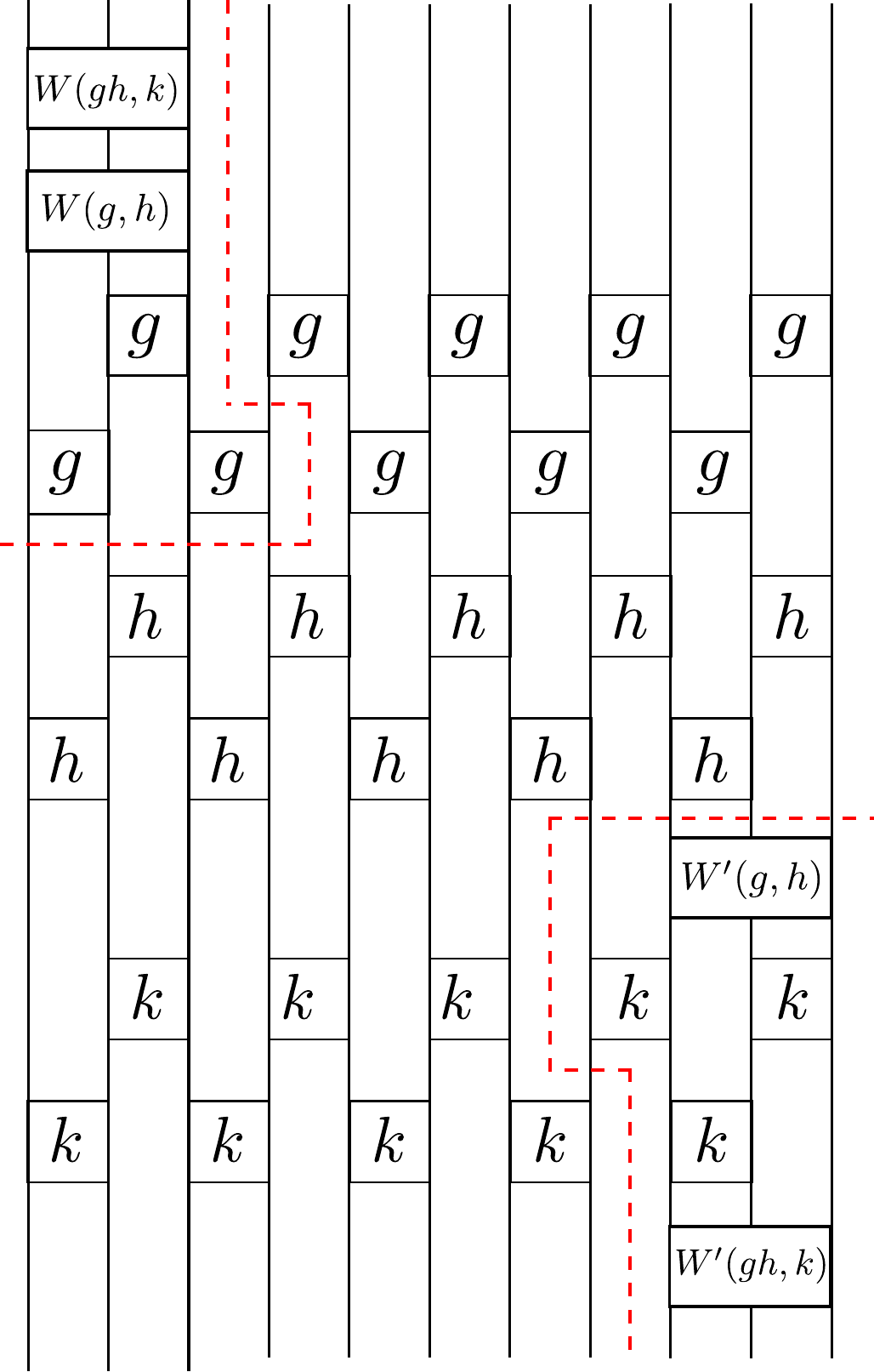} \nonumber\\ =\;\includegraphics[width=0.7\linewidth,valign=c]{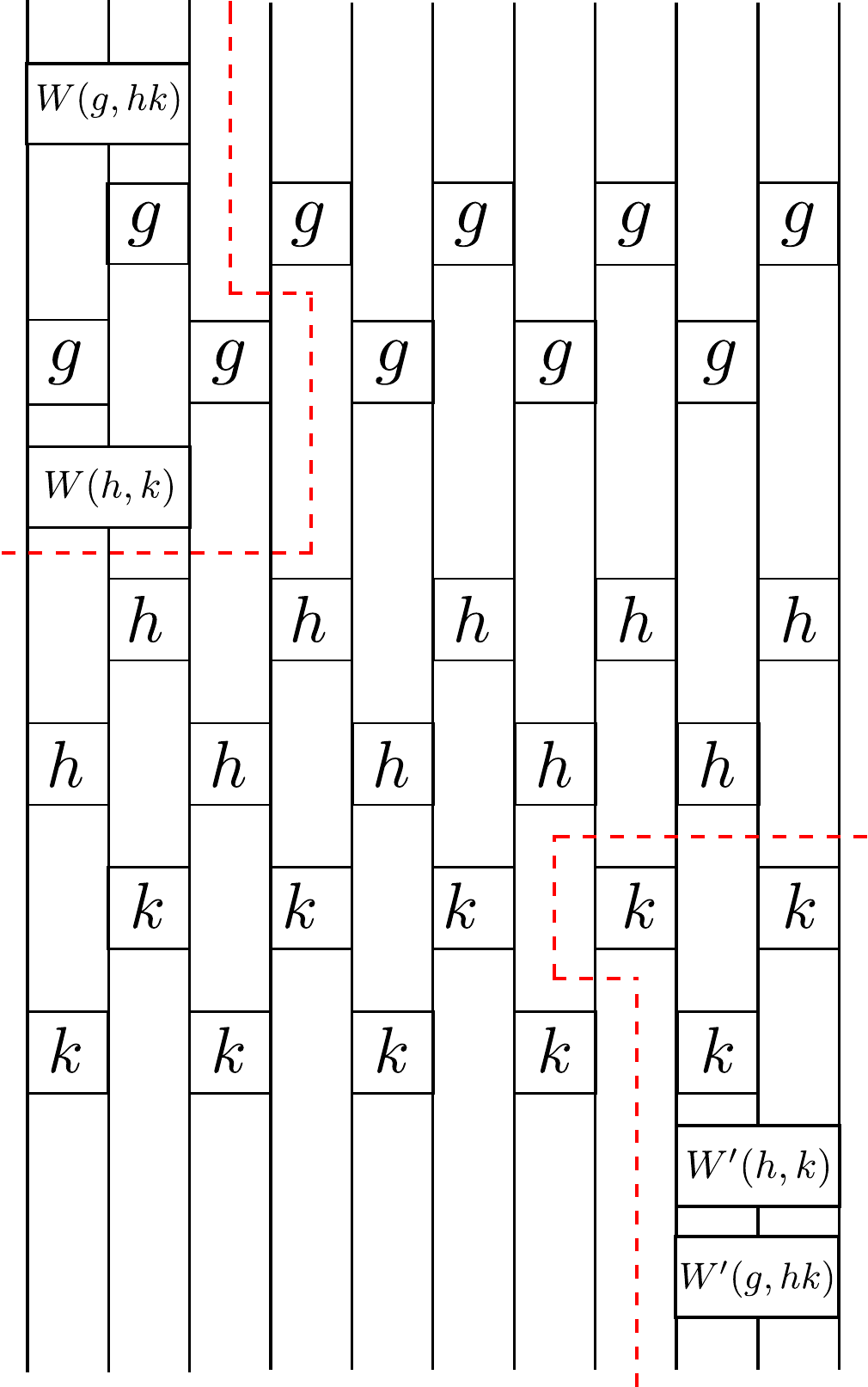}  \; .
\end{align}
Canceling out the middle sections as indicated by the red lines, the second equality implies
\begin{align}
   \includegraphics[width=0.2\linewidth,valign=c]{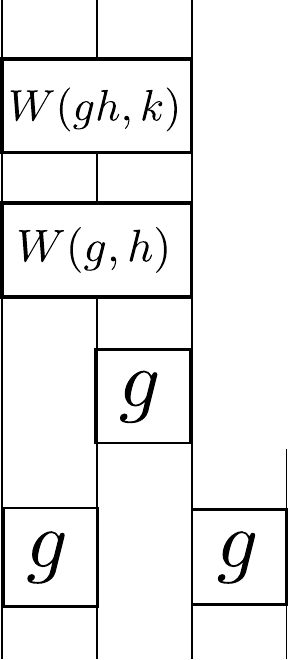} \;\;\bigotimes\;\;\; \includegraphics[width=0.2\linewidth,valign=c]{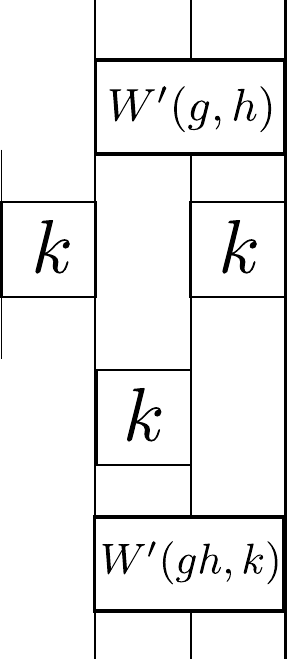} \nonumber \\ =\;\;\includegraphics[width=0.2\linewidth,valign=c]{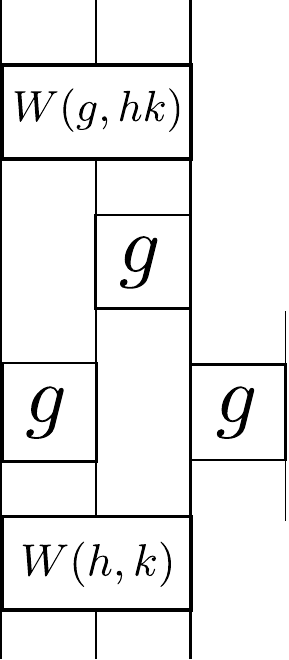} \;\;\;\bigotimes \;\; \includegraphics[width=0.2\linewidth,valign=c]{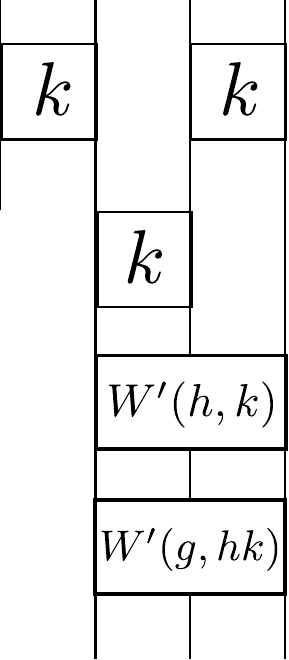}  \; ,
\end{align}
which means there must be some phase factor $\alpha(g,h,k)$ such that 
\begin{equation}
\label{eq:leftedge}    
    \includegraphics[width=0.2\linewidth,valign=c]{drawing20.eps}
  \;  =\; \;\alpha(g,h,k)\;\;\;\includegraphics[width=0.2\linewidth,valign=c]{drawing22.eps}
\end{equation}
\begin{equation}
    \label{eq:rightedge}
    \includegraphics[width=0.2\linewidth,valign=c]{drawing21.eps}\; \;= \;\; \alpha(g,h,k)^{-1} \;\; \includegraphics[width=0.2\linewidth,valign=c]{drawing23.eps} \ .
\end{equation}
$\alpha(g,h,k)$ is the function of three group variables that we are looking for.  Before proceeding further let first simplify the notation. Define
\begin{equation}
    X(g)\;\; =\;\;\; \includegraphics[width=0.2\linewidth,valign=c]{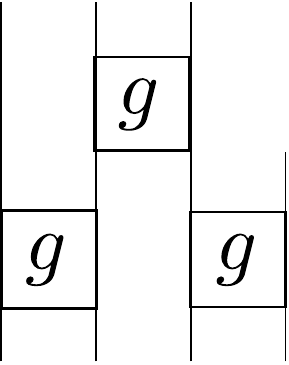} \; ,
\end{equation}
and, with a slight abuse of notation, we write, for example, Eq.~\eqref{eq:leftedge} algebraically as
\begin{equation}
    \label{eq:walpha}
    W(gh,k)W(g,h)X(g) = \alpha(g,h,k)W(g,hk)X(g)W(h,k).
\end{equation}
$W(g,h)$ inherits the gauge degree of freedom of the old $W_L(g,h)$, so it is invariant up to a transformation $W(g,h)\rightarrow \chi(g,h)W(g,h)$ for $\chi(g,h) \in U(1).$ 
After the transformation, we have Eq.~\eqref{eq:walpha} but with $\alpha(g,h,k)$ replaced by
\begin{equation}
    \alpha'(g,h,k) = \alpha(g,h,k)\frac{\chi(g,hk)\chi(h,k)}{\chi(g,h)\chi(gh,k)}.
\end{equation}
Thus $\alpha(g,h,k)$ is defined up to a 3-coboundary.  

Now we show that $\alpha(g,h,k)$ satisfies an analogue of Eq.~\eqref{eq:penta}, and therefore is a 3-cocycle. Consider the following expression involving four group elements,
\begin{align}
     \label{eq:nameofthisequation}
        &W(ghk,l)W(gh,k)W(g,h)X(g)X(h) \nonumber\\
         &= \alpha(g,h,k)W(ghk,l)W(g,hk)X(g)W(h,k)X(h) \nonumber\\
         &= \alpha(g,h,k)\alpha(g,hk,l)W(g,hkl)X(g)W(hk,l)W(h,k)X(h) \nonumber\\
         &=\alpha(g,h,k)\alpha(g,hk,l)\alpha(h,k,l)\times\nonumber\\
         & \quad \quad \quad W(g,hkl)X(g)W(h,kl)X(h)W(k,l)
\end{align}
where we have used Eq.~\eqref{eq:walpha} repeatedly. Let us introduce a new shorthand notation where, for example, Eq.~\eqref{eq:WWprime} is written as
\begin{equation}
    \label{eq:shorthand}
   \includegraphics[width=0.25\linewidth,valign=c]{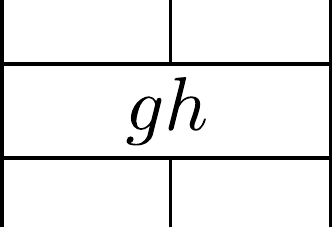} \;\; =\;\;\includegraphics[width=0.38\linewidth,valign=c]{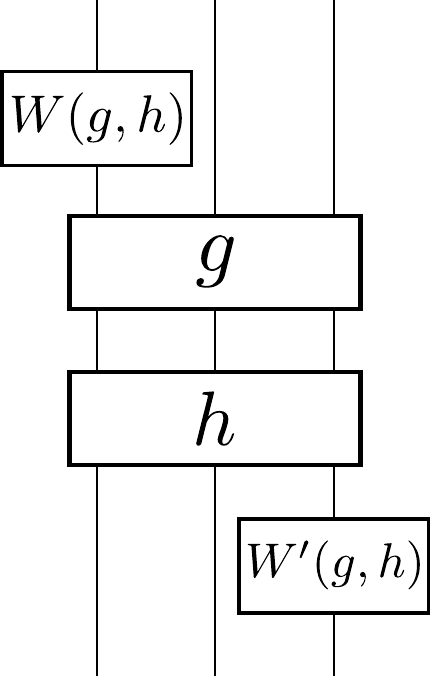} \ .
\end{equation}
Using this notation, we have the following expression involving four group elements:
\begin{equation}
    \label{eq:4groupelements}
   \includegraphics[width=0.32\linewidth,valign=c]{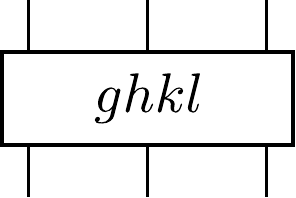} \;\;\;  = \;\;\includegraphics[width=0.38\linewidth,valign=c]{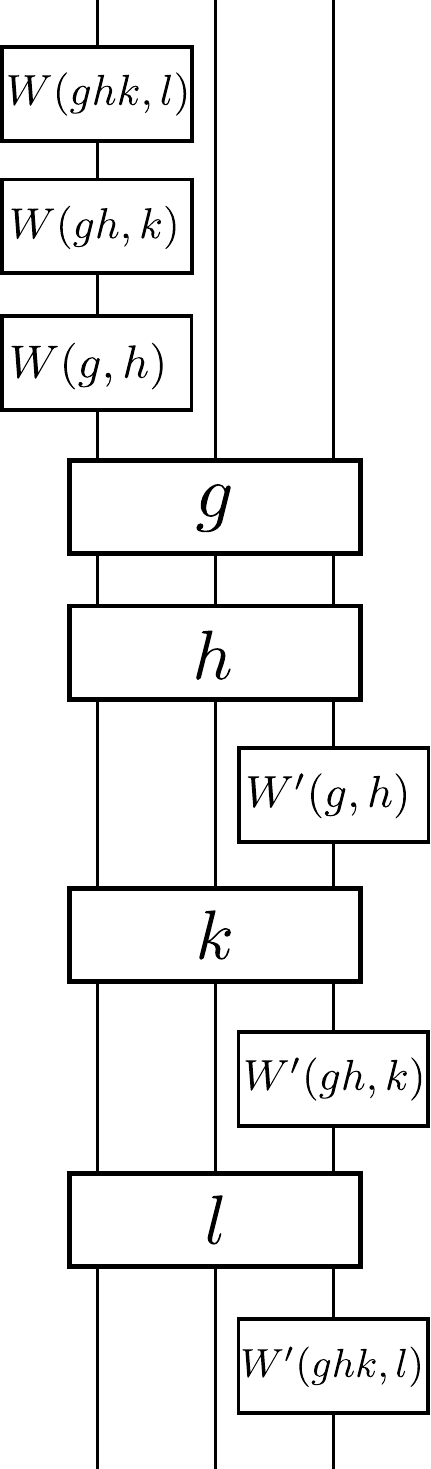} \ .
\end{equation}
To show that  $\alpha(g,h,k)$ is a 3-cocycle, we only need to consider an expression consisting of the top parts of the RHS of the above equation.  We then repeatedly apply Eq.~\eqref{eq:walpha} to the left edge of that expression.  There are two ways to do this.  First, we can apply Eq.~\eqref{eq:nameofthisequation} (converted back into diagrammatic form) and immediately obtain
\begin{equation}
   \includegraphics[width=0.35\linewidth,valign=c]{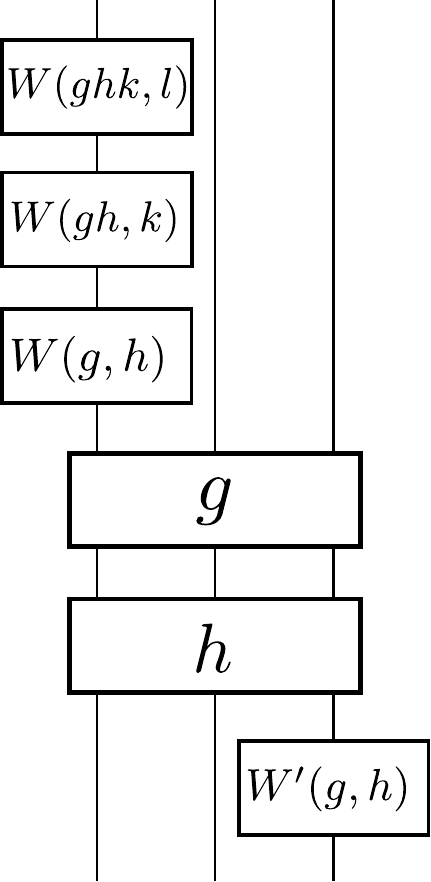} \;  =\;
   \begin{matrix}
   \alpha(g,h,k)\times\\\alpha(g,hk,l)\times\\\alpha(h,k,l)\times
   \end{matrix}
   \;\;\includegraphics[width=0.35\linewidth,valign=c]{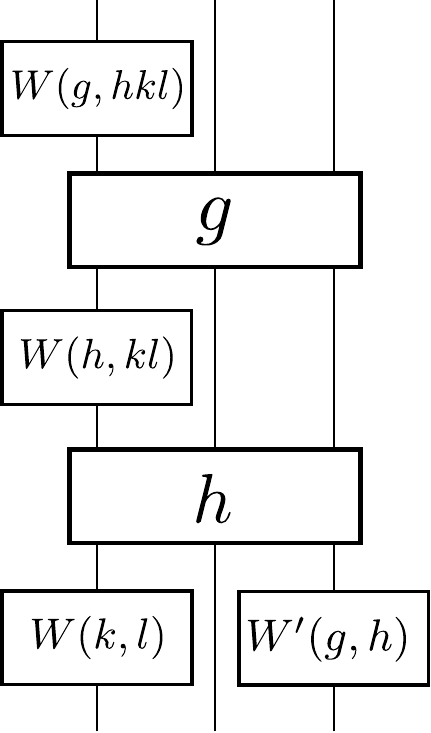}.
\end{equation}
Alternatively, we can calculate via a different route
\begin{align}
 \includegraphics[width=0.35\linewidth,valign=c]{drawing33.eps} \; = \;\;\; \includegraphics[width=0.35\linewidth,valign=c]{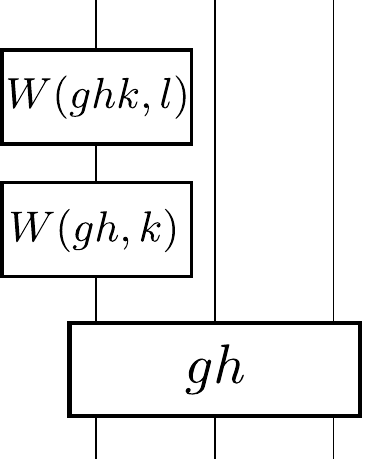}\nonumber\\
 =\;\alpha(gh,k,l)\; \includegraphics[width=0.35\linewidth,valign=c]{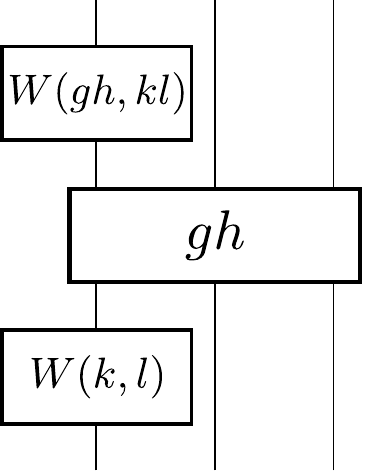} \;\;=\;\;\alpha(gh,k,l)\times \nonumber\\ \includegraphics[width=0.35\linewidth,valign=c]{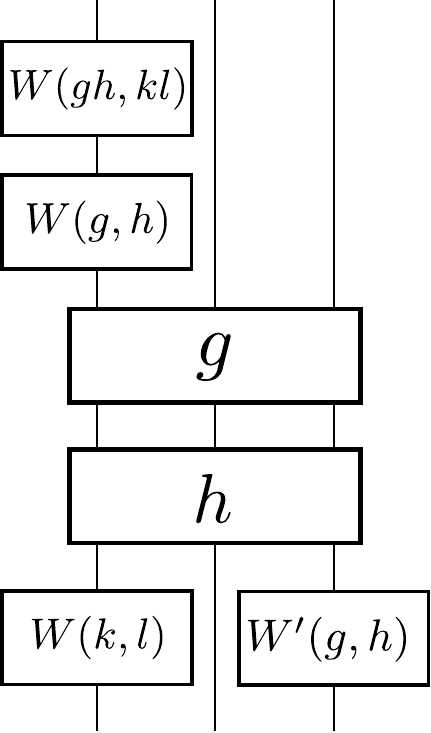}
 =\;\begin{matrix}\alpha(gh,k,l)\times\\\alpha(g,h,kl)\times\end{matrix}\; \includegraphics[width=0.35\linewidth,valign=c]{drawing34.eps} \ .
\end{align}
Comparing the above two final expressions we find that indeed
\begin{equation}
    \frac{\alpha(g,h,k)\alpha(g,hk,l)\alpha(h,k,l)}{\alpha(gh,k,l)\alpha(g,h,kl)} = 1.
\end{equation} 
Note that we have only considered $W(g,h)$ but the same argument applies to the right edge and $W'(g,h)$, which from the $\alpha^{-1}$ in Eq.~\eqref{eq:rightedge} is associated with the inverse element of $H^3(G,U(1))$. 

\begin{figure}[h]
    \centering
    \includegraphics[width=0.75\linewidth]{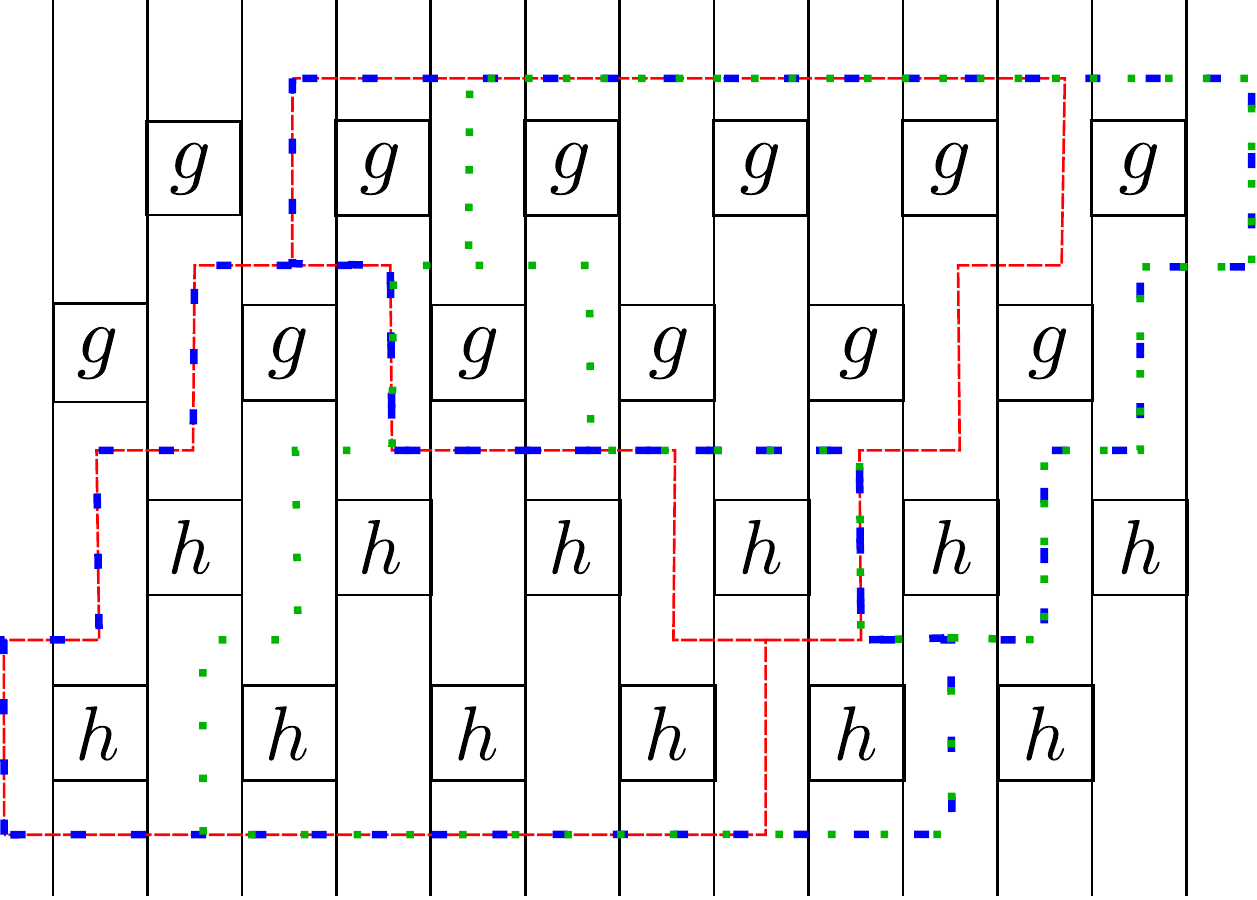}
    \caption{Three ways of blocking: original 4-blocking (red), 4-blocking shifted by one block (green), and 5-blocking (blue)}
    \label{fig:blockschemes}
\end{figure}

To complete the argument, we need to show that the cohomology class does not depend on the position of the block; recalling the non-translational-invariance of the original quantum circuit, we need to show that the $W(g,h)$ and $W'(g,h)$ of the adjacent block are associated with the same element of $H^3(G,U(1))$.  We also need to show that there is no dependence on the blocking scheme.  
For instance, we can use block sizes of larger than four (though it is easy to see that the above arguments would not work for block sizes of three or smaller).  Fig~\ref{fig:blockschemes} depicts different ways of blocking.

\begin{figure}[h]
    \centering
    \includegraphics[width=0.75\linewidth]{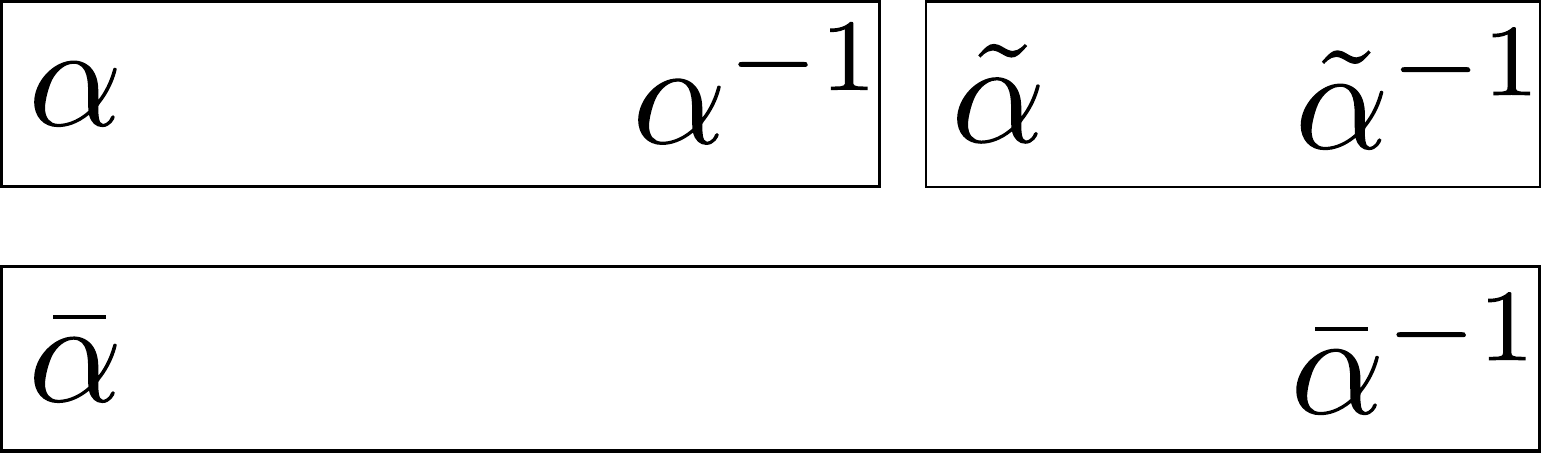}
    \caption{A schematic diagram depicting two adjacent blocks and a large block encompassing both of them, and their associated phase factors.  Note that the adjacent blocks do not have to be of the same length.}
    \label{fig:alphas}
\end{figure}

The argument is as follows. Suppose we are looking at a 4-blocking starting from a certain index, such as $4k-3$ as in Eq.~\eqref{blocking1}.  Then let us consider a larger blocking also starting from the same index.  (For example, we could consider the red 4-blocking and the blue 5-blocking in Fig~\ref{fig:blockschemes}.) Applying Eq.~\eqref{eq:gauge1}, we have 
\begin{equation}
    \includegraphics[width = 0.4\linewidth,valign=c]{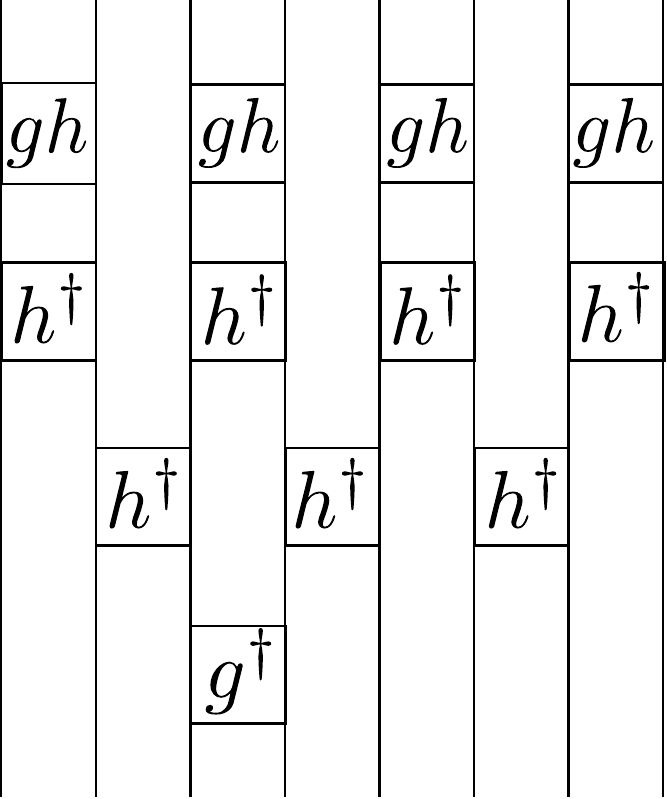}\;\; = \includegraphics[width = 0.35\linewidth,valign=c]{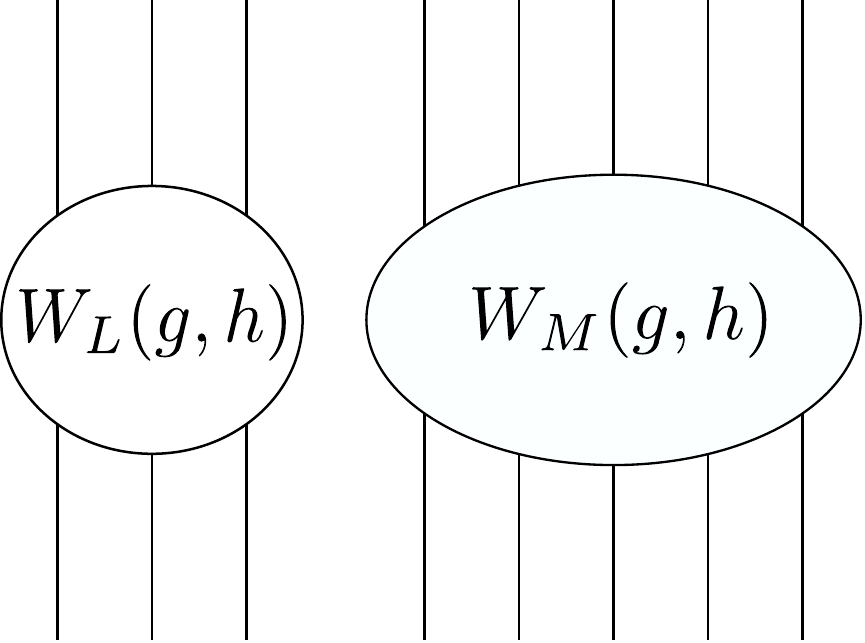}.
\end{equation}
From the larger blocking, we have, separately (using the ellipsis notation to we emphasize that this works for an arbitrarily large blocking):
\begin{equation}
    \includegraphics[width = 0.56\linewidth,valign=c]{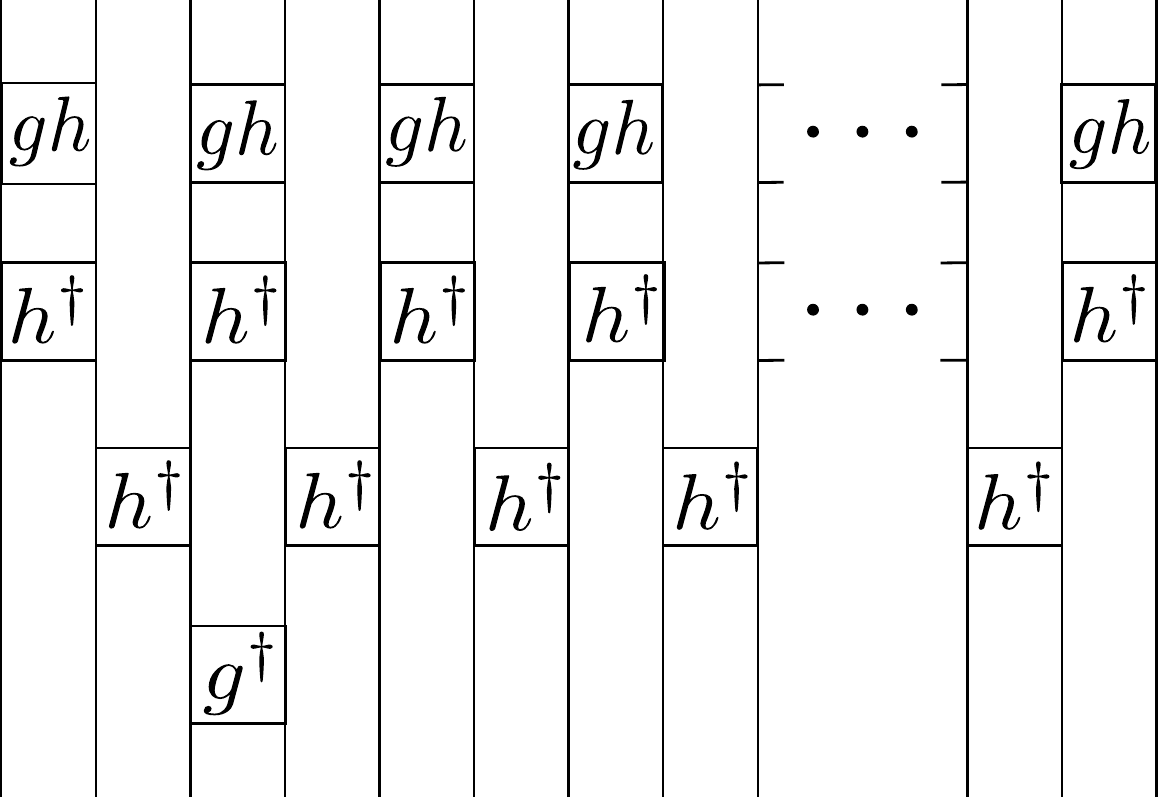}\;\; = \includegraphics[width = 0.35\linewidth,valign=c]{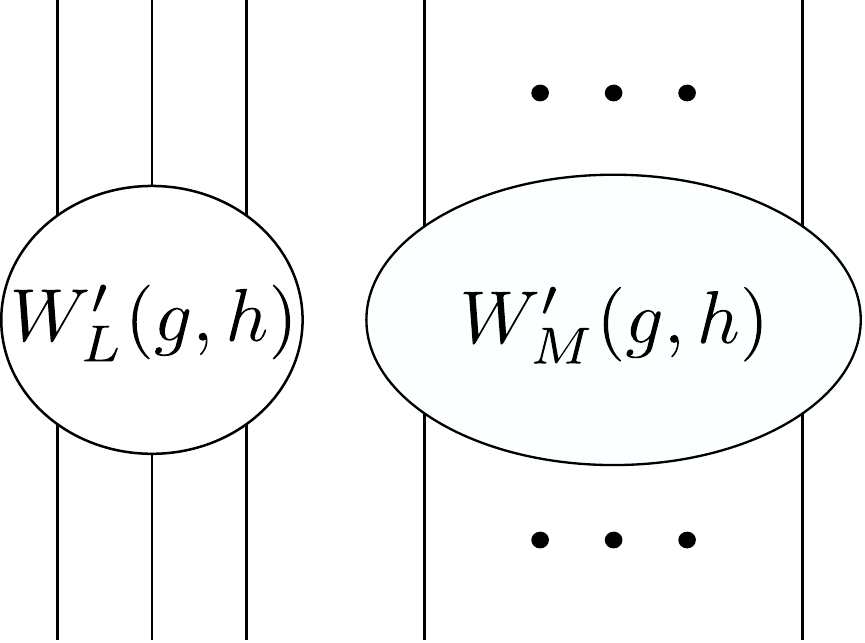}.
\end{equation}
The two above equations taken together imply that $W_L(g,h)$ and $W_L'(g,h)$ are the same up to a phase.  So the $W_L(g,h)$ and hence $W(g,h)$ are the same up to a phase in either blocking.  Hence two blocks of different sizes that start at the same point along the quantum circuit have the same cohomology class.  
The same argument applies to $W_R(g,h)$ and $W'(g,h)$ of different blocks that share the same right edge.    This then implies that the entire quantum circuit is associated with a single cohomology class $a \in H^3(G,U(1))$, 
because we can then use the above results to argue that \textit{any} any two blocks in the quantum circuit correspond to the same cohomology class: We may deduce from the schematic picture Fig.~\ref{fig:alphas} that $a = \bar{a} = \tilde a$, i.e. they are the same element of $H^3(G,U(1))$ while the corresponding functions  $\alpha(g,h,k)$, $\bar{\alpha}(g,h,k)$, and $\tilde{\alpha}(g,h,k)$ would be equal up to a 3-coboundary.  It is easy to see that this generalizes to show that any block (from any blocking scheme) produces $W(g,h)$ and $W'(g,h)$ labeled by the same $a$ and $a^{-1}\in H^3(G,U(1))$, respectively.  The entire quantum circuit representation of $G$ is associated with one particular element of $H^3(G,U(1))$, completing the proof of the lemma.

\subsection{Invariance of the topological index across the 2D system}\label{sec:wxwTopTrivial}

The above lemma applies separately to each $W_j^g$ that appears on the LHS of Eq.~\eqref{eq:wxw1} and Eq.~\eqref{eq:wxw2} and also to the overall quantum circuits of those equations.  To complete the argument for our two-dimensional MBL phase classification, we must show that the different  $W_j^g$ along the $x$-direction have the same 3rd cohomology class.  This can be done by showing that $W_j^g\otimes W_{j+1}^g$ is topologically trivial, that is, corresponds to the identity element of $H^3(G,U(1))$. 

Because $W_j^g$ takes a different form for odd and even $j$, we have two points to show, that $W_{2k-1}^g\otimes W_{2k}^g$ is topologically trivial, and that $W_{2k}^g \otimes W_{2k+1}^g$ is topologically trivial as well.

\subsubsection{$W_{2k-1}^g\otimes W_{2k}^g$ is topologically trivial}
From Eq.~\eqref{eq:wxw1}, we have
\begin{align} \label{eq:top_trivial}
        &\includegraphics[width=0.6\linewidth,valign=c]{ww1-1.eps}\;\; \nonumber\\ &= \underbrace{\includegraphics[width = 0.7\linewidth,valign=c]{2019_07_ww2_v2}}_{\mathrm{indices\;can\;be\;fixed,\;see\;next\;section}} \nonumber\\ 
      &= \; \; \includegraphics[width=0.9\linewidth,valign=c]{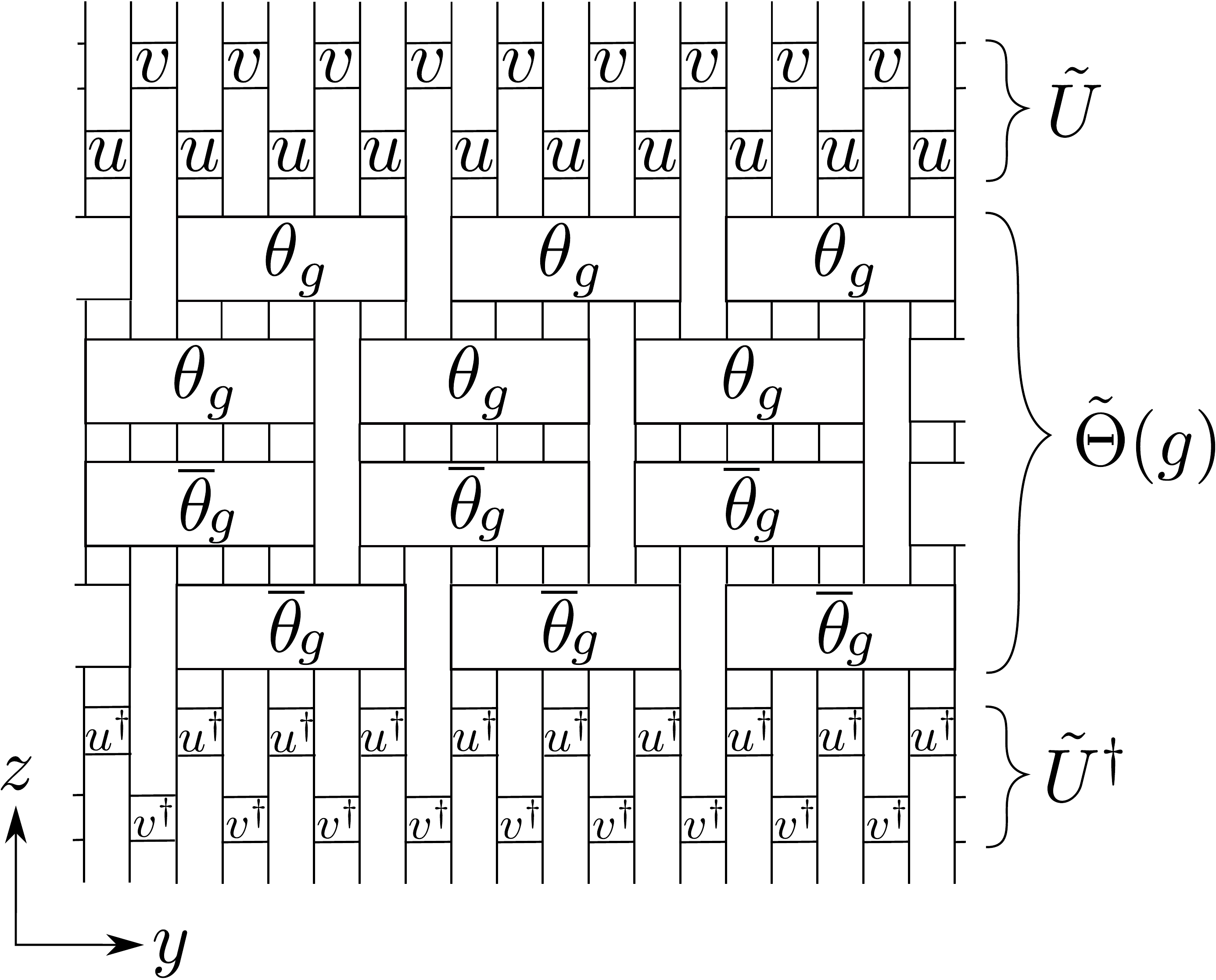} \nonumber \\
      &= \tilde U \tilde \Theta(g) \tilde U^\dagger \equiv \tilde W(g) 
\end{align}
$\Theta_j^g$ can be chosen in such a way that~\cite{1DSPTMBL}  $\Theta_j^g \Theta_j^h = \Theta_j^{gh}$, i.e., it is a linear representation of the group $G$. Since $\tilde W(g)$ is unitarily equivalent to a product of $\Theta^g$-quantum circuits, $\tilde W(g)$ must be a linear representation, too,
\begin{align}
\tilde W(g) \tilde W(h) = \tilde W(gh).    
\end{align}
The third cohomology class is a topological label of quantum circuits which are a projective representation of the group $G$. Hence, two quantum circuits corresponding to different third cohomology classes cannot be continuously connected while preserving the fact that they projetively represent the group $G$. Keeping that in mind, we note that $\tilde W(g)$ can be continuously connected to $\tilde \Theta(g)$ by defining $\tilde W_\lambda(g)$, $\lambda \in [0,1]$ via
\begin{align}
    u_{j, \lambda} &= e^{i L_j (1-\lambda)}, \\ 
    v_{j, \lambda} &= e^{i M_j (1-\lambda)} 
\end{align}
with $L_j = L_j^\dagger$, $M_j = M_j^\dagger$ and the original unitaries $u_j = e^{i M_j}$. Hence, $\tilde W_0(g) = \tilde W(g)$ and $\tilde W_1(g) = \tilde \Theta(g)$ and since for all $\lambda$
$\tilde W_\lambda(g) \tilde W_\lambda(h) = \tilde W_\lambda(gh)$, $\tilde W(g)$ and $\tilde \Theta(g)$ must correspond to the same element of the third cohomology group. Finally, we show that $\tilde \Theta(g)$ corresponds to the identity of the third cohomology group. This can be most easily seen by combining $\theta_g$'s and $\overline \theta_g$'s by commuting them through each other and combining four and two adjacent legs to respectively one. We call the newly obtained unitaries $\theta_g^u$ and $\theta_g^v$. 
The  $\theta_g^u$ and $\theta_h^v$ commute with each other, that is
\begin{equation}
    \includegraphics[width=0.17\linewidth,valign=c]{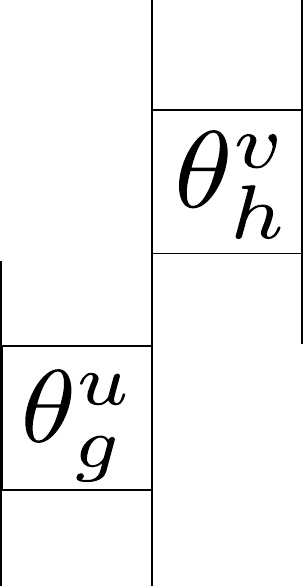}\;\;\;\;=\;\;\;\;\includegraphics[width=0.17\linewidth,valign=c]{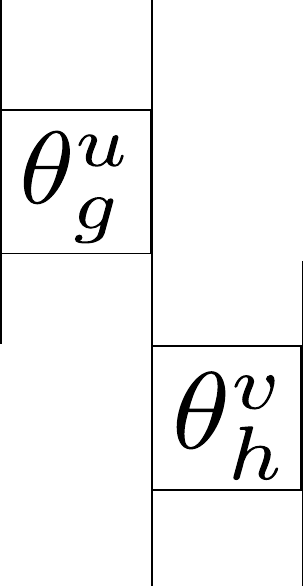}
    \label{eq:commu_ug}
\end{equation}
and 
\begin{equation}
    \includegraphics[width=0.17\linewidth,valign=c]{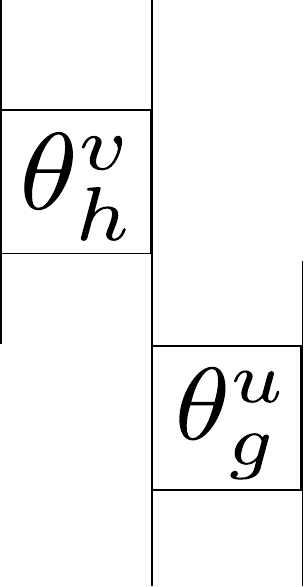}\;\;\;\;=\;\;\;\;\includegraphics[width=0.17\linewidth,valign=c]{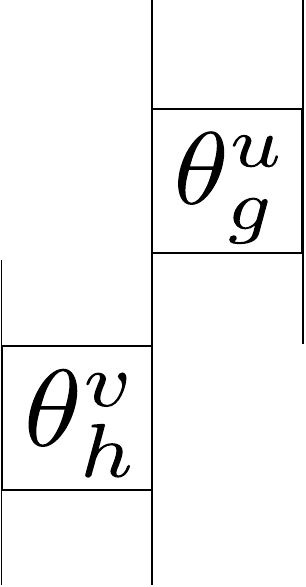}. 
    \label{eq:commu_vg}
\end{equation}
Furthermore, $\Theta_j^g \Theta_j^h = \Theta_j^{gh}$ can be used to show in the same way as in Ref.~\onlinecite{1DSPTMBL} that also the $\theta_g$'s (and $\overline \theta_g$'s) can be gauged such that $\theta_{g} \theta_{h} = \theta_{gh}$, which implies $\theta_g^u \theta_h^u = \theta_{gh}^u$ and $\theta_g^v \theta_h^v = \theta_{gh}^v$. Using this and Eqs.~\eqref{eq:commu_ug} and~\eqref{eq:commu_vg}, reveals via Eq.~\eqref{eq:WWprime} that $W(g,h) = W'(g,h) = \mathbb{1}$, i.e., after the deformation $\alpha(g,h,k) = 1$. Thus, $W_{2k-1}^g \otimes W_{2k}^g$ is topologically trivial, as claimed.

%
%
%
%

\subsubsection{$W_{2k}^g\otimes W_{2k+1}^g$ is topologically trivial}

From Eq.~\eqref{eq:wxw2}, we have 
\begin{align}\label{eq:ww2k2k+1}
        \includegraphics[width=0.6\linewidth,valign=c]{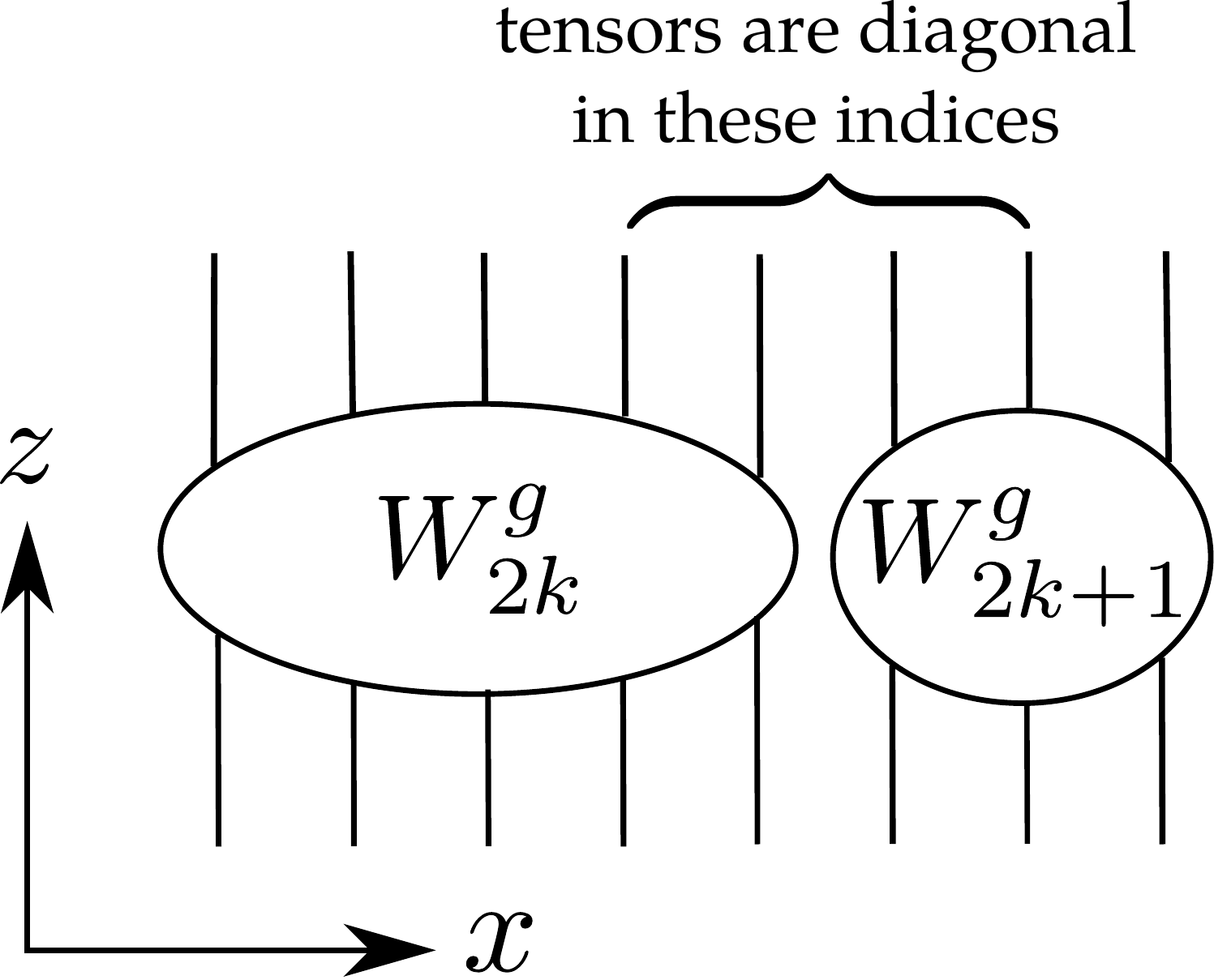} \nonumber\\ = \;\; \includegraphics[width=0.8\linewidth,valign=c]{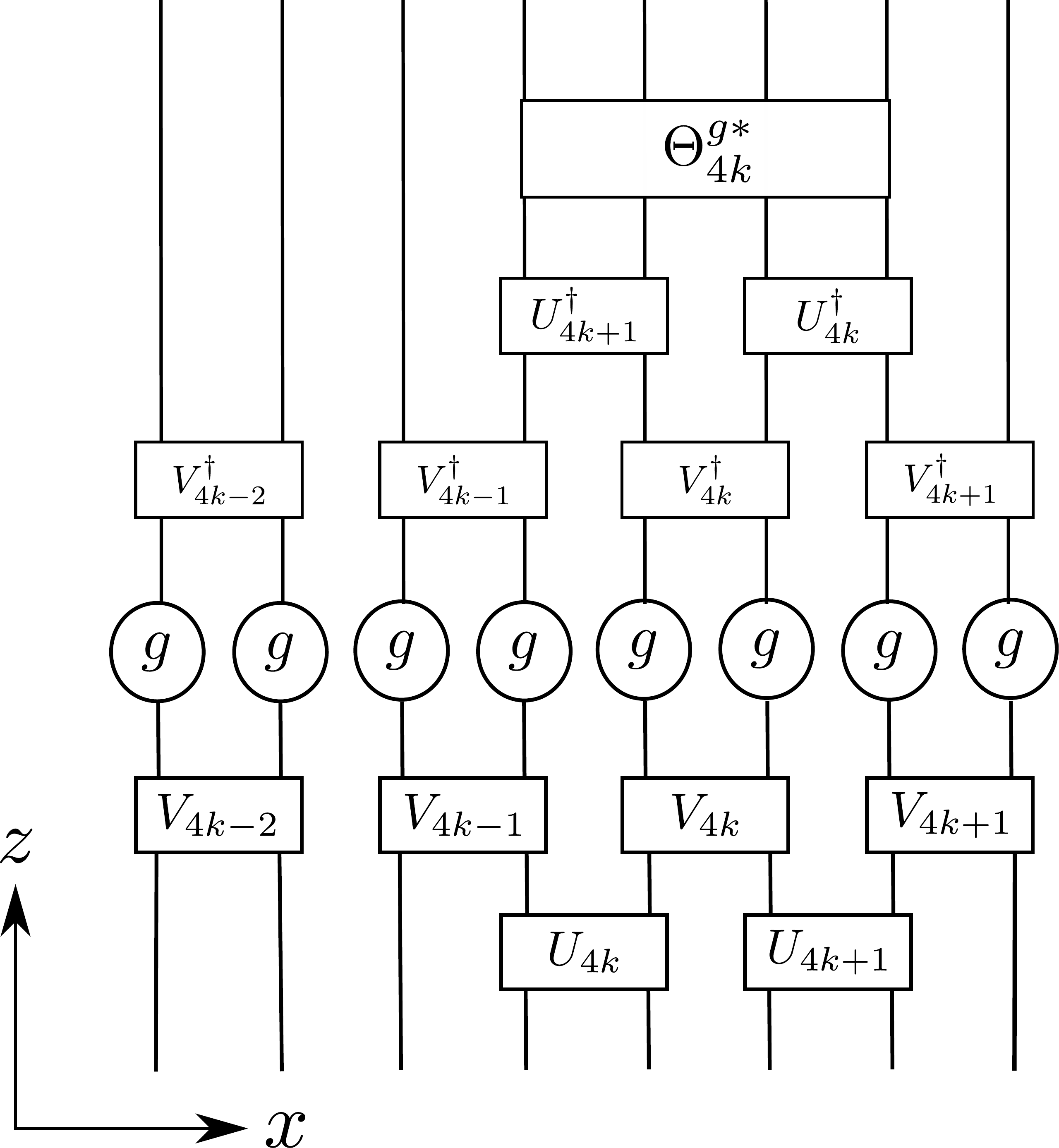} \nonumber\\ = \;\, \includegraphics[width=0.9\linewidth,valign=c]{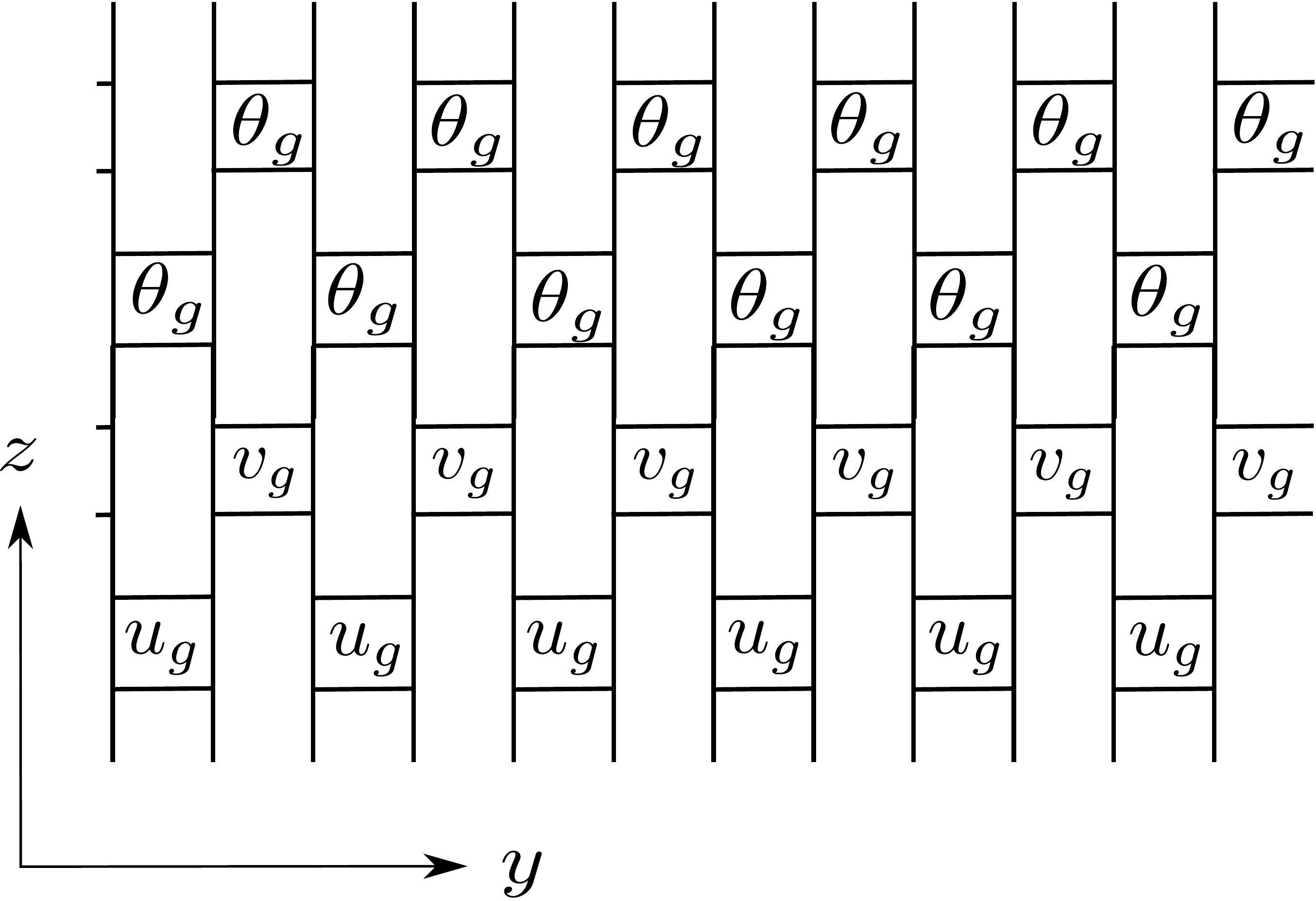}  \nonumber\\ = \;\, \includegraphics[width=0.9\linewidth,valign=c]{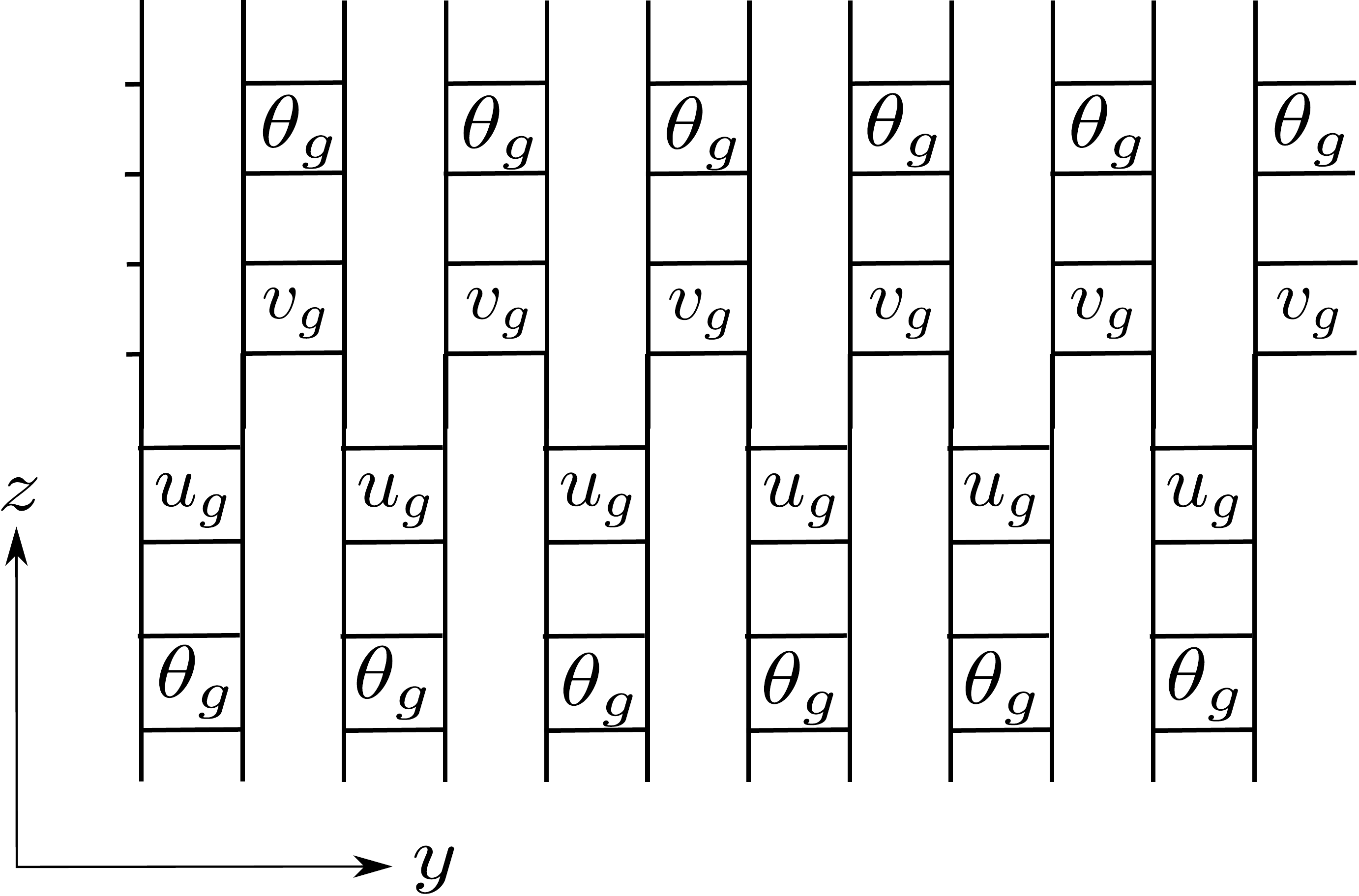} 
\end{align}
where the boxes labeled by $u_g$, $v_g$ indicate blocks of unitaries $u_j$, $v_j$ and $\mathpzc{v}_g$, and we have combined legs. In the last part of the equation, we used that due to the diagonality of the corresponding indices, the $\theta_g$'s commute with the quantum circuit comprised of the unitaries $u_g$ and $v_g$. Due to its local structure, this implies also
\begin{align}
        \includegraphics[width=\linewidth,valign=c]{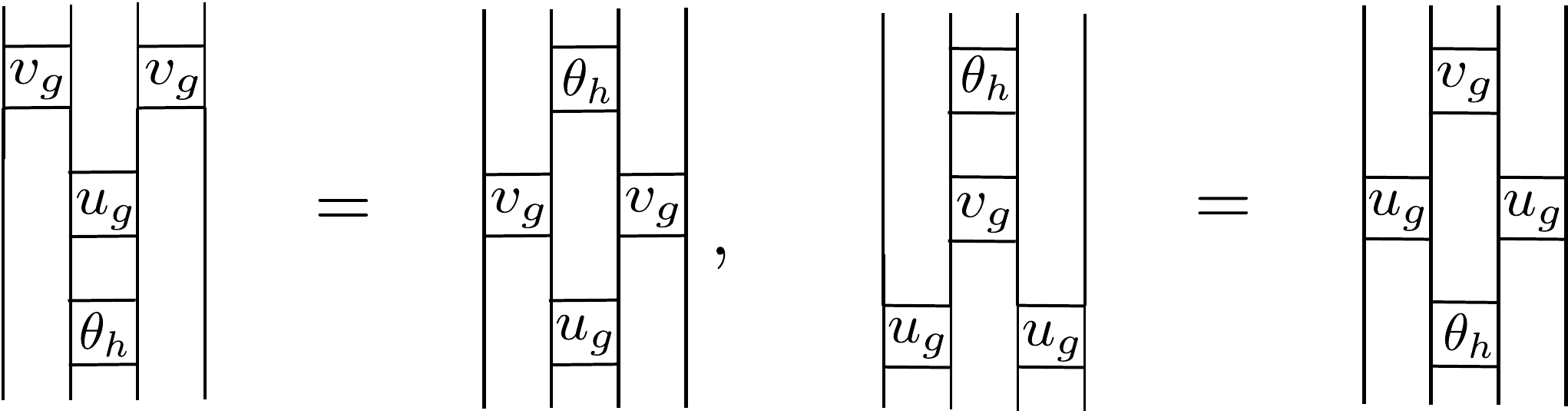} \, .
        \label{eq:commutation}
\end{align}
We now show that this implies that $u_g$ and $v_g$ can also be gauged in such a way that they individually commute with $\theta_h$. From the previous equation it follows that
\begin{align}
        \includegraphics[width=0.6\linewidth,valign=c]{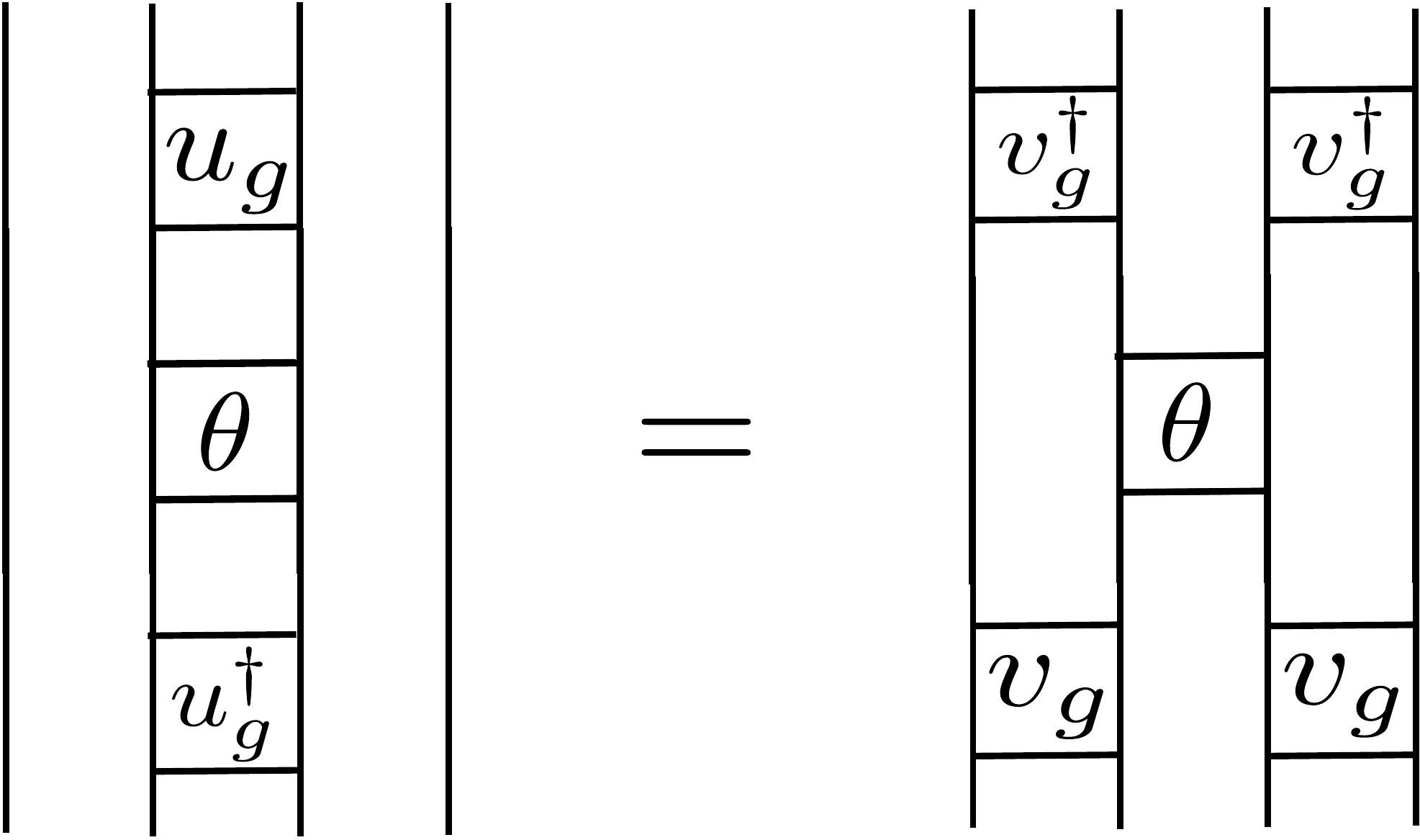} \, ,
        \label{eq:cool}
\end{align}
where we have replaced $\theta_h$ by a diagonal matrix $\theta$, which has the diagonal structure common to all $\theta_h$'s, but whose non-trivial phase factors can be chosen arbitrarily. These correspond to the indices of the forth to seventh leg from the left in the second part of Eq.~\eqref{eq:ww2k2k+1}. We now choose $\Theta = \vartheta \otimes \mathbb{1}$, such that Eq.~\eqref{eq:cool} simplifies to 
\begin{align}
        \includegraphics[width=0.6\linewidth,valign=c]{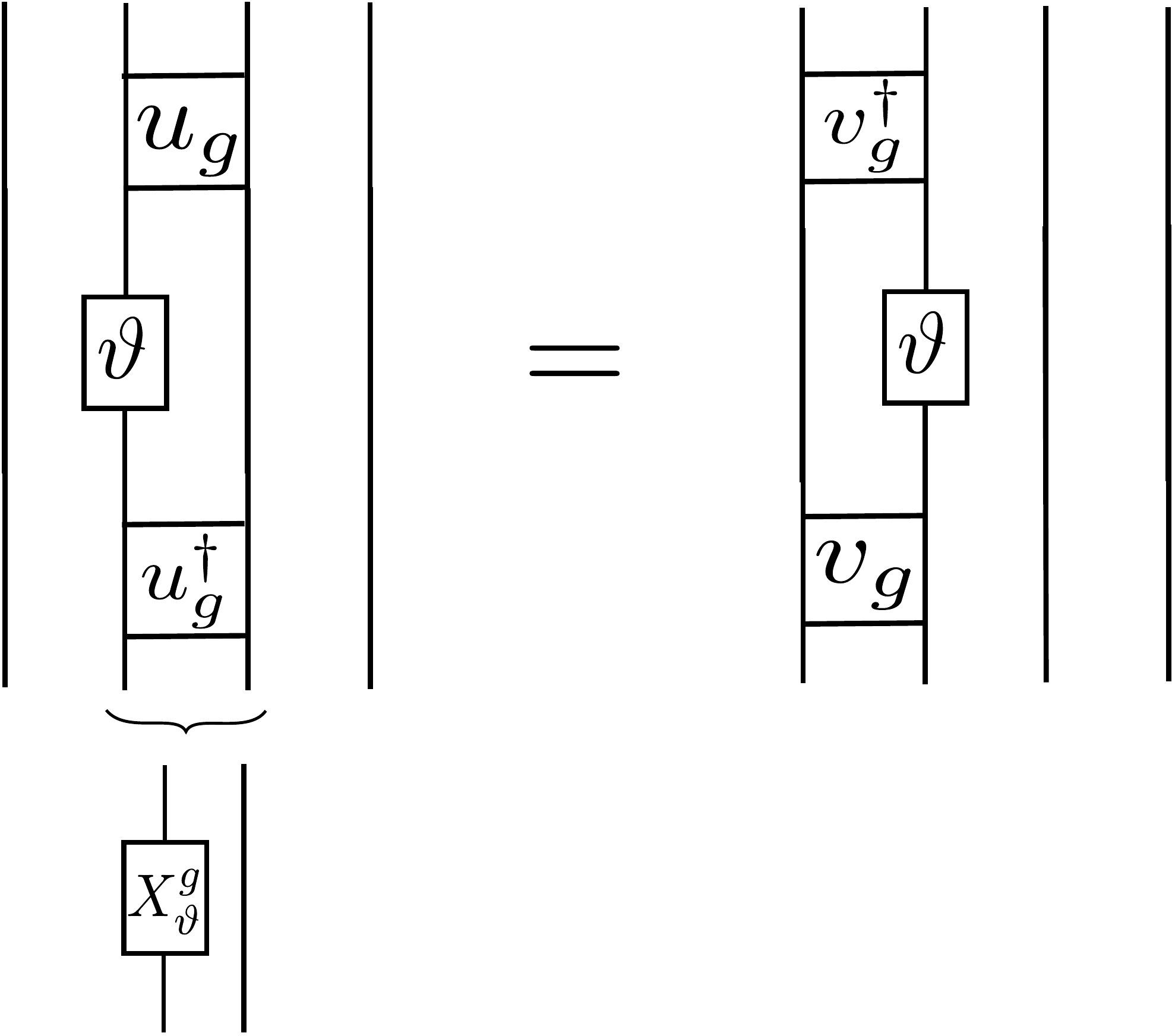} \, .
        \label{eq:simply_cool}
\end{align}
This implies that $[X_\vartheta^g \otimes \mathbb{1}, X_{\vartheta'}^g \otimes \mathbb{1}] = 0$, i.e., $[X_\vartheta^g, X_{\vartheta'}^g] = 0$. Since $X_\vartheta^g$ and $X_{\vartheta'}^g$ are unitaries, they can be diagonalized by the same matrix $w^g$. The result of the diagonalization would be $\vartheta$, i.e., $X_\vartheta^g = w_g \vartheta w_g^\dagger$. Hence, if we use a gauge transformation as in Eqs.~\eqref{eq:gauge1} and~\eqref{eq:gauge2} to replace $u_g$ by $(w_g \otimes \mathbb{1}) u_g$, the RHS of Eq.~\eqref{eq:simply_cool} is $\mathbb{1} \otimes \vartheta \otimes \mathbb{1} \otimes \mathbb{1}$. Moreover, in Eq.~\eqref{eq:cool}, we could instead have set $\theta = \mathbb{1} \otimes \vartheta$ leading to
\begin{align}
        \includegraphics[width=0.6\linewidth,valign=c]{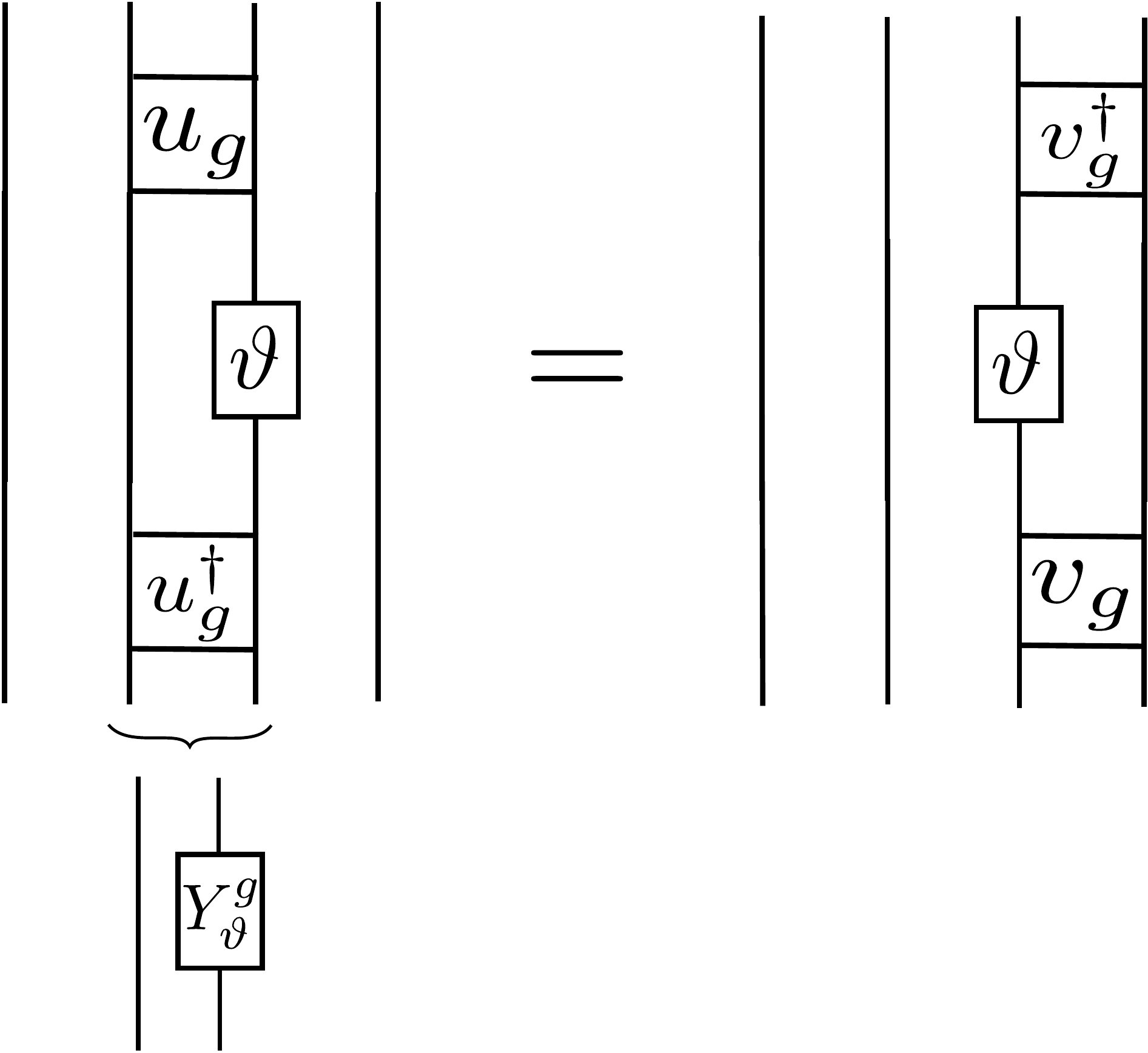} \, .
        \label{eq:simply_cool2}
\end{align}
Similarly, it follows that $Y_\vartheta^g$ can be diagonalized by a unitary matrix $\tilde w_g$ which does not depend on $\vartheta$. Hence, the gauge transformation $u_g \rightarrow (w_g \otimes \tilde w_g) u_g$ (and the corresponding one for $v_g$) ensures that the new $u_g$ commutes with $\vartheta \otimes \vartheta'$ for all $\vartheta, \vartheta'$. Hence, it must also commute with $\theta$ (which could be written as $\sum_i \vartheta_i \otimes \vartheta_i'$ if we relax the condition that $\vartheta$ and $\vartheta'$ have diagonal elements of magnitude 1, which is not needed for the above derivation). In the new gauge, $[u_g, \theta_h] = 0$ and the second part of Eq.~\eqref{eq:commutation} implies that in that gauge $[v_g, \theta_h] = 0$ as well. In other words, we can choose $u_g$ and $v_g$ such that they all commute with $\theta_h$, i.e., the $\theta_h$'s can be moved through them in all the diagrams. We now take advantage of the fact that the last expression of Eq.~\eqref{eq:ww2k2k+1} can be written as a two-layer quantum circuit after blocking unitaries, such that Eq.~\eqref{eq:WWprime} implies
\begin{align}\label{eq:ww2k2k+1}
        \includegraphics[width=0.9\linewidth,valign=c]{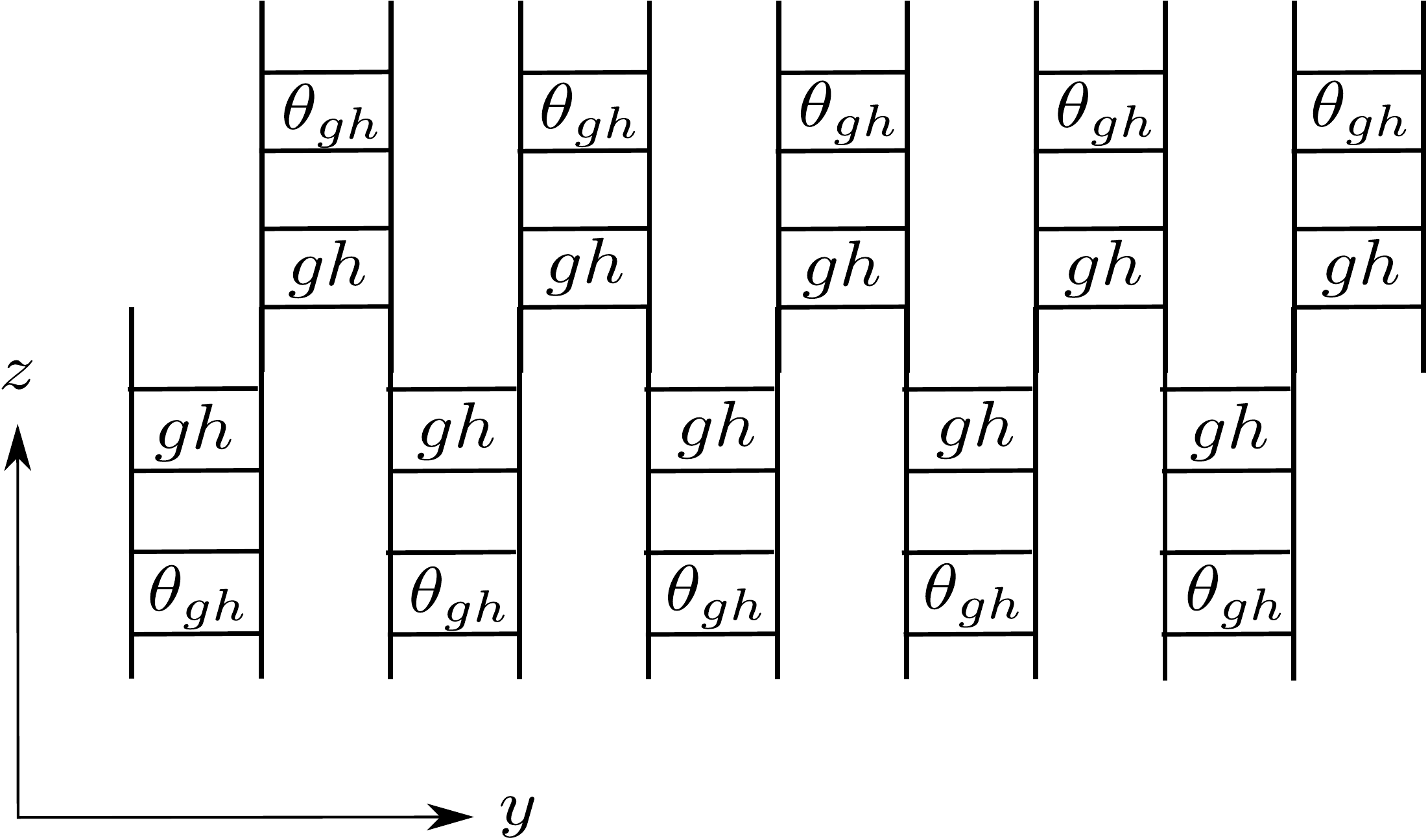}  \nonumber\\ = \;\, \includegraphics[width=0.9\linewidth,valign=c]{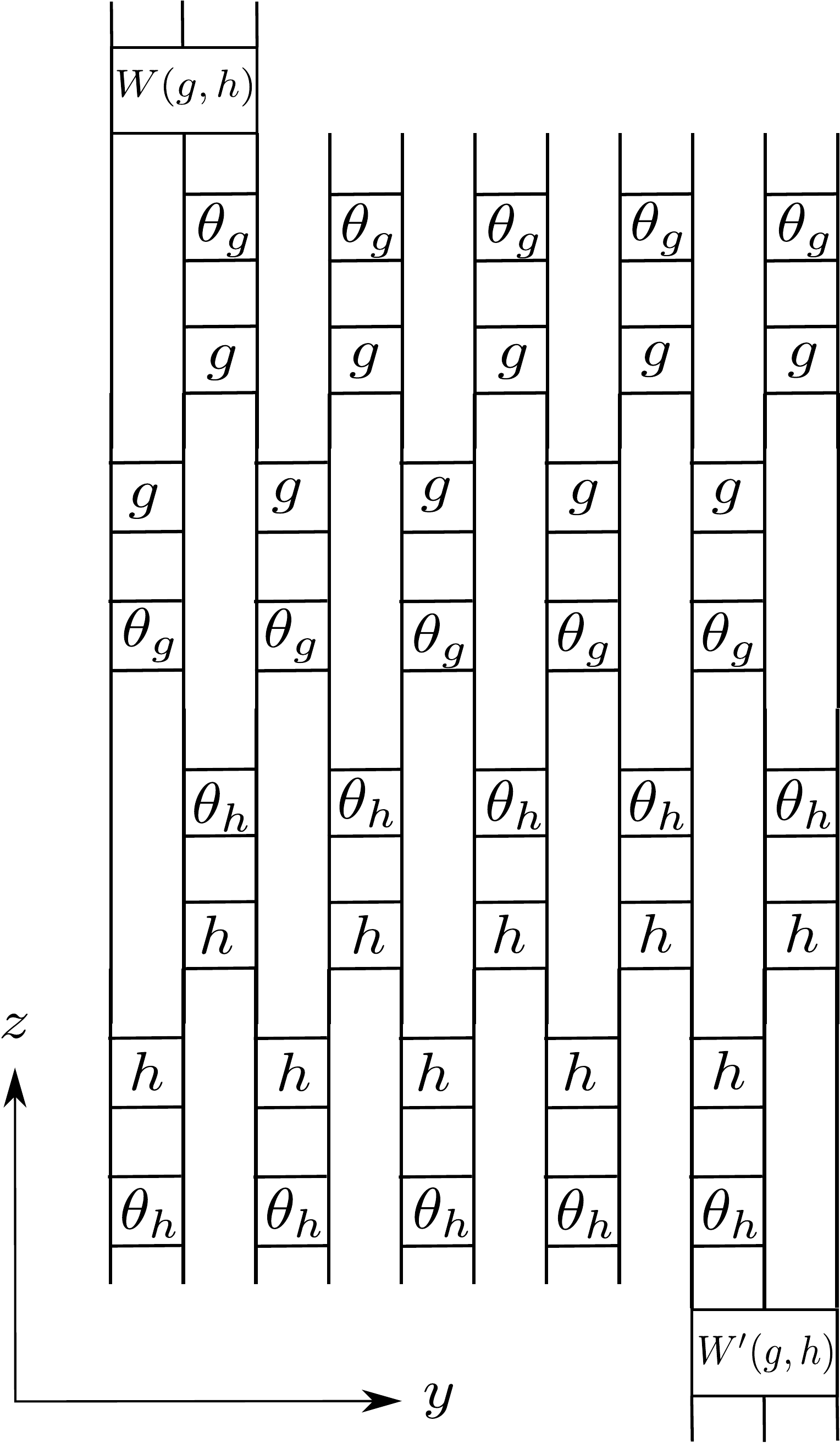} .
\end{align}
We can gauge $\theta_g$ such that $\theta_g \theta_h = \theta_{gh}$ (see above), i.e., in the new gauge of $u_g$ and $v_g$, all $\theta$'s can be canceled out, leading to
\begin{align}
   \includegraphics[width=0.6\linewidth,valign=c]{drawing7.eps} \;  \nonumber \\ =\;\includegraphics[width=0.6\linewidth,valign=c]{drawing11.eps} \; .
\end{align}
Hence, the $W(g,h)$ and $W'(g,h)$ are the same (up to a phase) as the ones corresponding to the quantum circuit~\eqref{eq:ww2k2k+1} without the $\theta_g$'s. That is, the third cohomology group of $W_{2k}^g \otimes W_{2k+1}^g$ is the same as the one of
\begin{align}
   \includegraphics[width=0.7\linewidth,valign=c]{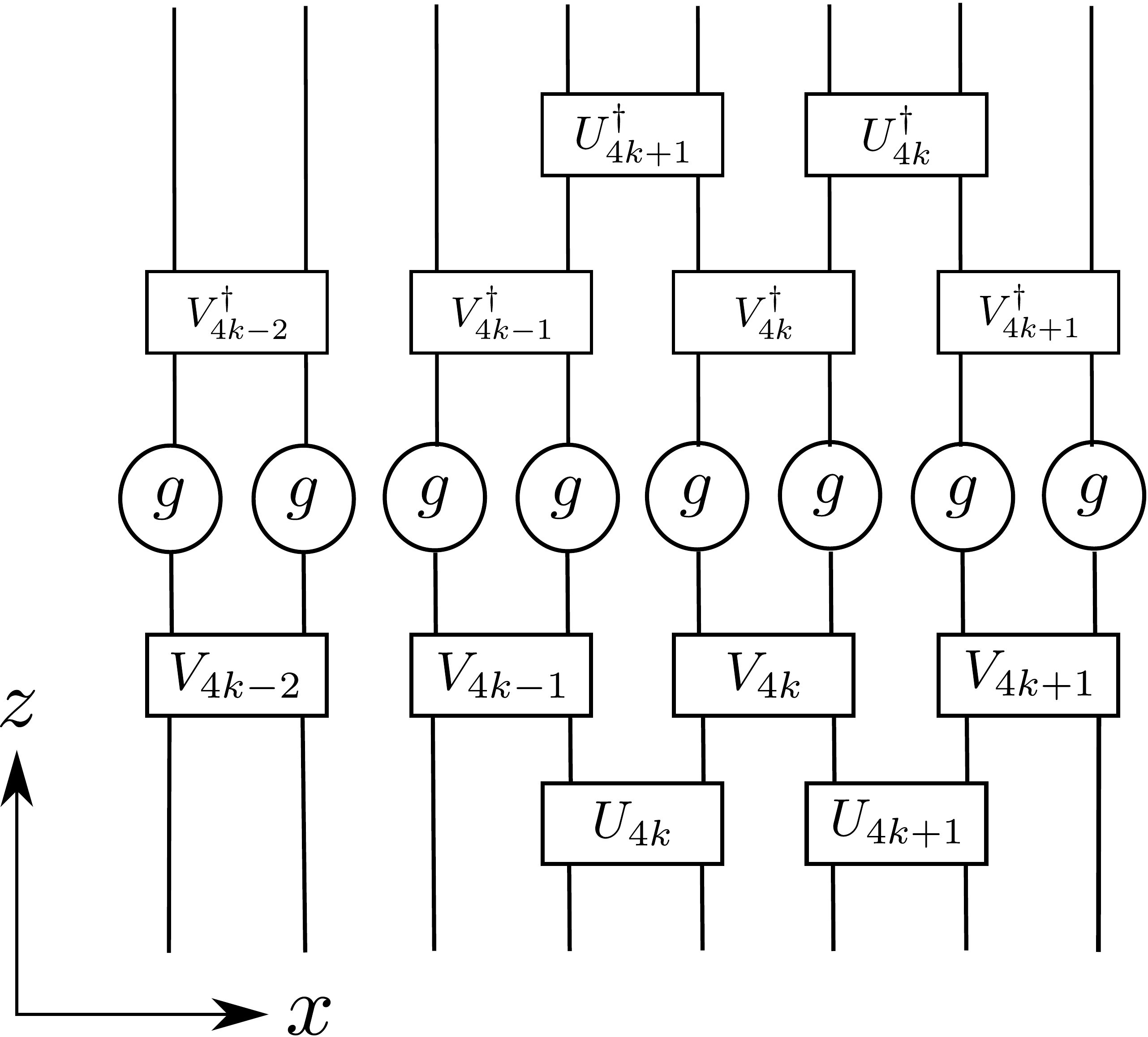} \;  . 
\end{align}
For this quantum circuit, we can use the same approach as in the previous subsection and continuously deforming the $u$'s and $v$'s to $\mathbb{1}$ while preserving the property that it forms a linear representation of the group $G$ due to $\mathpzc{v}_g \mathpzc{v}_h = \mathpzc{v}_{gh}$. Eventually, one is left with $\mathpzc{v}_g^{\otimes N^2}$, which is topologically trivial. Thus, $\alpha(g,h,k) = 1$ after the deformation, and $W_{2k}^g \otimes W_{2k+1}^g$ is topologically trivial, too.

\subsection{Equivalence of the topological label across eigenstates}

One important point is that the three-leg-wide $W_{2k-1}^g$ as in Eq.~\eqref{eq:wxw1} or the first expression in Eq.~\eqref{eq:top_trivial} is actually diagonal in its first (left) two indices, and the five-leg-wide $W_{2k}^g$ is likewise diagonal in its last (right) two indices.  
This follows immediately from Eq.~\eqref{eq:wxw2}. 

Say in the second expression of Eq.~\eqref{eq:top_trivial}, we fix the first two and last two indices to $L_1$, $L_2$, $L_3$, and $L_4$. These indices correspond to the l-bit configuration of the eigenstates which are being approximated, since those indices are lower indices in Eq.~\eqref{eq:1d1}, which according to Eq.~\eqref{eq:psi_lbit} are eigenstate labels. Hence, \textit{a priori}  $W_{2k-1}^{g, L_1 L_2}$ has cohomology class $a_{L_1 L_2}$ depending on the indices $L_1$, $L_2$ (and thus on the eigenstates). Similarly,  $W^{g, L_3 L_4}_{2k}$ has cohomology class $a_{L_3 L_4}$ again depending on the l-bits.  However, since together they are topologically trivial, we must have $a_{L_1 L_2} a_{L_3 L_4} = 1$.  By fixing $L_1$, $L_2$ we conclude that the cohomology class cannot depend on $L_3$, $L_4$, and by fixing $L_3$, $L_4$ we conclude that the cohomology class cannot depend on $L_1$, $L_2$.  Hence the topological label must be the same for all eigenstates.

\section{Anti-unitary symmetries}\label{sec:anti}

The above treatment can be generalized by allowing as well for anti-unitary symmetries. That is, for some group elements $g \in G$ we have
\begin{align}
    H = \mathpzc{v}_g^{\otimes N^2} H^* (\mathpzc{v}_g^\dagger)^{\otimes N^2},
\end{align}
which analogously leads to
\begin{align}
    \Theta_g = U^\dagger \mathpzc{v}_g^{\otimes N^2} U^*.
\end{align}
Other group elements $g'$ may still satisfy Eqs.~\eqref{eq:Ham_symmetry} and~\eqref{eq:symmetry}. A special case is the one of simple time-reversal symmetry, where $G = \mathbb{Z}_2 = \{e,z\}$ and the group element $z$ comes with a complex conjugation. The classification will be given by the elements of the generalized third cohomology group defined below, which is trivial for the case of simple time-reversal symmetry~\cite{Chen2013}. 

We define~\cite{Bultnick2017} $\gamma(g)$ such that $\gamma(g) = 1$ ($\gamma(g) = 0$) if the symmetry operation does (not) involve complex conjugation. Hence,
\begin{align}
    \lfloor X \rceil^{\gamma(g)} = \left\{\begin{matrix}
X&\mathrm{if} \ \gamma(g) = 0 \\
X^*&\mathrm{if} \ \gamma(g) = 1. \\
\end{matrix}\right.
\end{align}
The on-site operators $\mathpzc{v}_g$ must thus fulfill $\mathpzc{v}_g \lfloor \mathpzc{v}_h \rceil^{\gamma(g)}= \mathpzc{v}_{gh}$.  Eqs.~\eqref{eq:1d1}, \eqref{eq:wxw2} and~\eqref{eq:wxw2} read now
\begin{align}
        \includegraphics[width=\linewidth,valign=c]{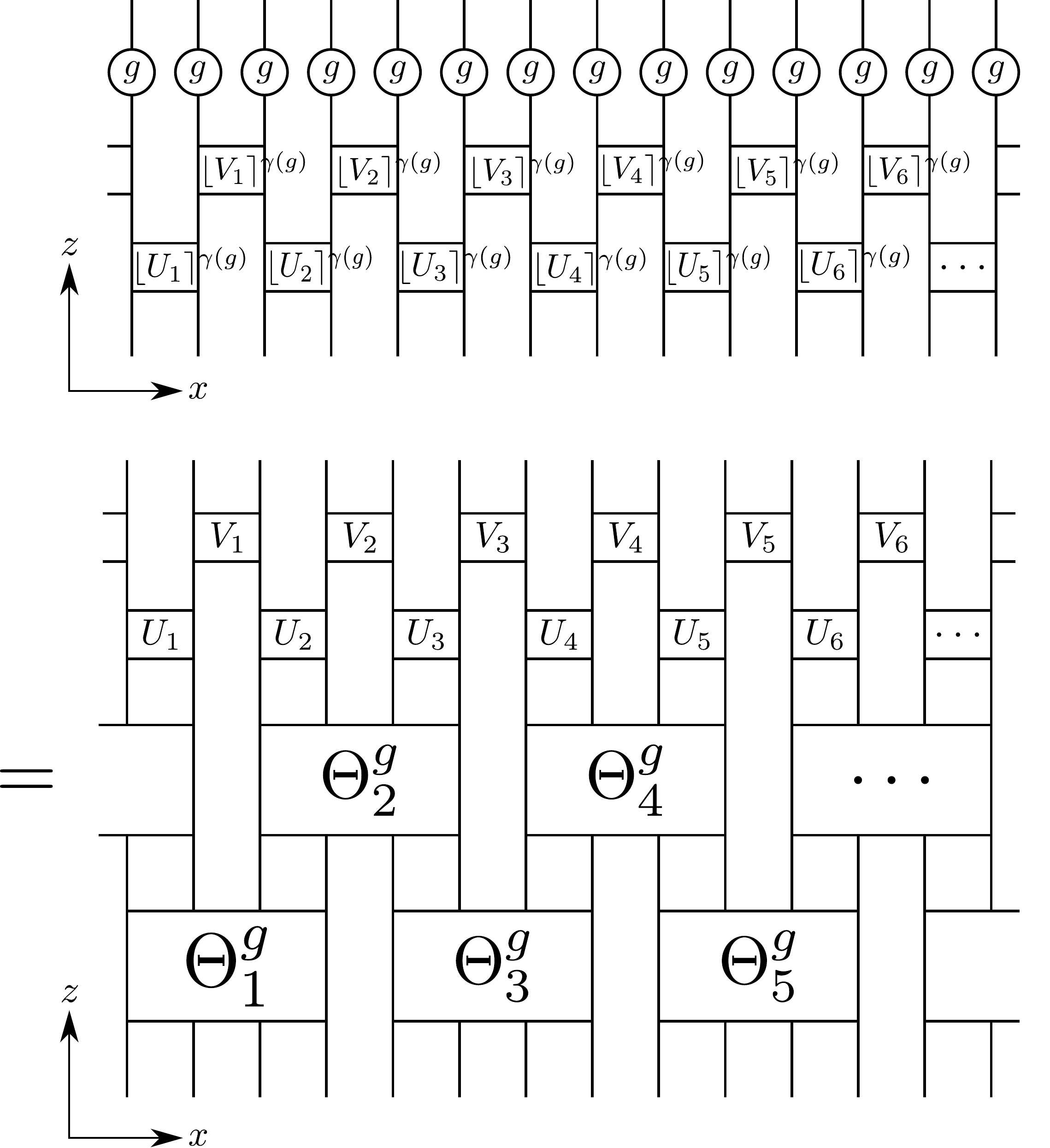} \; ,
\end{align}
\begin{align}
   \includegraphics[width = 0.45\linewidth,valign=c]{ww1-1.eps} \;\;=\nonumber\\  \includegraphics[width = 0.7\linewidth,valign=c]{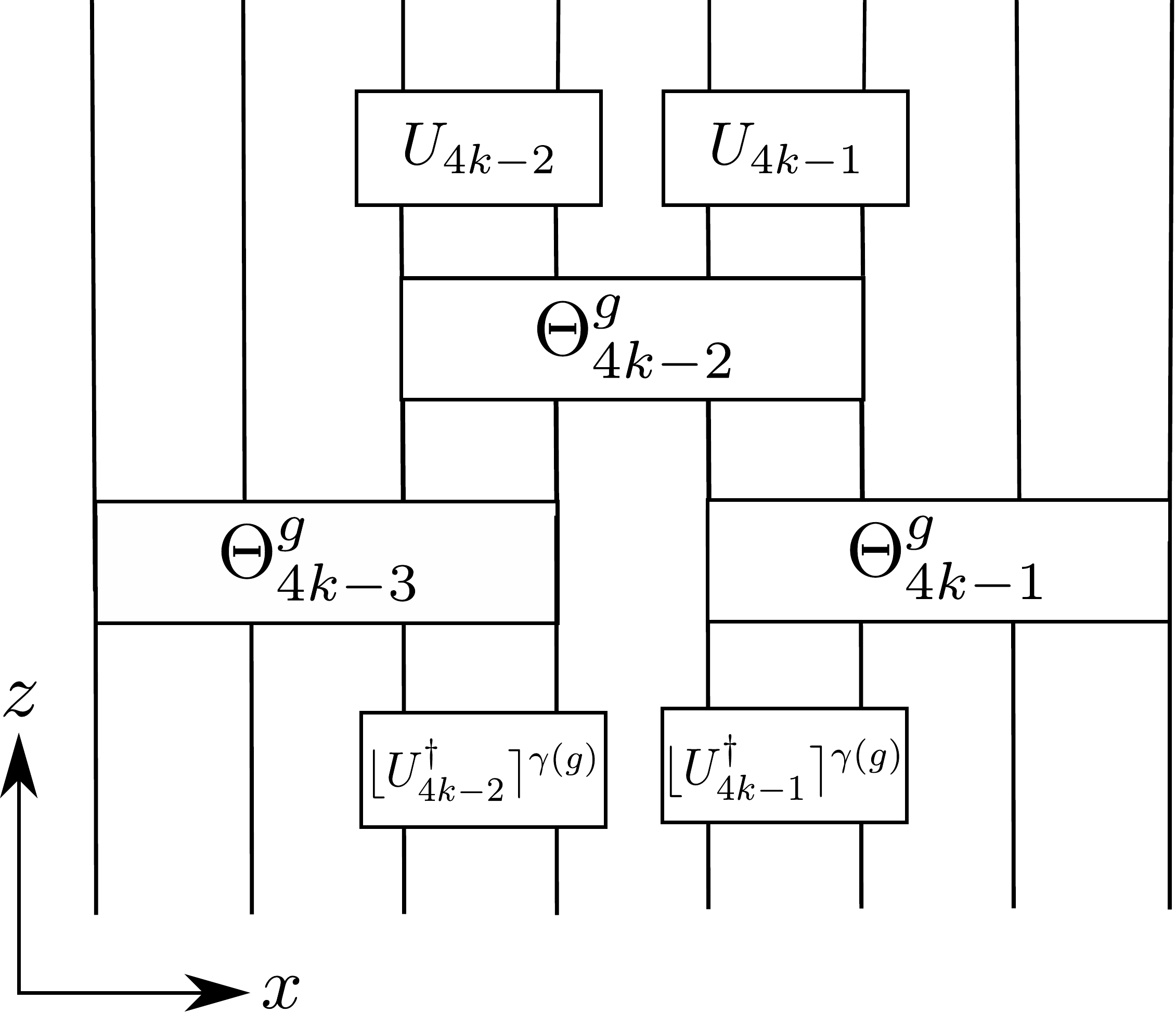}  \; \;,
\end{align}
and
\begin{align}
   \includegraphics[width = 0.45\linewidth,valign=c]{ww3-1-1.eps} \;\;=\nonumber\\  \includegraphics[width = 0.7\linewidth,valign=c]{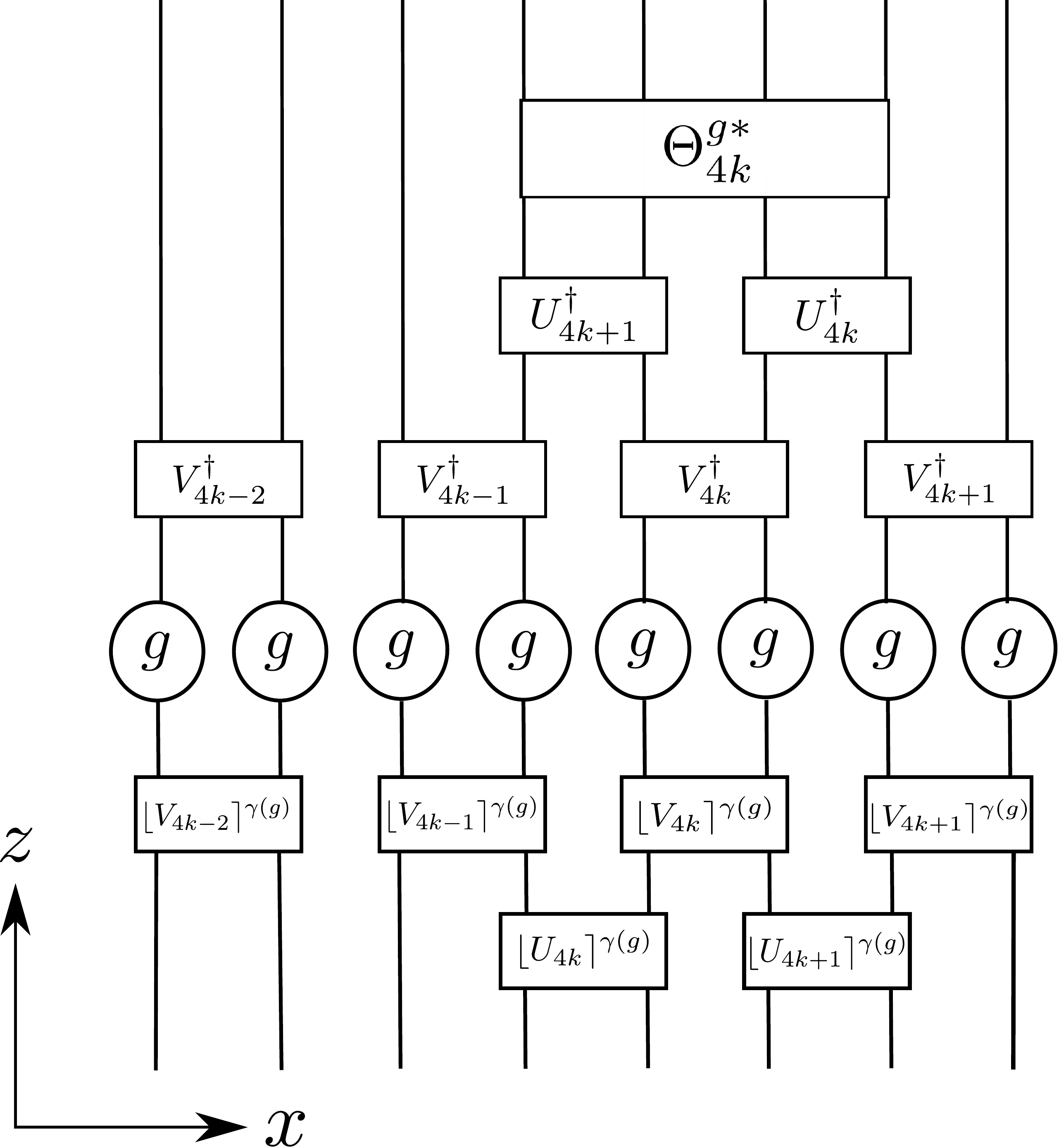} \; \; .
\end{align} 
Due to $\mathpzc{v}_g \lfloor \mathpzc{v}_h \rceil^{\gamma(g)} = \mathpzc{v}_{gh}$, we thus have $W_j(g) \lfloor W_j(h) \rceil^{\gamma(g)} = \beta_k(g,h) W_j(gh)$. Therefore, when approximating them by quantum circuits, we have (cf. Eq.~\eqref{blocking}) 
\begin{equation}\label{eq:qcrep}
  \cdots  \includegraphics[width=0.25\linewidth,valign=c]{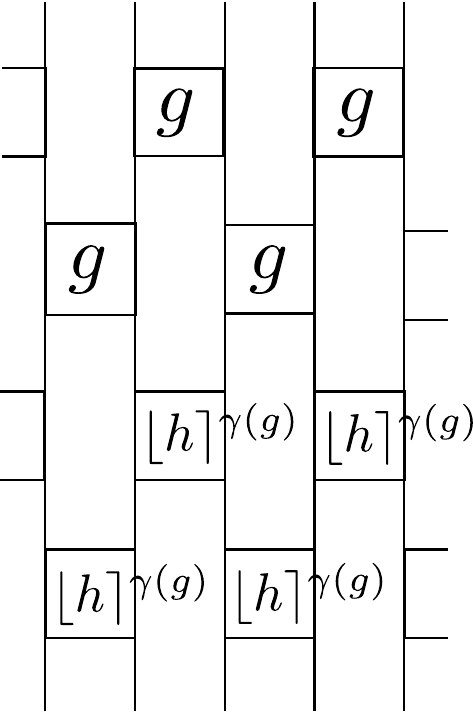}\cdots\;= \;\beta(g,h)\Bigg(\cdots\includegraphics[width=0.25\linewidth,valign=c]{drawing00.eps}\cdots\Bigg),
\end{equation}
$\beta(g,h) \in U(1)$. Using the same line of reasoning as in Sec.~\ref{sec:qcreps}, we obtain for a patch of the quantum circuit 
\begin{align}
   \includegraphics[width=0.6\linewidth,valign=c]{drawing7.eps} \;  \nonumber \\ =\;\includegraphics[width=0.6\linewidth,valign=c]{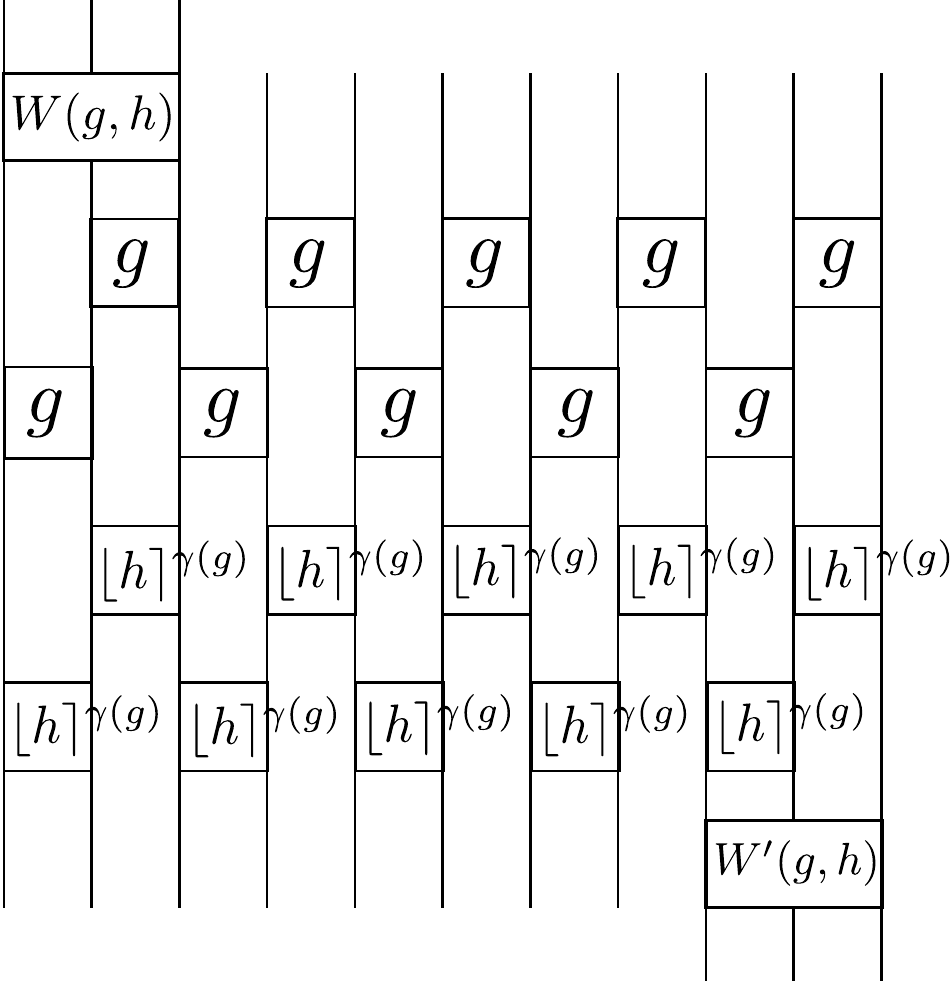} \; .
\end{align}
The concatenation of three group elements $g,h,k$ thus takes the form
\begin{align}
    \label{eq:ghk}
   \includegraphics[width=0.7\linewidth,valign=c]{drawing15.eps} \; \nonumber\\ =\;\includegraphics[width=0.7\linewidth,valign=c]{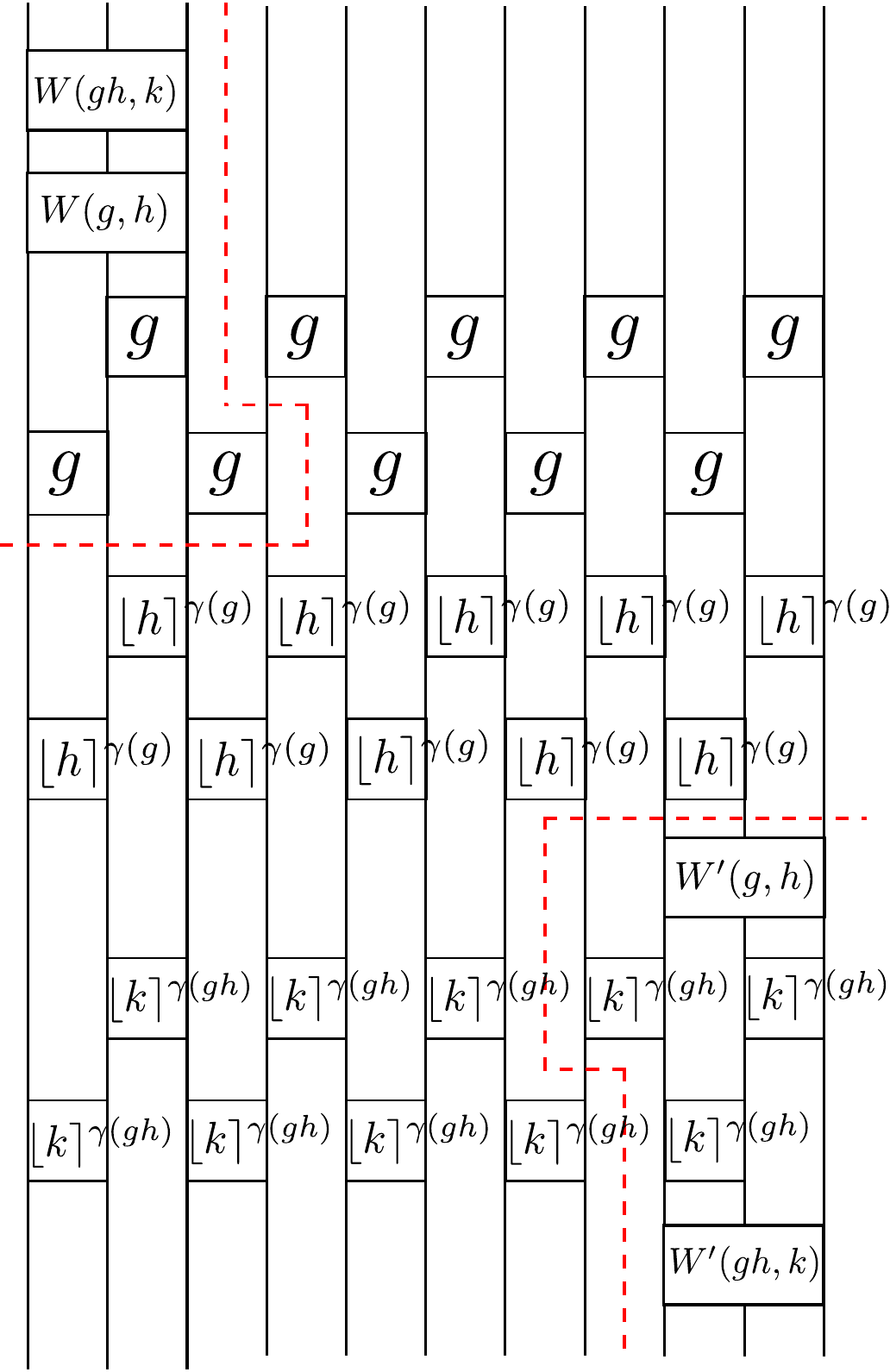} \nonumber\\ =\;\includegraphics[width=0.7\linewidth,valign=c]{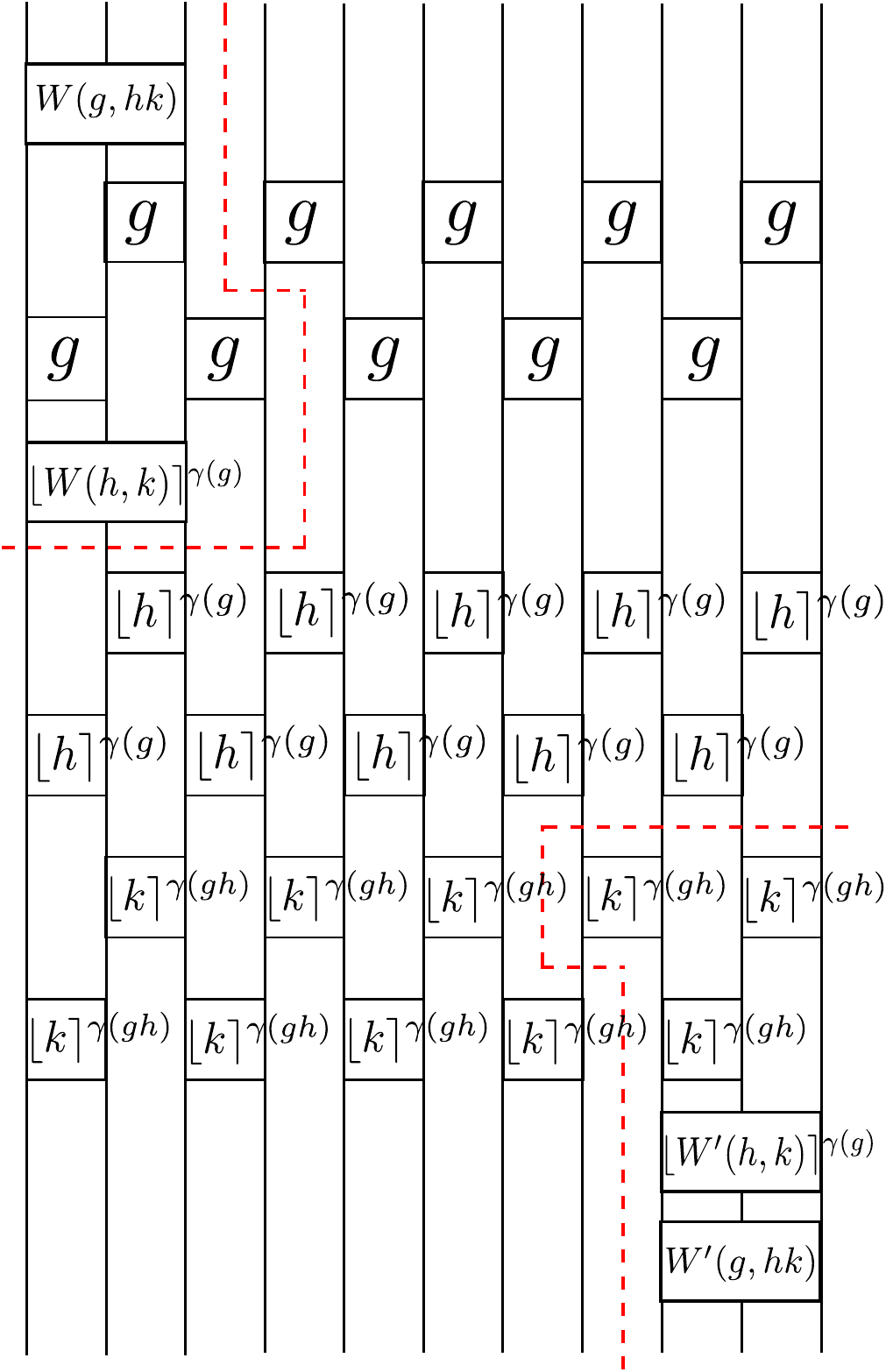}  \; .
\end{align}
This finally results in
\begin{align}
    &W(gh,k)W(g,h) X(g) \notag \\ 
    = &\alpha(g,h,k) W(g,hk)X(g) \lfloor W(h,k) \rceil^{\gamma(g)}.
    \label{eq:alpha_conj}
\end{align}
Thus, the gauge transformation $W(g,h) \rightarrow \chi(g,h) W(g,h)$, $\chi(g,h) \in U(1)$ corresponds to
\begin{align}
    \alpha'(g,h,k) = \alpha(g,h,k) \frac{\chi(g,hk) \lfloor \chi(h,k) \rceil^{\gamma(g)}}{\chi(g,h) \chi(gh,k)}, \label{eq:gauge_conj}
\end{align}
which is a redefinition of $\alpha(g,h,k)$ by a generalized 3-coboundary. Eq.~\eqref{eq:alpha_conj} implies
\begin{align}
        &W(ghk,l)W(gh,k)W(g,h)X(g)\lfloor X(h) \rceil^{\gamma(g)} \nonumber\\
         &= \alpha(g,h,k)W(ghk,l)W(g,hk)X(g)\lfloor W(h,k) \rceil^{\gamma(g)} \lfloor X(h) \rceil^{\gamma(g)} \nonumber\\
         &= \alpha(g,h,k)\alpha(g,hk,l)W(g,hkl)X(g)\lfloor W(hk,l) \rceil^{\gamma(g)} \times \notag \\ 
         & \ \ \ \ \lfloor W(h,k) \rceil^{\gamma(g)} \lfloor X(h) \rceil^{\gamma(g)} \nonumber\\
         &=\alpha(g,h,k)\alpha(g,hk,l) \lfloor \alpha(h,k,l) \rceil^{\gamma(g)} 
         W(g,hkl)X(g) \times \notag \\
         & \ \ \ \ \lfloor W(h,kl) \rceil^{\gamma(g)} \lfloor X(h) \rceil^{\gamma(g)} \lfloor W(k,l) \rceil^{\gamma(g)},
\end{align}
which in the shorthand notation of Eq.~\eqref{eq:shorthand} leads to 
\begin{equation}
   \includegraphics[width=0.32\linewidth,valign=c]{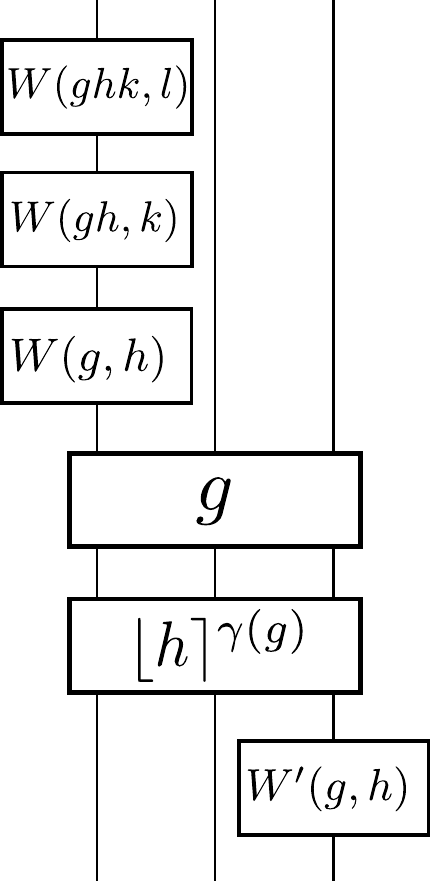} \;  =\;
   \begin{matrix}
   \alpha(g,h,k)\times\\\alpha(g,hk,l)\times\\ \lfloor \alpha(h,k,l) \rceil^{\gamma(g)} \times
   \end{matrix}
   \;\;\includegraphics[width=0.32\linewidth,valign=c]{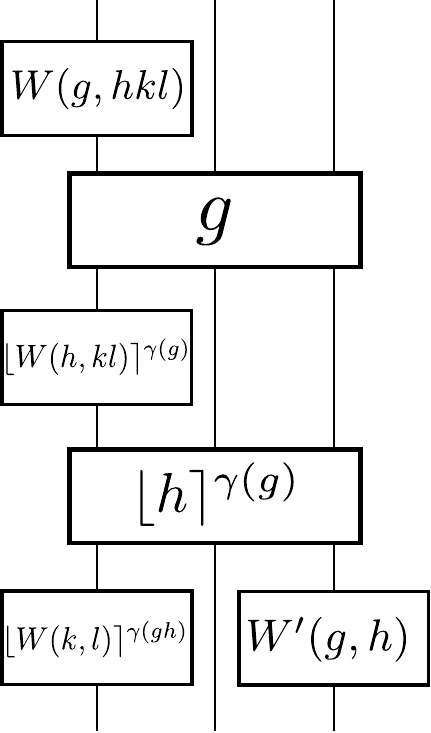}.
\end{equation}
Again, we can reach a similar relation using a different sequence of manipulations, 
\begin{align}
 \includegraphics[width=0.35\linewidth,valign=c]{drawing33_conj.eps} \; = \;\;\; \includegraphics[width=0.35\linewidth,valign=c]{drawing35.eps}\nonumber\\
 =\;\alpha(gh,k,l)\; \includegraphics[width=0.35\linewidth,valign=c]{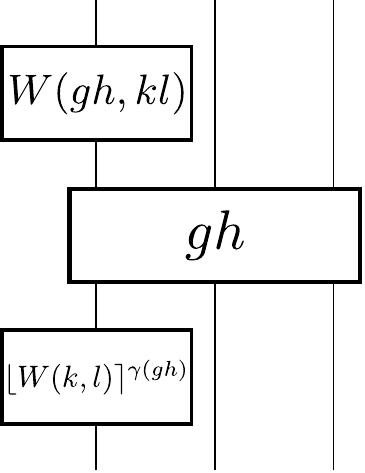} \;\;=\;\;\alpha(gh,k,l)\times \nonumber\\ \includegraphics[width=0.35\linewidth,valign=c]{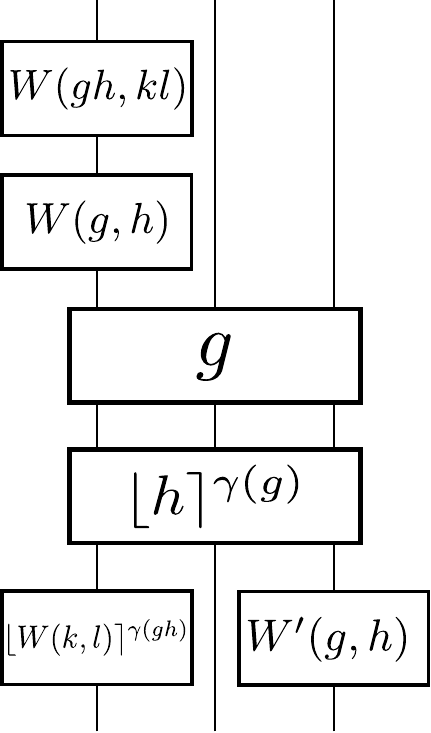}
 =\;\begin{matrix}\alpha(gh,k,l)\times\\\alpha(g,h,kl)\times\end{matrix}\; \includegraphics[width=0.35\linewidth,valign=c]{drawing34_conj.eps} \ .
\end{align}
Comparing the above two final expressions leads to 
\begin{equation}
    \frac{\alpha(g,h,k)\alpha(g,hk,l) \lfloor \alpha(h,k,l) \rceil^{\gamma(g)}}{\alpha(gh,k,l)\alpha(g,h,kl)} = 1.
\end{equation} 
Together with Eq.~\eqref{eq:gauge_conj}, this defines elements $\alpha(g,h,k)$ of the generalized third cohomology group. 

Keeping in mind that $W_j^g \lfloor W_j^h\rceil^{\gamma(g)} = \beta(g,h) W_j^{gh}$, one can use a similar line of reasoning as in Sec.~\ref{sec:wxwTopTrivial} to show that $W_{2k-1}^g \otimes W_{2k}^g$ and $W_{2k}^g \otimes W_{2k+1}^g$ are also topologically trivial in the current setup. This shows that for anti-unitary symmetries, all eigenstates of SPT MBL-like phases in two dimensions share the same topological label, which corresponds to an element of the generalized third cohomology group.

\section{Robustness to perturbations}\label{sec:robustness}

In the following, we show that the cohomology class is invariant under a local symmetry-preserving perturbation.  The discussion here is very similar to the argument for the one-dimensional  case~\cite{Thorsten,1DSPTMBL}.  Let us consider two FMBL Hamiltonians $H(0)$ and $H(1)$ connected via an FMBL-preserving path $H(\lambda)$ for a finite but large system. $H(\lambda)$ is required to continuously depend on the parameter $\lambda \in [0,1]$ and to be invariant under the symmetry. 
We represent the unitary which diagonalizes $H(\lambda)$ by a quantum circuit $U(\lambda)$, neglecting losses of topological properties over time scales of the order Eq.~\eqref{eq:time-scale}. The constituting unitaries of the best representation $U(\lambda)$ as defined in Sec.~\ref{sec:assumptions} might not be continuous as a function of $\lambda$. We now compare the topological properties of $U(\lambda - \epsilon)$ and $U(\lambda + \epsilon)$. In the limit $\epsilon \rightarrow 0$, the two unitaries might differ, but by assumption they equally well diagonalize the Hamiltonian $H(\lambda)$. 
As the system is FMBL for all $\lambda \in [0,1]$, we are allowed to alter the path by a small but non-zero local perturbation keeping the end points $H(0)$ and $H(1)$ fixed  (cf.~Section \ref{sec:1DMBL}). We choose the perturbation such that $H(\lambda)$ is analytic (which can be done since the Hamiltonian is bounded) and that degeneracies only appear at isolated points $\lambda_k \in [0,1]$. (Note that there are no protected degeneracies for abelian symmetries.) 
Hence, according to perturbation theory (up to corrections vanishing for $N \rightarrow \infty$) $U(\lambda-\epsilon)$ and $U(\lambda + \epsilon)$ can only differ by a permutation matrix in the limit $\epsilon \rightarrow 0$, i.e.,
\begin{align}
U(\lambda - \epsilon) P(\lambda) = U(\lambda + \epsilon).
\end{align}
$P(\lambda)$ is a permutation matrix whose non-vanishing entries have magnitude 1 and arbitrary phases. Since all approximate eigenstates encoded in $U(\lambda - \epsilon)$ have the same topological label, it must thus be the same as the one of the approximate eigenstates contained in $U(\lambda + \epsilon)$. We have thus shown that the cohomology class cannot change discontinuously along the path $\lambda \in [0,1]$. As it is discrete, it cannot change continuously either, demonstrating that it is unchanged along the evolution $H(\lambda)$ between two Hamiltonians $H(0)$ and $H(1)$ in the same SPT MBL phase. Choosing $H(1)$ as a small local perturbation of $H(0)$ (which always preserves FMBL) then shows that this topological index is robust to local symmetry-preserving perturbations. For truly randomly disordered systems and if the avalanche scenario is correct, the obtained topological properties persist on time scales of the order Eq.~\eqref{eq:time-scale}.

\section{Conclusion}\label{sec:conclusion}

We have shown that given a two-dimensional FMBL spin system invariant under an on-site symmetry, the SPT phases are classified by the elements of the third cohomology group of the symmetry group. 
Though we have only considered the bosonic case, the ideas and results from the one-dimensional version of this problem\cite{1DSPTMBL} imply that the classification is likely to be the same as for ground states also for fermionic systems.  

One potential direction for further research is to investigate whether the method presented here can be adapted to rigorously show the correctness of the classification of ground state SPT phases in two-dimensional gapped systems, which is currently an open problem~\cite{ciracOpenProblems}. 

Another potential direction for further investigation is the extension of our classification to three and higher dimensions, though an obvious difficulty would be the challenge of working with higher dimensional tensor network diagrams.  This case would also be particularly interesting as the cohomology classification is not complete in $d \geq 3$ dimensions~\cite{Xiong2018,Gaiotto2019}.

Finally, we note that the calculations presented here do not preclude the existence of additional topological indices in the 2D MBL case that do not exist for ground states.  Specifically, although we have shown that quantum circuits belonging to different elements of the third cohomology group cannot be continuously connected, we have not shown the converse of this statement.  In other words, we have not demonstrated the completeness of our classification, i.e., there may exist additional SPT MBL phases.

\section*{Acknowledgements}

This project has received funding from the European Union’s Horizon 2020
research and innovation programme under the Marie Sk\l{}odowska-Curie grant agreement No. 749150 and under the ERC Starting Grant No. 678795 TopInSy. AC received support by the EPSRC Grant No. EP/N01930X/1. TBW acknowledges support from the European Commission through the ERC Starting Grant No.678795 TopInSy. The contents of this article reflect only the authors' views and not the views of the European Commission.

\appendix

\section{Projective and gerbal representations} \label{app:endix}
An important idea in the study of one dimensional SPT phases is the relation between second cohomology groups and projective representations.  Here we briefly introduce gerbal representations, the third cohomology analogue, which are relevant in the context of two-dimensional SPT phases.  For an introduction to group cohomology as applied to the physics of SPT phases and definitions of cocycles and coboundaries, see Ref.~\onlinecite{Chen2013}. 

A projective representation satisfies 
\begin{equation}
\label{projrepmain}
    u(g)u(h) = \omega(g,h)u(gh),
\end{equation}
where $\omega(g,h)\in U(1)$ is the \textit{factor system}. Two projective representations are equivalent if their factor systems are related by
\begin{equation}
\label{projrepequiv}
    \omega'(g,h) = \frac{\chi(g)\chi(h)}{\chi(gh)}\omega(g,h),
\end{equation}
$\chi \in U(1)$, i.e. if they differ by a 2-coboundary.  We also observe that an expression like $u(g)u(h)u(k)$ can written in two different ways, namely
\begin{align}
u(g)u(h)u(k) = \omega(g,h)\omega(gh,k)u(ghk)\nonumber\\ = \omega(g,hk)\omega(h,k)u(ghk).
\end{align}
So we obtain the result that the factor system of a projective representation must satisfy the following rule:
\begin{equation}
\label{prep}    
\frac{ \omega(g,h)\omega(gh,k)}{\omega(g,hk)\omega(h,k)} = 1,
\end{equation}
i.e. it must be a 2-cocycle.
While elements of the second cohomology group $H^2(G,U(1))$ correspond to projective representations of $G$, elements of the third cohomology group $H^3(G,U(1))$ correspond to \textit{gerbal} representations\cite{gerbal} of $G$. A gerbal representation associates an operator $w(g,h)$ to each pair of group elements $g,h$ rather than to a single group element.  $w(g,h)$ does not act on a vector space, but on a space of functors of an abelian category.  (A \textit{category} consists of \textit{objects} linked by \textit{arrows}, also known as \textit{morphisms}.  There exists an identity arrow for each object, and a binary operation $\circ$ to compose arrows associatively.  An abelian category is one in which the objects and morphisms can be added.  A \textit{functor} is a homorphism between categories.)

First, we need to consider another, ``auxiliary" representation of $G$ that is a representation over an abelian category.  In that representation, $g\in G$ is associated with a functor $f_g$.  The functor $f_g$ essentially behaves as function $f_g(\;\,)$ with the peculiar feature that the composition $f_g\circ f_h (\;\,) = f_g(f_h(\;\,))$ does not live in the same space as $f_g$.  $f_e$ is the identity map ($e$ being the identity element of $G$). Since we cannot demand $f_g\circ f_h$ be equal to $f_{gh}$, we instead demand that they be related by an isomorphism.  The gerbal representation operator $w(g,h)$ is then defined to be the isomorphism map, i.e. $w(g,h): f_g\circ f_h\mapsto f_{gh}$.  When acting on compositions of many functors the action is defined to be $w(g,h): f_g\circ f_h \circ f_k \mapsto f_{gh}\circ f_k$, and $w(h,k): f_g\circ f_h \circ f_k \mapsto f_g\circ f_{hk}$, and so on.  From this we see that $w(a,b)$ commutes with $w(c,d)$ if $a,b,c,d$ are different group elements, since 
\begin{equation}
\begin{split}
w(a,b)w(c,d)(f_a\circ f_b \circ f_c \circ f_d) = \\ w(c,d)w(a,b)( f_a\circ f_b \circ f_c \circ f_d) = f_{ab}\circ f_{cd}
\end{split}
\end{equation}
Now let us see how $w(g,h)$ represents the group $G$.  For that, consider $ f_g\circ f_h \circ f_k$ which is isomorphic to $f_{ghk}$.  We can get $f_{ghk}$ by acting on $ f_g\circ f_h \circ f_k$ with either $w(gh,k)w(g,h)$ or $w(g,hk)w(h,k)$.  We can demand they be equal, or we can relax that slightly and instead demand
\begin{equation}
\label{gerbrepmain}
    w(gh,k)w(g,h) = \alpha(g,h,k)w(g,hk)w(h,k),
\end{equation}
for some $\alpha(g,h,k) \in U(1)$ acting as the factor system.

We can derive a relation similar to Eq.~\eqref{prep} that the $\alpha$ must satisfy, by considering that $f_{abcd}$ can be obtained by acting on $ f_a\circ f_b \circ f_c \circ f_d$ with either $w(a,b)w(ab,c)w(abc,d)$ or $w(c,d)w(b,cd)w(a,bcd)$.  We can go $w(a,b)w(ab,c)w(abc,d) \longrightarrow w(c,d)w(b,cd)w(a,bcd)$ by repeatedly applying Eq.~\eqref{gerbrepmain} via two different routes (i.e. start from the left, or start from the right).  In each case we obtain a prefactor, and we require both of them to be equal, leading to
\begin{equation}
\label{gerbalcocycle}
    \frac{\alpha(a,b,c)\alpha(a,bc,d)\alpha(b,c,d)}{\alpha(ab,c,d)\alpha(a,b,cd)} = 1,
\end{equation}
i.e. $\alpha$ is a 3-cocycle.  As with projective representations, two gerbal representations are equivalent if they differ by only a phase, i.e. $w$ and $v$ are equivalent if $v(g,h) = \chi(g,h)w(g,h)$ for some $\chi \in U(1)$.  The analogue of Eq.~\eqref{projrepequiv}, then, is that two gerbal representations are equivalent if their factor systems are related by a 3-coboundary,
\begin{equation}
\label{gerbalgauge}
    \alpha'(g,h,k) = \frac{\chi(g,h)\chi(gh,k)}{\chi(g,hk)\chi(h,k)}\alpha(g,h,k) \, .
\end{equation}

\bibliography{biblioMBL}{}

\end{document}